\documentclass{aa}  

\usepackage{graphicx}
\usepackage{txfonts}
\usepackage{lscape}
\usepackage{booktabs}
\usepackage{wasysym}
\usepackage{arydshln}
\usepackage[bookmarks=false,colorlinks=true,linkcolor=black,citecolor=cyan,filecolor=black,urlcolor=cyan]{hyperref}
\usepackage{array}

\newcolumntype{C}[1]{>{\centering\arraybackslash}p{#1}}

\begin{document} 

   \title{Do stellar-mass and super-massive black holes have similar dining habits?}

   \author{R. Arcodia
          \inst{1}
          \and
		  G. Ponti \inst{2,1}
          \and
          A. Merloni \inst{1}
          \and
          K. Nandra \inst{1}
          }

   \institute{Max-Planck-Institut f\"ur extraterrestrische Physik (MPE), Giessenbachstrasse 1, 85748 Garching bei M\"unchen, Germany\\
              \email{arcodia@mpe.mpg.de}
         \and
		INAF-Osservatorio Astronomico di Brera, Via Bianchi 46, I-23807 Merate (LC), Italy
         }

   \date{Received ; accepted }

 
  \abstract{Through the years numerous attempts have been made to connect the phenomenology and physics of mass accretion onto stellar-mass and super-massive black holes in a scale-invariant fashion. In this paper, we explore this connection at the radiatively-efficient (and non-jetted) end of accretion modes by comparing the relationship between the luminosity of the accretion disk and corona in the two source classes. Motivated by the apparently tight relationship between these two quantities in AGN, we analyse 458 RXTE-PCA archival observations of the X-ray binary (XRB) GX 339-4, using this object as an exemplar for the properties of XRBs in general. We focus on  the soft and soft-intermediate states, which have been suggested to be analogous to radiatively efficient AGN. The observed scatter in the $\log L_{disk}- \log L_{corona}$ relationship of GX 339-4 is high ($\sim0.43\,$dex) and significantly larger than in a representative sample of radiatively-efficient, non- or weakly-jetted Active Galactic Nuclei (AGN, $\sim0.30\,$dex). On the face of it, this would appear contrary to the hypothesis that the systems simply scale with mass. On the other hand we also find that GX339-4 and our AGN sample show different accretion rate and power-law index distributions, with the latter in particular being broader in GX 339-4 (dispersion of $\sim0.16$ cf. $\sim0.08$ for AGN). GX339-4 also shows an overall softer slope, with mean value of $\sim2.20$ as opposed to $\sim2.07$ for the AGN sample. Remarkably, once similarly broad $\Gamma$ and $\dot{m}$ distributions are selected, the AGN sample overlaps nicely with GX 339-4 observations in the mass-normalised $\log L_{disk}- \log L_{corona}$ plane, with a scatter of $\sim0.30-0.33\,$dex in both cases. This indicates that a mass-scaling of properties might hold after all, with our results being consistent with the disk-corona systems in AGN and XRBs exhibiting the same physical processes, albeit under different conditions for instance in terms of temperature, optical depth and/or electron energy distribution in the corona, heating-cooling balance, coronal geometry and/or black hole spin. 
}
   \keywords{}

   \titlerunning{}
   \authorrunning{R. Arcodia et al.} 
   \maketitle

%
\section{Introduction}

It is an intriguing and long-standing question whether the accretion flow around black holes (BHs) is similar among masses ($m=M_{BH}/M_{\astrosun}$) that are orders of magnitude apart, ranging from X-ray binaries \citep[XRBs, $m\sim5-15$;][]{Casares+2017:masses} to active galactic nuclei \citep[AGN, $m \sim 10^6-10^{10}$;][]{Padovani+2017:agnreview}. There are evidences that the phenomenology of how BHs accrete matter is indeed somewhat analogous \citep[e.g.,][]{Ruan+2019:analogous_struct} between AGN and XRBs, suggesting that they lack not only "hair" \citep{Ruffini+Wheeler1971:nohair} but also diversity in dining habits. What is yet to be established is the extent of this analogy between supermassive and stellar BHs and the impact of a different matter reservoir (i.e. density, temperature, ionisation and consequently pressure support) and surrounding environment (i.e. a single star with respect to the center of a galaxy) on the physical processes behind the observed phenomenology.

The description of the accretion flow structure around BHs is typically simplified with a more \citep[e.g.][]{Shakura+Sunyaev1973:accretion,Pringle1981:accr_disc} or less \citep[e.g.][]{Narayan+1994:ADAF} radiatively-efficient disk, which in the former case peaks in the soft X-rays for XRBs and in the UV for AGN, a "corona" \citep{Galeev1979:coronae,Haardt+Maraschi1991:twophase1,Svensson&Zdziarski94:corona_f}, responsible for the X-ray emission between fractions and hundreds of keV in both systems, and, possibly, a relativistic jet \citep{Fender2001:jets_xrb,Blandford+2019ARA&A:jets_AGN}. Similarities in a scale-invariant fashion between XRBs and AGN have always been hunted for and they were found, for instance, in the break of the power spectrum hence in the X-ray variability amplitudes \citep[e.g.][]{Uttley+2002:PSD_AGN_1cmp,McHardy+2004:PSD_NGC4051_comp,Uttley+2005:variabSey,McHardy+2006:variab} or by exploring possible correlations among observational proxies of these different spectral components \citep{Heinz+2003:FP_theory}.

In particular, an evidence of a common accretion-ejection paradigm emerged from the so-called "Fundamental Plane" \citep{Merloni+2003:fund_plane}, which connects radio and X-ray luminosity with the BH mass: low-luminosity \citep[which are found to be more radio loud\footnote{See \citep{Hao+2014:RLdef} for a few definitions of radio loudness, usually defined as a flux or luminosity ratio between the radio with respect to another band.}, e.g.][]{Ho2002:R_vs_mdot,Sikora+2007:RL_vs_mdot} AGN were shown to be scaled-up hard-state XRBs \citep[e.g.,][for a review on XRBs states]{Belloni+2016:review} with a prominent jet component and a radiatively-inefficient accretion flow \citep{Merloni+2003:fund_plane,Falcke+2004:unify} while moderately- and high-accreting AGN \citep[both combined spanning $\lambda_{edd}=L/L_{edd}\simeq0.02-1$; e.g.][]{Noda2018:CLAGN_mdot0.02,VahdatMotlagh+2019:transitions_XRBs} were connected to XRBs in the soft states \citep[SS;][]{Maccarone+2003:soft_vs_modAGN} and soft-intermediate states \citep[SIMS;][]{Sobolewska+2009:alphaox_GBH}. This picture has been confirmed and expanded by \citet{Koerding:2006:HIMS_RL} including the analogy between hard-intermediate states in XRBs, where there is some disk contribution but the radio jet is present as well, and radiatively-efficient radio-loud quasars; and by analysing AGN caught in the (very slow) act of transitioning between these states \citep[e.g.,][]{Marchesini+2004:transi,Marecki+2011:agn_trans}. Further, this scale-invariant accretion-ejection scenario has been proven to hold using simultaneous UV and X-ray observations of AGN \citep{Svoboda+2017:simult}.

We here aim to improve on this connection in the radiatively-efficient (and non-jetted) end of accretion modes, comparing AGN and XRBs in such regime and test whether they share the same phenomenology and physics based on their disk-corona energetic output. Regarding the phenomenology, the disk-corona connection was studied in AGN for decades \citep[e.g.,][and references therein]{Arcodia+2019:LxLuv} via the X-ray loudness parameter $\alpha_{OX}$ \citep{Tananbaum+1979:alphaOX} and it was also tested in XRBs with an analogous proxy \citep[e.g.][]{Sobolewska+2009:alphaox_GBH}. For the case of AGN, there are many indications that the physical scatter in X-ray coronae for a given disk luminosity (once excluding variability and non-simultaneous observations) is very small \citep[$\lessapprox0.19-0.20$\,dex;][]{Lusso&Risaliti2016:LxLuvtight,Chiaraluce+2018:dispandvariab_Lx_Luv}.
However, in SSs and SIMSs of XRBs (i.e. supposedly scaled-down radiatively-efficient AGN) the relative strength of the X-ray corona with respect to the disk shows a large scatter in a relatively narrow range of soft X-ray monochromatic (i.e. disk) flux \citep{Sobolewska+2009:alphaox_GBH}. What is more, one should keep in mind that when a single XRB is used, any possible issue arising from non-simultaneity of the data probing the two components is circumvented and there is no additional scatter coming from a mixed bag of masses, distances and inclinations. Then, under the assumption of a scale-invariant accretion paradigm, one would rather expect that scatter in XRBs to be smaller.

A more thorough study of the source of the scatter in the XRBs disk-corona plane may help in shedding light on the putative analogy between accretion flows around stellar-mass and supermassive BHs. 
This highlights the importance of our work, since the $\alpha_{OX}-L_{disk}$ relation in AGN revealed itself to be a powerful tool to study the physics of accretion \citep{Lusso&Risaliti2017:toymodel,Kubota+Done2018:model_lx_luv,Arcodia+2019:LxLuv} up to high redshift \citep{Nanni+2017:highz,Vito+2019:highz,Salvestrini+2019:highzlxluv} and its scatter represents an important factor in the now rejuvenated role of quasars as cosmology probes \citep{Risaliti&Lusso2015:Hubble_diagram,Risaliti+2019:hubble19,Melia2019:cosmoqso,Khadka2019:cosmoqso,Lusso+2019:tensioncosmo,Yang+2019:tensioncosmo,Velten+2019:cosmoQSO,Zheng+2020:qsocosmo}. 

In summary, in this work we tested the disk-corona emission in SSs and SIMSs of the X-ray binary GX 339-4 (Section~\ref{sec:gx339}), with a more quantitative focus on the phenomenology and physics hidden in the scatter of the $L_{disk}-L_{cor}$ relation (Section~\ref{sec:relation_binaries}) to investigate how it compares to the one that we observe in their putative scaled-up analogous, namely AGN in their efficient accretion mode (Section~\ref{sec:relation_XRB_vs_AGN} and~\ref{sec:discussion}).


\section{Our sandbox: GX339-4}
\label{sec:gx339}

GX 339-4 was discovered almost five decades ago \citep{Markert+1973:discovery} and it is one of the most studied Galactic BH candidates \citep{Zdziarski+1998:spec98,Hynes+2003:dyn}. It has since undergone several X-ray outbursts, which were also simultaneously detected and monitored at almost all wavebands \citep[e.g.,][]{Homan+2005ApJ:IR_2002,Coriat+2009:OIR, CadolleBel+2011:multiw_2010,Dincer+2012:OIR_11_decay,Buxton+2012:OIR_02-10,Corbel+2013:radio_X_15yrs,Vincentelli+2018:IR_X_2008}, with a particularly good coverage during the Rossi X-ray Timing Explorer (RXTE) era. 

We are interested in comparing XRBs in regions of the q-plot \citep[i.e. hardness-intensity or -luminosity diagram, HID, or HLD; e.g.,][]{Fender+2004:HID} where the analogy with bright radiatively-efficient quasars might hold \citep[e.g.][]{Maccarone+2003:soft_vs_modAGN,Koerding:2006:HIMS_RL,Sobolewska+2009:alphaox_GBH}. We conservatively selected both SS and SIMS states, namely including also spectra in which the hard component can be almost as strong as the soft component. This selection criterion was then confirmed a posteriori with our control AGN sample (Section~\ref{sec:relation_XRB_vs_AGN} and~\ref{sec:soft}).

We selected the 2002-2003, 2004-2005, 2006-2007 and 2010-2011 outbursts, which are the ones with the highest coverage in the RXTE archive for GX 339-4. We referred to the extensive literature on GX 339-4 to select SSs and SIMSs in the above-mentioned outbursts (hereafter SS02, SS04, SS07 and SS10, respectively), from both spectral (i.e. low hardness-ratio) and timing analysis (i.e. low fractional rms) constraints: in SS02 we included all RXTE observations between MJD=52411.60 and 52694 \citep{Belloni+2005:OTB02}, 116 in total; SS04 started at MJD$\sim$53235 \citep{Belloni+2006:begSS04} and ended at MJD$\sim$53456 \citep[using color constraints from][]{Dunn+2008:SS04}, with 78 observations in total; our SS07 selection started in MJD=54147 and ended around MJD=54230, including only observations marked as high-SSs or SIMSs from timing analysis constraints \citep[][]{Motta+2009:SS07}, 69 in total\footnote{\citep{Motta+2009:SS07} performed the timing analysis up to MJD=54208. We included all observations up to MJD=54230 with a cut in HR corresponding to the value at the start of the SIMS as reported by \citep{Motta+2009:SS07}.}; SS10 contains observations within MJD$=55331-55565$ \citep{Debnath+2010:ss10_part,Nandi+2012:SS10}, 195 in total. This adds up to 458 observations, covering almost 10 years of RXTE data. 

\section{Data Analysis}
\label{sec:analysis}
RXTE observations during SS02, SS04, SS07 and SS10 include data from the Proportional Counter Array \citep[PCA,][]{Jahoda+1996:PCA}. Data from the High Energy X-ray Timing Experiment \citep[HEXTE,][]{Rothschild+1998:HEXTE} were not included in the analysis, since the background of the instrument dominates over the (faint) hard spectral component in the SSs and SIMSs. In the PCA, we analysed only energies in the range $3-25\,$keV, where the Effective Area of the instrument is at its best.
We reduced the selected observations following the standard procedure outlined in the RXTE cookbook\footnote{\href{https://heasarc.gsfc.nasa.gov/docs/xte/recipes/cook_book.html}{RXTE cookbook}}. PCA spectra were extracted from the top layer of the Proportional Counter Unit (PCU) 2, which is reported to be the best calibrated. A systematic uncertainty of $0.5\%$ was added to all channels to account for calibration uncertainties. 

In this work, the spectral analysis on each individual observation was performed using v2.8 of the Bayesian X-ray Analysis software \citep[BXA,][]{Buchner+2014:BXA}\footnote{\href{https://github.com/JohannesBuchner/BXA/}{https://github.com/JohannesBuchner/BXA/}}, which connects a nested sampling algorithm \citep{Feroz+2009:multinest} with a fitting environment. For the latter, we used \texttt{Sherpa} v4.11.0 \citep{Freeman+2001:sherpa,Doe+2007:sherpa_py} for the spectral fits and \texttt{XSPEC} v12.10.1 \citep{Arnaud+1996:xspec}, with its Python oriented interface\footnote{\href{https://heasarc.gsfc.nasa.gov/docs/xanadu/xspec/python/html/index.html}{pyXSPEC}}, for the flux calculations and spectral simulations (see Appendix~\ref{sec:appendix_robustness}). 

Our approach was to first model for each observation the PCA background spectrum empirically in the $3-25\,$keV band with a mixture of broken power-law and Gaussian components. Once satisfactory residuals were obtained, this background model was included as a model component in the spectral fit of source plus background spectra, with a single free background parameter being the power-law normalisation. This ensured a more solid statistical treatment of the counts \citep[e.g.][]{Loredo1992:springerbook,VanDyk:2001:lowcounts}, since all the background-subtracted spectra would have had several bins with negative counts close to the high-energy end of the adopted $3-25\,$keV range. Moreover, the free background normalisation was allowed to span along the $3\sigma$ errors of the value obtained in the background fit alone, which excludes an overestimation of our knowledge of the PCA background. Unless stated otherwise, we quote and plot median values with 16th and 84th percentiles of the BXA posterior distributions.

\subsection{The spectral model}
\label{sec:model}

\begin{figure}[tb]
	\centering
	\includegraphics[width=\columnwidth]{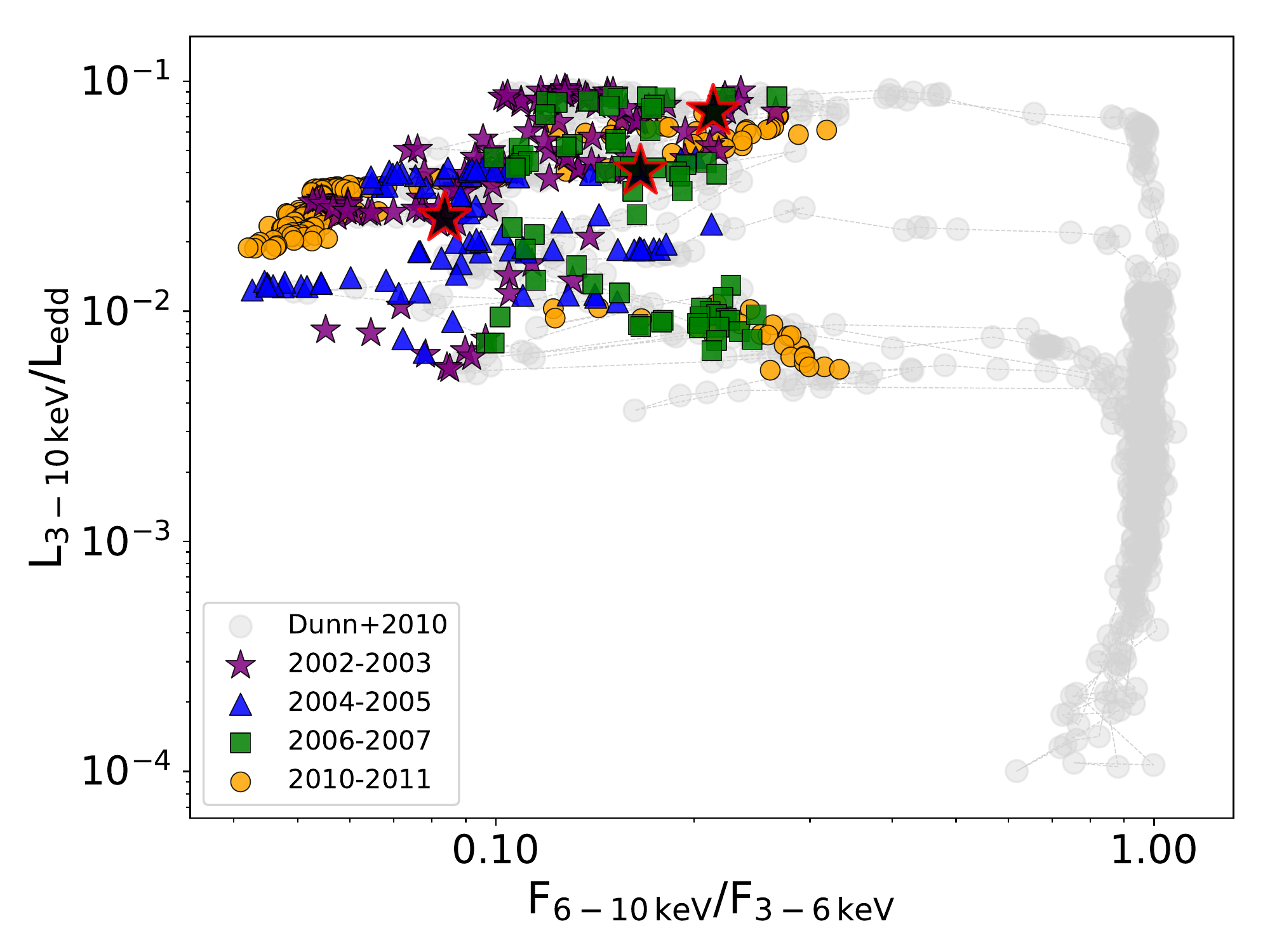}
	\caption{Hardness-luminosity diagram (HLD) of GX 339-4 in its four outbursts used in this analysis (i.e. 2002-2003, 2004-2005, 2006-2007 and 2010-2011). \emph{Grey} circles are archival data taken from \citet{Dunn+2010:global}, to which our own data points for SS02, SS04, SS07 and SS10 are superimposed, color coded following the legend in the Figure. Black symbols with red contours refer to the examples in Fig.~\ref{fig:spectra_examples}.}
	\label{fig:qplot}
\end{figure}

Each X-ray spectrum was fit with a source model consisting of an accretion disk \citep[\texttt{DISKBB};][]{Mitsuda1984:diskbb} plus a Comptonisation component \citep[\texttt{NTHCOMP};][]{Zdziarski+1996:nthcomp,Zycki+1999:nthcomp}\footnote{\label{note:Ecut}The model \texttt{compps} \citep{Poutanen+1996:compps} is reported to be more accurate for Comptonisation \citep[e.g. see comparison figures in][]{Niedewiecki+2019:models_for_rel}, although the major problems of \texttt{NTHCOMP} lie in the estimate of the high-energy cutoff. First, our analysis is restricted in the $3-25\,$keV band; moreover, we leave the cutoff energy parameter free to vary and we are not interested in using the (likely unconstrained) fit values. We are confident that the impact on our analysis would be minor and we use \texttt{NTHCOMP} for its simplicity.}, with the complex features of the reflection spectrum approximated with a Gaussian component\footnote{We also explored the \texttt{laor} model \citep{Laor1991:laor} to exclude that this simplified treatment of the reflection features had a significant impact on our results (see Section~\ref{sec:bias_refl}).}. The source model was then absorbed by a Galactic column density free to vary in a $\pm15\%$ uncertainty interval \citep[see][Sec. 5.7.2]{Arcodia+2018:nh15} around the tabulated value including the molecular component \citep[$N_H=5.18\times10^{21}\,$cm$^{-2}$;][]{Willingale+2013:nhmol}. This source model corresponds to \texttt{xstbabs*(xsdiskbb+xsnthcomp+xsGaussian)} in \texttt{Sherpa} and \texttt{tbabs*(diskbb+nthcomp+Gaussian)} in \texttt{XSPEC}, to which a complex background spectral model was added, with only a free normalisation parameter.

\texttt{DISKBB} free parameters are the temperature at the inner disk radius $T_{in}$ and the normalisation, which is a function of the inner radius $R_{in}$, the distance $d$ of the source and inclination $i$ of the disk. The \texttt{NTHCOMP} free parameters were the asymptotic photon index $\Gamma$, the normalisation and the electron temperature $kT_e$ (see footnote~\ref{note:Ecut}), while we tied the seed photons temperature (i.e. the low-energy rollover) to the typical disk temperature as fit by \texttt{DISKBB}. The multi-color black-body approximation in the \texttt{DISKBB} model was chosen over more rigorous accretion disk models as, for instance, \texttt{BHSPEC} \citep{Davis&Hubeny2006:bhspec} due to its simplicity and easier coupling with the Comptonisation emission of \texttt{NTHCOMP}. In Appendix~\ref{sec:bhspec_test} we further discussed our choice and we presented our tests with \texttt{BHSPEC} performed in order to verify the impact of a different disk model on our results. All the parameters in the Gaussian line model were left free to vary within the following intervals: a line with $E_{line}=6.4-6.966\,$keV, width $\sigma_{line}=0-1.5\,$keV and free normalisation.

In BXA, we adopted uninformative priors for all the 10 free parameters\footnote{In a few rare cases we observed a bimodality in some of the Photon Index posterior distributions, or a posterior pegged at one of the extremes ($\Gamma=1-4$). For those, a broad Gaussian prior centered at the peak of the observed distribution with a sigma of $0.3-0.5$ was adopted. This avoids un-physical posterior distributions without strongly affecting the spectral fit, being the prior very broad.}. The Bayesian methodology allowed us to use this complex model for all spectra, even for the few ones in which the Gaussian component might not have been needed. In this cases, the procedure would yield a flat posterior distribution for (i.e. a correct marginalisation over) the free parameters of that component.

\subsection{Results of the spectral fits}
\label{sec:fit_results}
\begin{figure*}[tb]
	\centering
	\includegraphics[width=0.663\columnwidth]{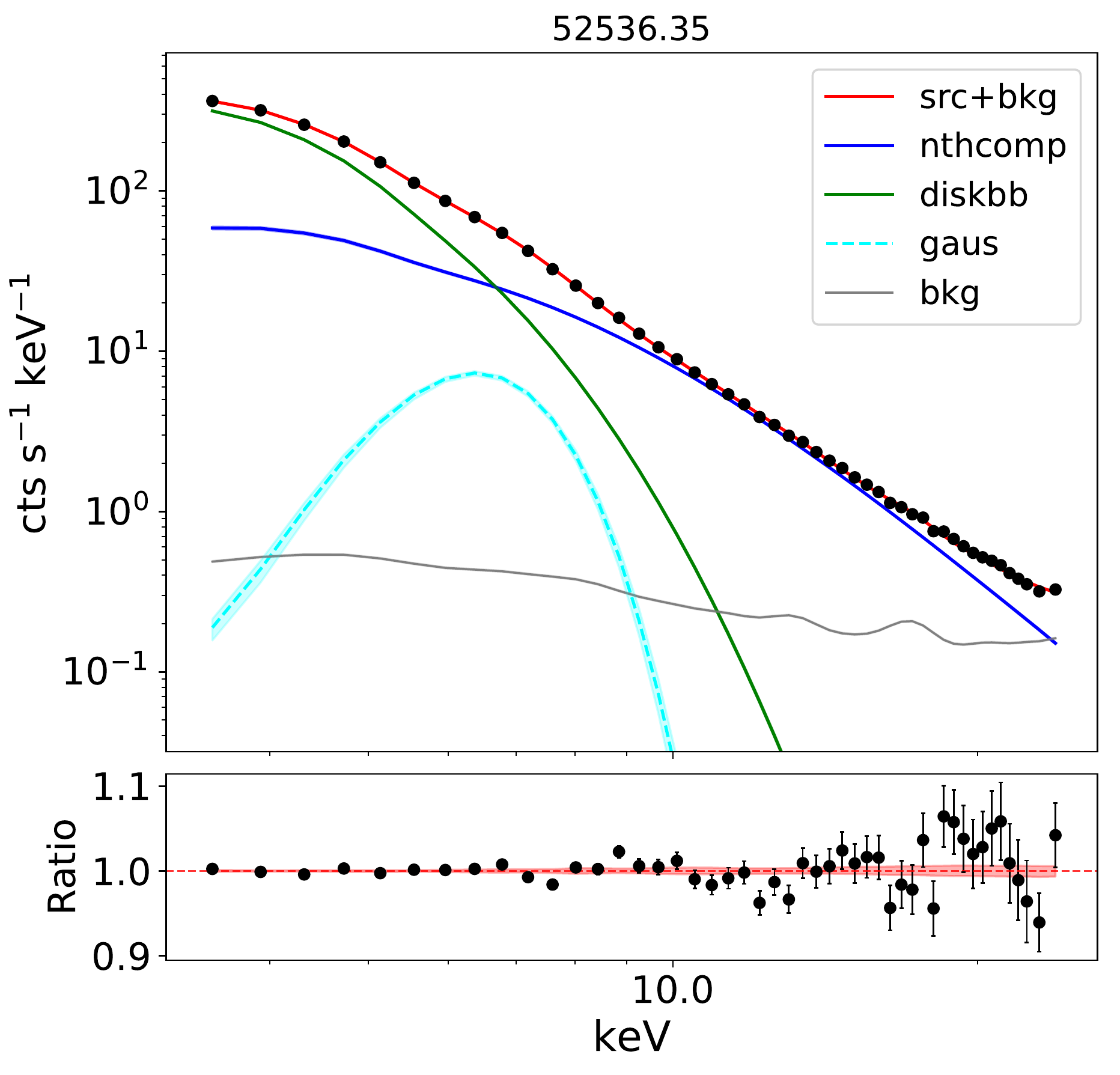}
	\includegraphics[width=0.663\columnwidth]{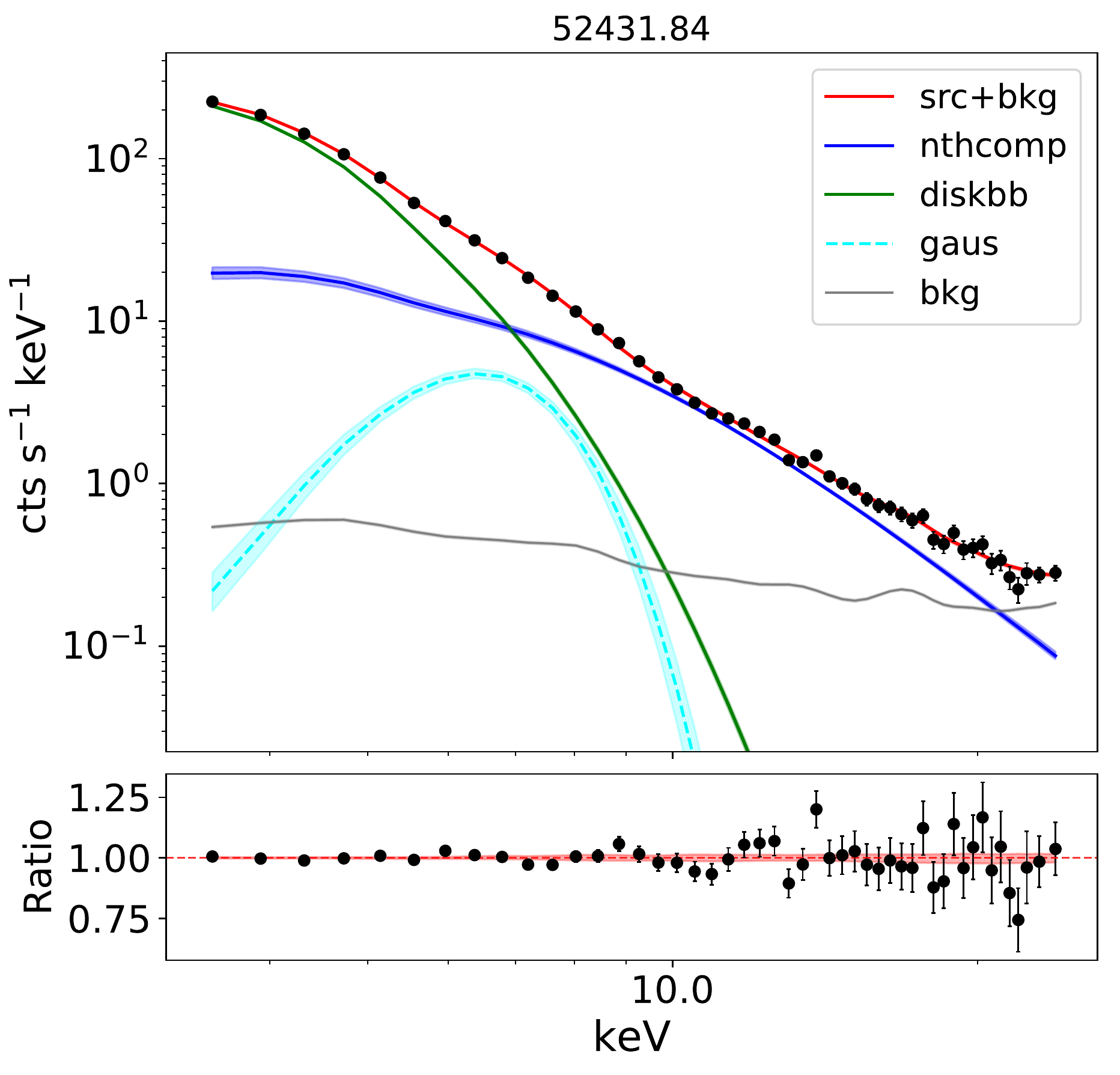}	\includegraphics[width=0.663\columnwidth]{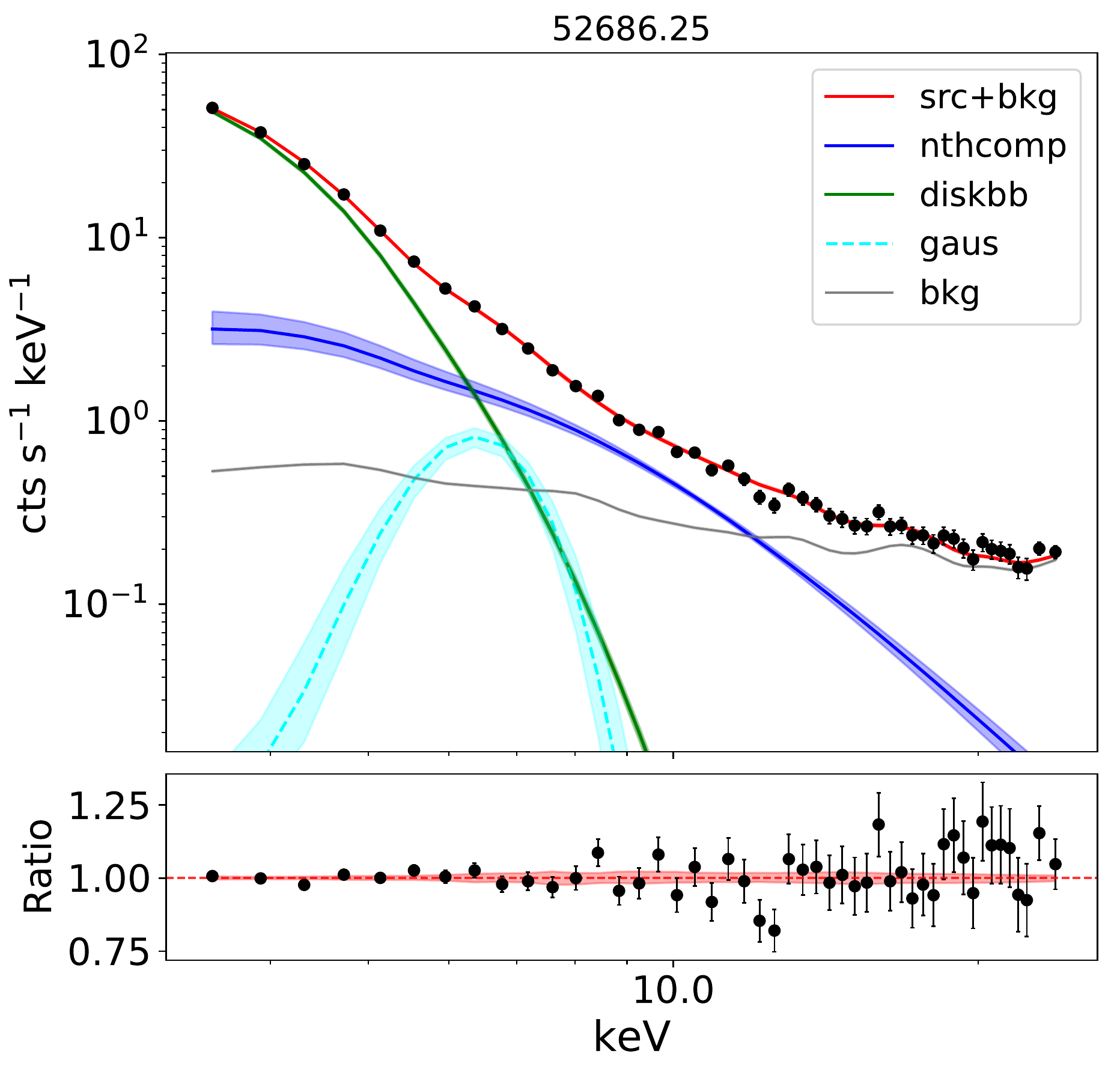}	
	\caption{Three examples of source plus background spectra (black dots, error bars included), with related data-model ratios in the lower panels. The three observations were selected taking the 84th percentile, median and 16th percentile of the total $3-25\,$keV flux distribution of the full XRB sample used in Section~\ref{sec:relation_XRB_vs_AGN}, shown from left to right respectively. They correspond to the three bigger black symbols in Fig.~\ref{fig:qplot}, going downwards in the q-plot. All additive model components are shown and defined in the legend, with the total source plus background model shown in red. For each component, the solid lines represents the median of the model distribution computed from the posteriors of the fit parameters (with 16th-84th percentile colored contours around them, in some cases smaller than the thickness of the line).}
	\label{fig:spectra_examples}
	\end{figure*}

The overall behavior of GX 339-4 in its SSs and SIMSs is studied with 458 observations. We show in Fig.~\ref{fig:qplot} the HLD, in which the complete four outbursts are shown in grey \citep[data from][]{Dunn+2010:global} and our data from SS02, SS04, SS07 and SS10 are represented with purple stars, blue triangles, green squares and yellow circles, respectively. Three examples of source plus background spectra are reported in Fig.~\ref{fig:spectra_examples}, selected taking the 84th percentile, median and 16th percentile of the total $3-25\,$keV flux distribution of the total XRB sample used in Section~\ref{sec:relation_XRB_vs_AGN}. They correspond to the three bigger black symbols in Fig.~\ref{fig:qplot}, going downwards in luminosity along the HLD.

\begin{figure*}[tb]
	\centering
	\includegraphics[width=2\columnwidth]{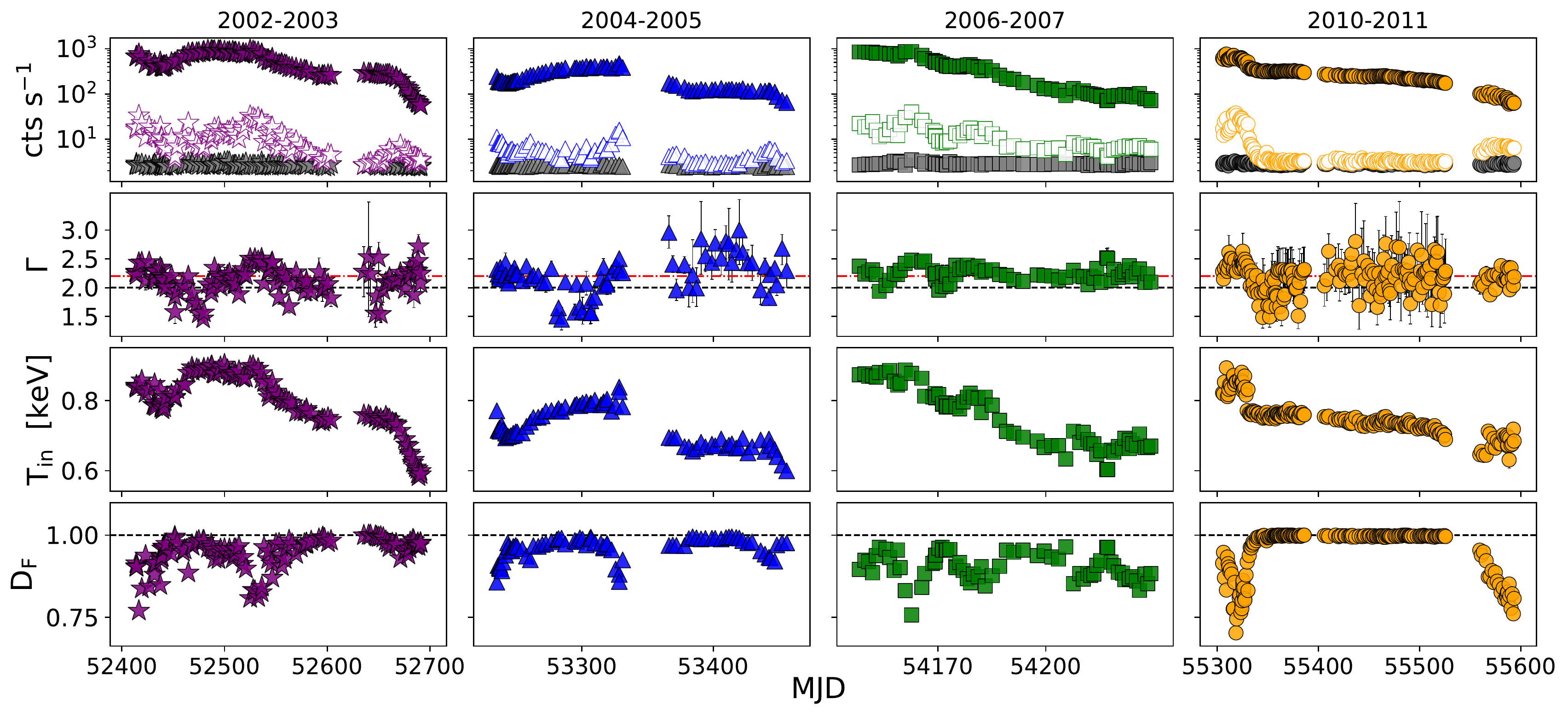}
	\caption{The evolution of the $3-25\,$keV count rate, of the fit Photon Index, disk temperature and of the disk fraction $D_F$ (see Eq.~\ref{eq:Df}) is shown along the four different outbursts (same color coding and symbols as in Fig.~\ref{fig:qplot}). In the top panels, filled colored symbols represent the total (source plus background) $3-25\,$keV count rates, the colored empty symbols the total count rates in the $10-25\,$keV band and the grey symbols the background count rates in the $10-25\,$keV band. 
	In the middle-top panel, a black dashed line at $\Gamma=2$ is shown to guide the eye, whereas the red dot-dashed line highlights the median $\Gamma=2.20$ of the whole XRB sample used in Section~\ref{sec:relation_XRB_vs_AGN}.}
	\label{fig:stuff_vs_mjd}
\end{figure*}

In Fig.~\ref{fig:stuff_vs_mjd} we show the evolution of spectral quantities with time along the four outbursts, namely the source plus background count rate in the $3-25\,$keV, the X-ray Photon Index $\Gamma$, the disk temperature $T_{in}$ (i.e. a proxy of the mass accretion rate) and the disk fraction $D_F$. The latter is defined as in \citet{Dunn+2010:global}:
\begin{equation}
\label{eq:Df}
D_F=\frac{F_{0.001-100\,keV,\,disk}}{F_{0.001-100\,keV,\,disk} + F_{1-100\,keV,\,cor}}
\end{equation}  

The \texttt{NTHCOMP} parameter $kT_e$ was as expected, given the RXTE-PCA bandpass, completely unconstrained (see also footnote~\ref{note:Ecut}). It was left free to vary to avoid, as much as possible, systematics on the estimate of $\Gamma$, which with the used nested sampling algorithm is marginalized over the unconstrained $kT_e$. Thus, the uncertainties on $\Gamma$ should include our lack of knowledge on the corona temperature.

We refer to Appendix~\ref{sec:appendix_robustness} for our spectral simulations and posterior predictive checks, that were made to investigate the robustness of our fit results.


\section{The disk-corona relationship in GX339-04}
\label{sec:relation_binaries}

\begin{figure}[tb]
	\centering
	\includegraphics[width=0.724\columnwidth]{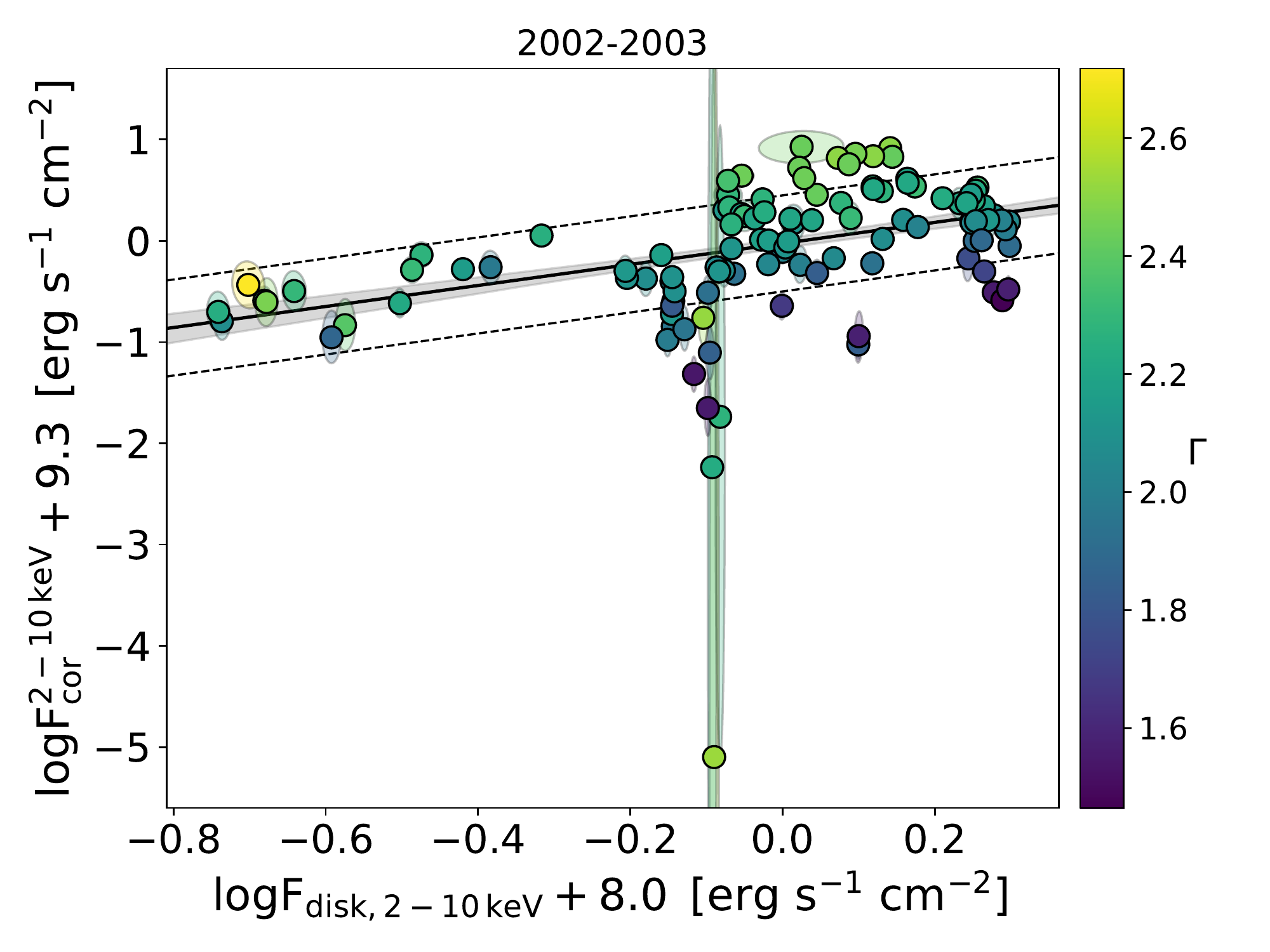}
	\includegraphics[width=0.724\columnwidth]{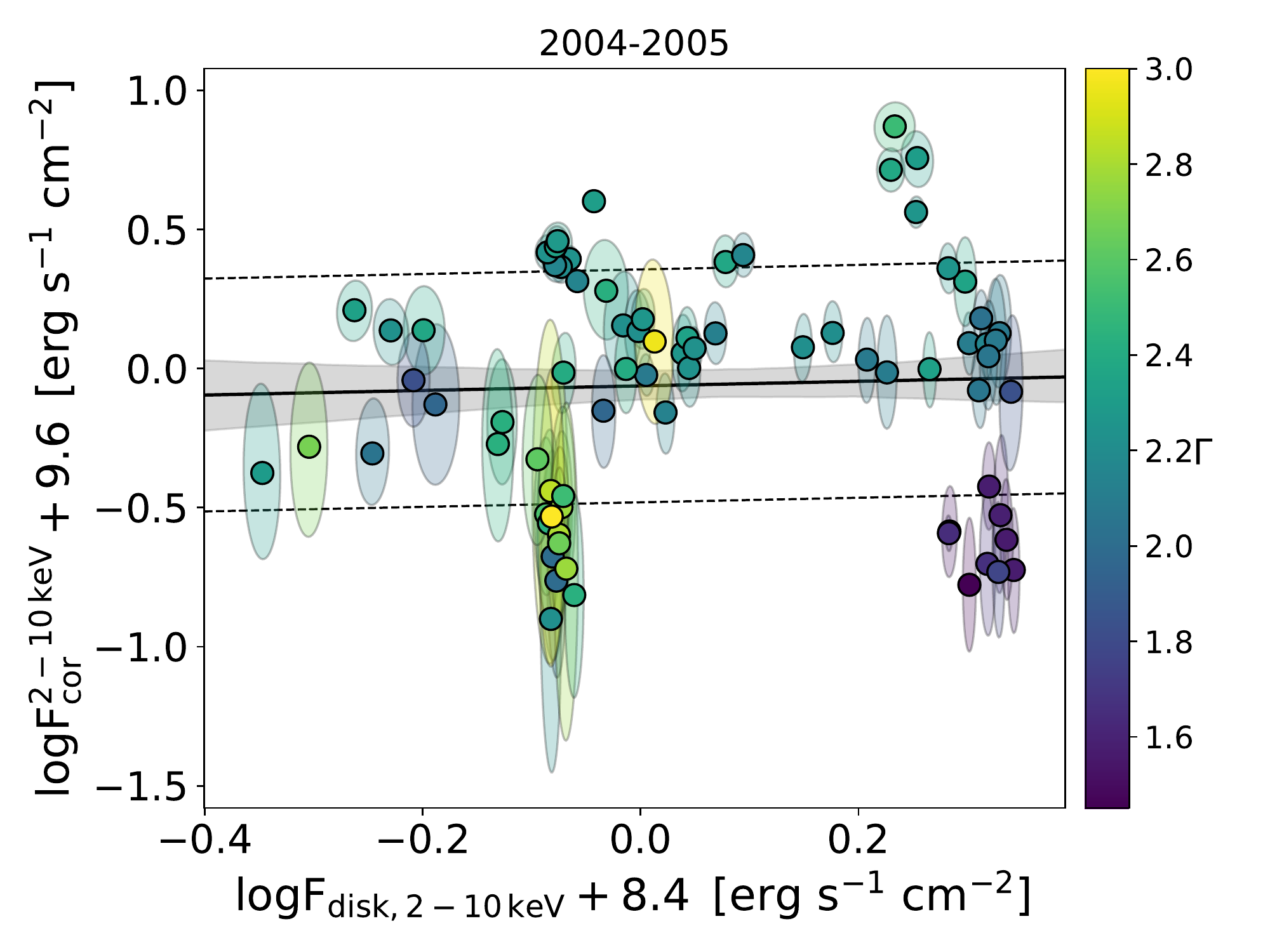}
	\includegraphics[width=0.724\columnwidth]{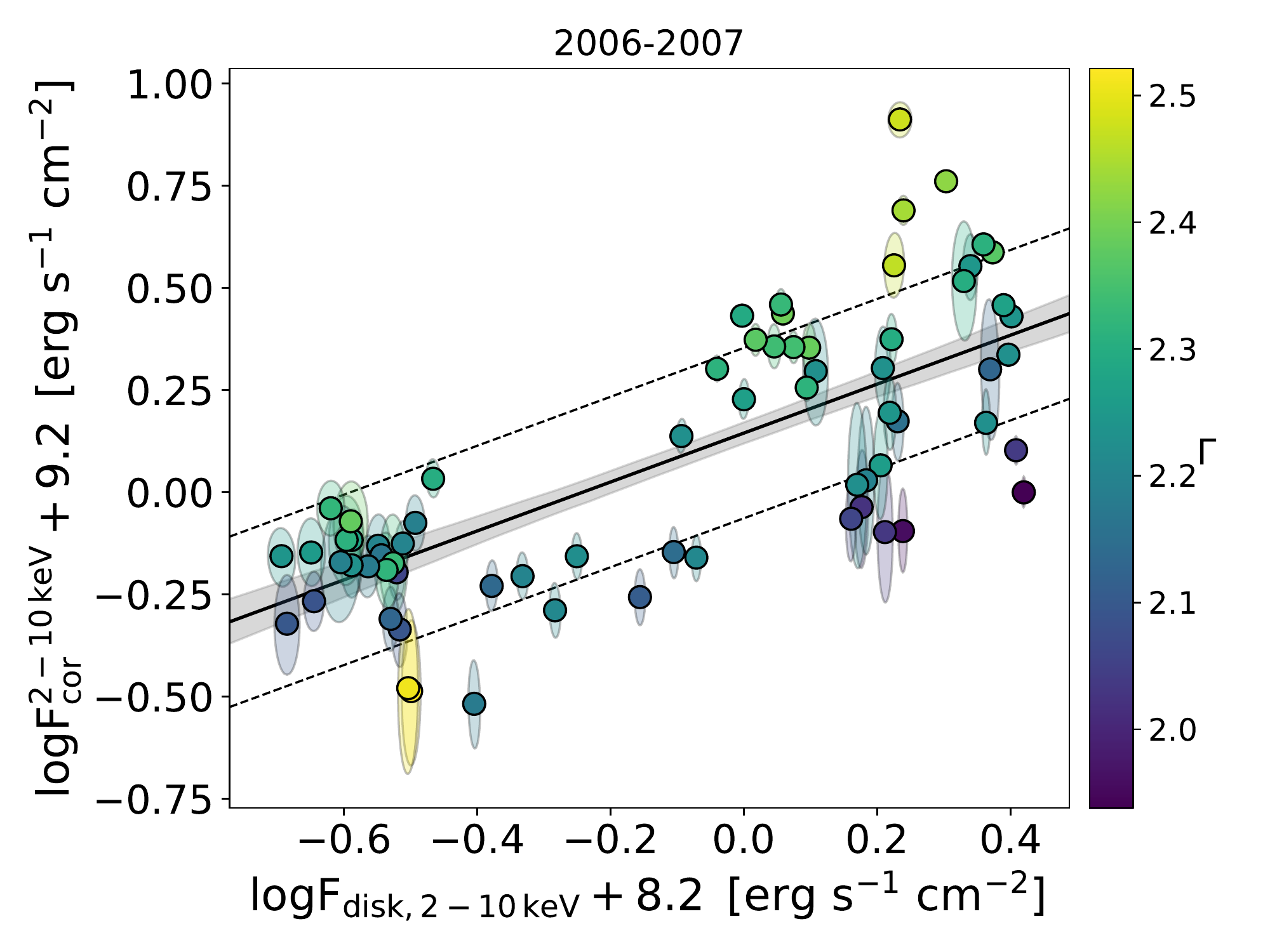}	
	\includegraphics[width=0.724\columnwidth]{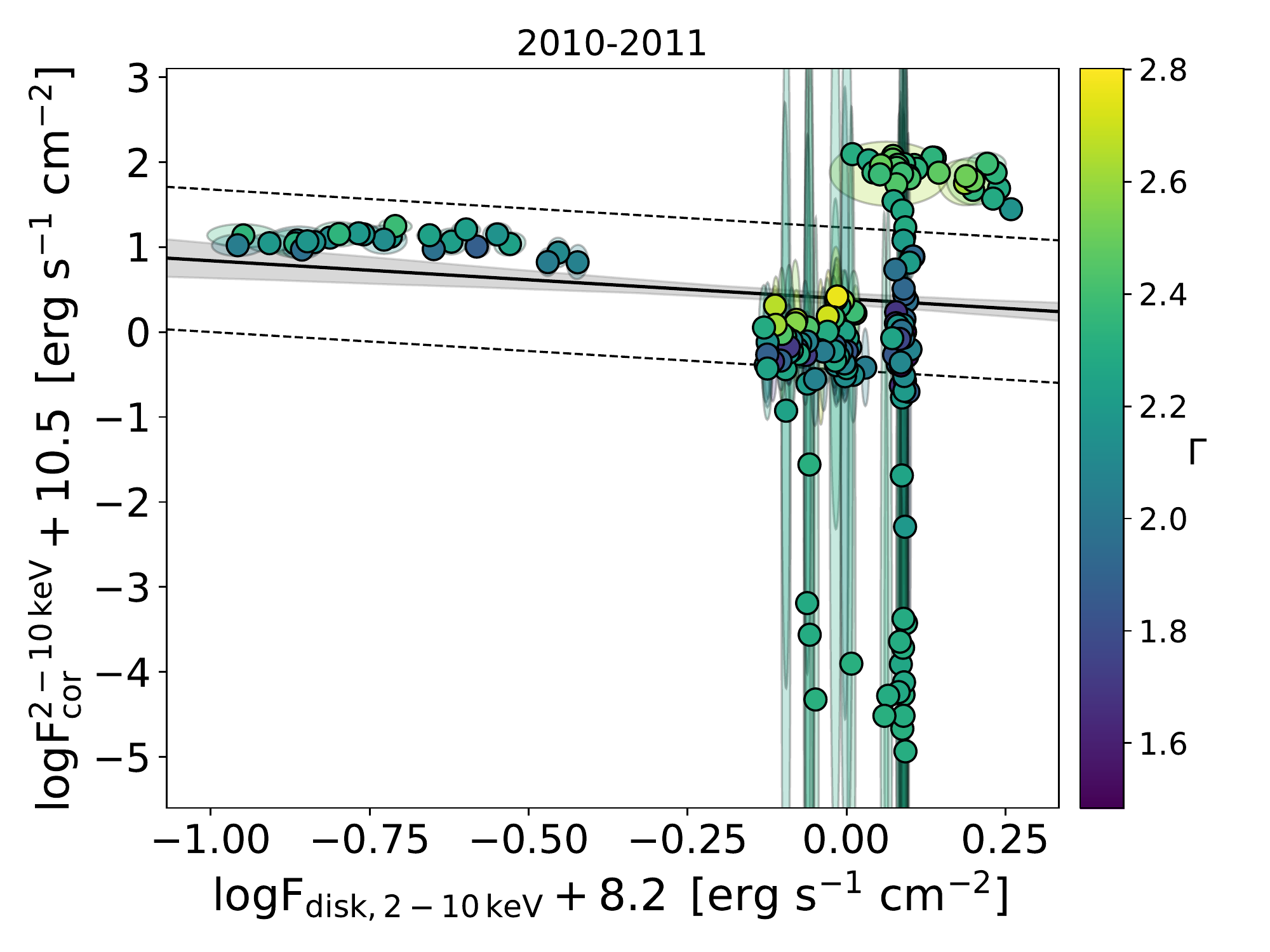}	
	\caption{$F_{disk}-F_{cor}$ plane for SS02, SS04, SS07 and SS10 (from top to bottom), with fluxes scaled with their median value. As uncertainties, we report 3-sigma contours of the 2D distribution of fluxes from the posterior chains, shown with ellipsoids. The solid black line is the median regression line obtained with emcee, with the corresponding 16th and 84th percentiles represented with the shaded gray area. The dashed black lines show the fit observed scatter around the median relation.}
	\label{fig:results_Fdisk_Fcor}
\end{figure}

In radiatively-efficient AGN we observe a tight correlation between monochromatic X-ray and UV luminosities. Its small physical intrinsic scatter \citep[$\sigma_{phys}\lessapprox0.19-0.20$\,dex; e.g., ][]{Vagnetti+2013:variab_onalphaOX,Lusso&Risaliti2016:LxLuvtight,Chiaraluce+2018:dispandvariab_Lx_Luv} defines the diversity in coronae emission for a given disk. Its slope, that is smaller than unity in log space, represents instead the evidence that going from fainter to brighter sources the coronal emission increases less than the disk emission \citep[see, e.g., ][]{Kubota+Done2018:model_lx_luv,Arcodia+2019:LxLuv}.

A similar disk-corona regulating mechanism might be also in place in SSs and SIMSs of XRBs, although previous comparisons have only been qualitative. For instance, the XRB analogous of the AGN $\alpha_{OX}$ parameter has been reported with a large scatter in a relatively narrow range of soft X-ray monochromatic (i.e. disk) flux \citep{Sobolewska+2009:alphaox_GBH,Sobolewska+2011:simul}. However, one would expect the scatter in XRBs to be smaller, since they are free from any non-simultaneity biases and the single source obviously comes with the same mass, distance and inclination. 

In this work we want to populate the $\log F_{disk}-\log F_{cor}$ plane (hereafter also simply referred to as $F_{disk}- F_{cor}$), which is the XRB equivalent of the $L_X-L_{UV}$ (or $\alpha_{OX}-L_{UV}$) relation in AGN \citep[see][and references therein]{Arcodia+2019:LxLuv}. With respect to earlier literature, we refined the choice of the observables going in the $F_{disk}-F_{cor}$. For instance, we refrained from using as a disk emission proxy a monochromatic flux in the soft band obtained with the full (soft plus hard component) model \citep[e.g. as in][]{Sobolewska+2009:alphaox_GBH}, as this would bias the estimate in a hardly predictable way moving along the HLD. For instance, in the bottom panels of Fig.~\ref{fig:stuff_vs_mjd} one can see the $D_F$ distribution: even conservatively selecting states above $D_F\sim0.8$ \citep[e.g.,][]{Dunn+2010:global} there is still up to $\sim20\%$ of the total flux coming from the Comptonisation component. Instead, we computed our disk and corona emission proxies with fluxes under the single \texttt{DISKBB} and \texttt{NTHCOMP} model, respectively. Moreover, the hard component in SSs and SIMSs can fluctuate down to the background level (e.g. see white symbols with respect to the grey in the top panel of Fig.~\ref{fig:stuff_vs_mjd}), thus our approach of modeling the background emission gave us a better handle on the disentanglement between the hard component and the background minimizing statistical problems related to the counts subtraction process \citep[e.g.][]{Loredo1992:springerbook,VanDyk:2001:lowcounts}.

In the next Sections, unless otherwise stated, we will use as proxy for the disk and corona components the $2-10\,$keV flux under the single \texttt{DISKBB} and \texttt{NTHCOMP} model, respectively. We adopted a non-monochromatic proxy since the corona emission estimate was found to be more stable against the variations of the putative disk-corona relation due to the $\Gamma$ distribution, and the $2-10\,$keV band simply for being more easily comparable with AGN (see Appendix~\ref{sec:appendix_proxies} for differences in the $F_{disk}-F_{cor}$ among the different proxies).

\subsection{The $F_{disk}-F_{cor}$ plane across the outbursts}

\begin{table}[tb]
	\footnotesize
	\caption{Summary of slope and scatter of the $\log F_{disk}-\log F_{cor}$ plane, computed in the $2-10\,$keV energy band, across the four outbursts. The second last rows refers to results with a joint analysis on the combined 458 states, the last is obtained on a subset of the total sample described in Section~\ref{sec:scatter}.}
	\label{tab:emcee_single_outb}
	\centering
	\begin{tabular}{C{0.07\columnwidth} C{0.4\columnwidth} C{0.25\columnwidth}}%
		\toprule
		\multicolumn{1}{c}{Outburst} &
		\multicolumn{1}{c}{Slope} &		
		\multicolumn{1}{c}{Scatter} \\
		\midrule
		SS02 & $1.04\pm0.18$ & $0.47\pm0.03$ \\
		SS04 & $0.08\pm0.25$ & $0.42^{+0.04}_{-0.03}$ \\
		SS07 & $0.60\pm0.07$ & $0.21\pm0.02$ \\
		SS10 & $-0.45\pm0.22$ & $0.84\pm0.05$ \\
		\midrule
		All & $0.34\pm0.12$ & $0.71^{+0.03}_{-0.02}$ \\
		All\_r1.3 & $0.47\pm0.07$ & $0.41\pm0.02$ \\
		\bottomrule	
	\end{tabular}
\end{table}

In this Section we focused on the $\log F_{disk}- \log F_{cor}$ plane across the four outbursts of GX339-4 separately, in order to see if and how they compare. We show the relations in Fig.~\ref{fig:results_Fdisk_Fcor} and report the related results of the linear regression performed with \texttt{emcee} \citep{Foreman-Mackey+2013:emcee} in Table~\ref{tab:emcee_single_outb}. The full relation used is $\log F_{cor}-c_1=a+b\,(\log F_{disk}-c_2)$, where $c_1$ and $c_2$ are the median value of $\log F_{cor}$ and $\log F_{disk}$, respectively (i.e. a different scaling for each regression). Uncertainties on all variables and an additional scatter term (hereafter also referred to as observed scatter) were accounted for using the likelihood provided in \citet{Dagostini2005:fits}.

The main conclusion from Fig.~\ref{fig:results_Fdisk_Fcor} is that, at first glance, the
four separate $\log F_{disk}- \log F_{cor}$ planes  do not appear the same. First, the linear correlations do not show evidence of a common slope, which instead spans positive to negative values. A possible reason for this might be that the dynamic range covered by $F_{disk}$ (i.e. the horizontal axis), one order of magnitude and even less, is too small for a solid estimate of the slope. Such a range is in fact not even close to the three-four orders of magnitude spanned by UV luminosities in bright AGN \citep[e.g.][]{Lusso&Risaliti2016:LxLuvtight}. This will be further addressed in Section~\ref{sec:relation_XRB_vs_AGN}, however we here conclude that the slope in the $F_{disk}-F_{cor}$ plane does not appear as a good proxy for the disk-corona physics in XRBs.

Furthermore, the path of an outburst in the HLD also somewhat reverberates on the $\log F_{disk}- \log F_{cor}$. It is particularly evident in SS10, where the both the HLD and the $\log F_{disk}- \log F_{cor}$ are populated by three clumps (e.g. see the bottom panel of Fig.~\ref{fig:results_Fdisk_Fcor} and the yellow points in Fig.~\ref{fig:qplot}). Moreover, in Fig.~\ref{fig:results_Fdisk_Fcor} data points seem to oscillate around the putative relation rather than sitting on it; this abrupt changes in $F_{cor}$ for a narrow range of $F_{disk}$ reflect the horizontal paths in the HLD commonly observed during SSs and SIMSs, in which a source can significantly change its hardness-ratio maintaining the same total (disk-dominated) luminosity (e.g. see Fig.~\ref{fig:qplot}). This can be either a peculiarity of XRBs or a trend that we would observe in more massive sources if the coverage was comparably high cadenced. However, a crude mass-scaling of this short day-level timescales would be around hundreds of thousands of years for AGN. 

Finally, also the observed scatter, which is in general very high (between $\sim0.2-0.8$\,dex), seems to be inconsistently different across the outbursts, although it appears to be clearly proportional to the range spanned by $\Gamma$ during the outburst (see color coding in Fig.~\ref{fig:results_Fdisk_Fcor}). Before comparing XRBs to AGN, XRB data need to be homogenised across the outbursts and the differences among them understood and addressed. In particular, the observed scatter seems a more promising and understandable proxy of the disk-corona relation in XRBs and it will be the focus of the next Section.
 
\subsection{The observed scatter of the disk-corona relation}
\label{sec:scatter}

\begin{figure}[tb]
	\centering
	\includegraphics[width=\columnwidth]{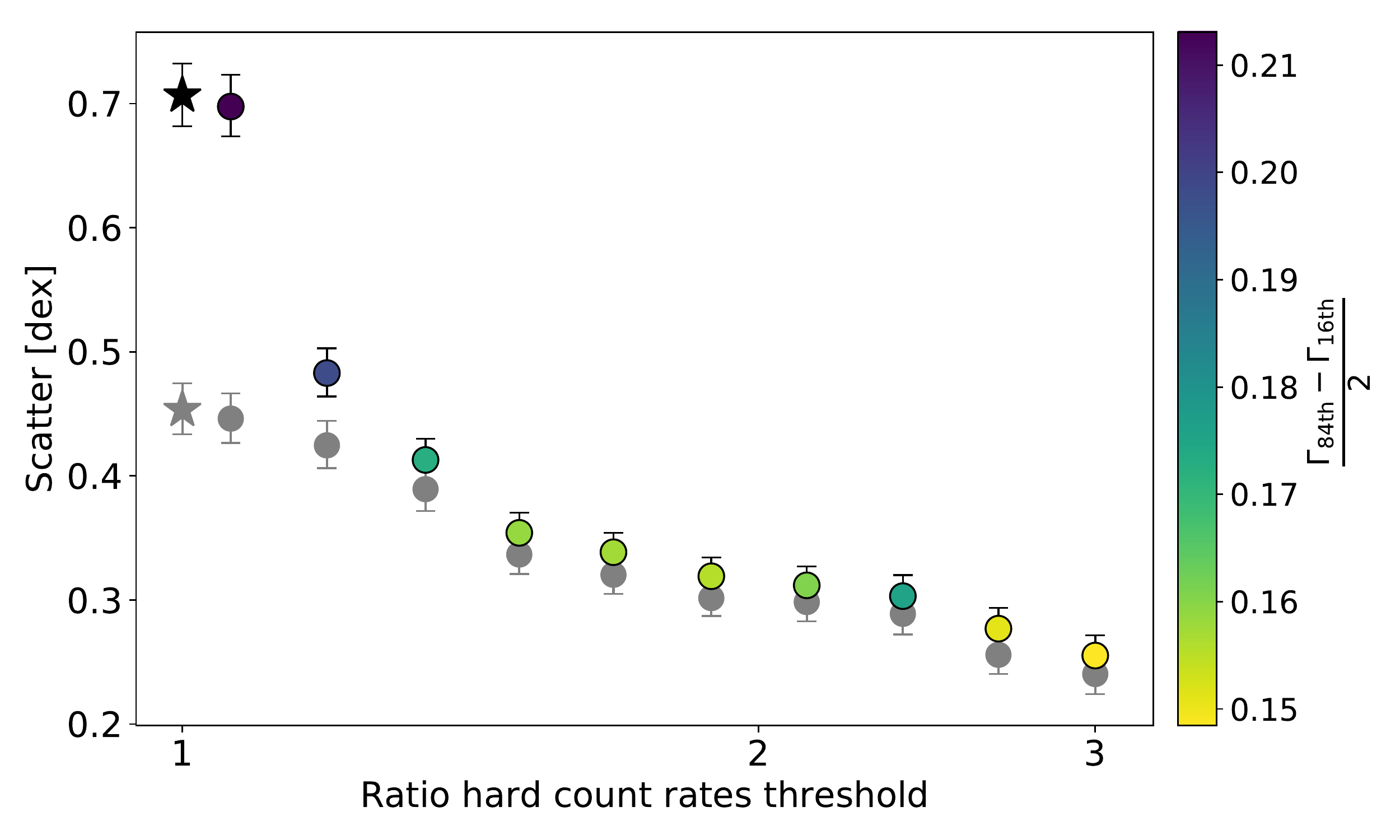}
	\includegraphics[width=\columnwidth]{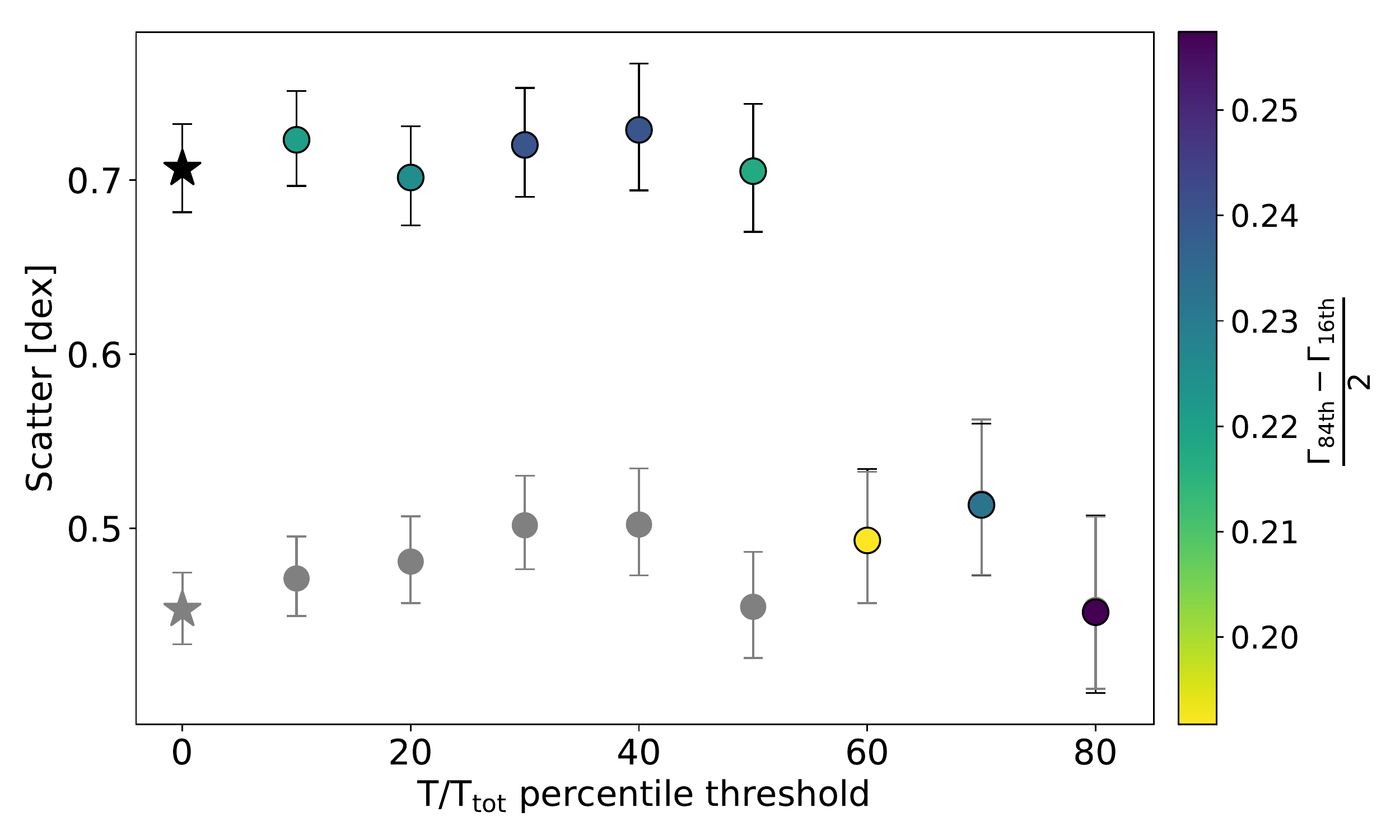}	\caption{\emph{Top panel}: the scatter of the $F_{disk}-F_{cor}$ relation as a function of a cut in the ratio between the total (source plus background) and background-only $10-25\,$keV count rates. Colored points are relative to the full sample and coded with the inter-quantile range in the related $\Gamma$ distributions, whereas grey points are obtained excluding SS10. Stars correspond to the starting sample with no filters. \emph{Bottom panel}: the scatter as a function of a cut in the fraction of time spent by GX339-4 in a region of the outburst, including SSs and SIMSs only (see Section~\ref{sec:scatter} for a detailed description). Color coding and symbols are the same as in the top panel.}
	\label{fig:scatter_vs_bkgcut}
\end{figure}

The scatter of the $F_{disk}-F_{cor}$ relation is likely due to a combination of factors and, before a comparison with AGN is performed, a more thorough test on our whole GX339-4 dataset is necessary since it spans rather different values across the outbursts. For instance, the scatter does not depend on the luminosity range covered by an outburst in the HLD or by the dynamic range in $F_{disk}$ (i.e. accretion rate). This can be evinced by the scatter in SS04 being compatible to the one in SS02, despite the former has a much lower spread in $F_{disk}$ (see Fig.~\ref{fig:results_Fdisk_Fcor}) and in $L_{3-10\,keV}/L_{edd}$ (see Fig.~\ref{fig:qplot}); and by the fact that SS02 and SS07 span roughly the same range in luminosity and disk temperature, somewhat related to accretion rate, despite the latter shows an incompatibly smaller scatter. Conversely, the scatter appears to be lower for outbursts with a narrower $\Gamma$ distribution and particularly higher in SS10, for which several states had hard count rates  ($\gtrsim10\,$keV) at background level (see top panel in Fig.~\ref{fig:stuff_vs_mjd}). 

The two quantities seem to be somewhat correlated, as $\Gamma$ seems to reach the extremes of its distribution mostly in these background-contaminated states (see Fig.~\ref{fig:gamma_vs_bkg}). Nonetheless, while a simple cut in $\Gamma$ is rather arbitrary, as there is no physical reason to remove the flat or steep end of the corona emission a priori, it is experimentally meaningful to test the impact of the background influence on the scatter of the $F_{disk}-F_{cor}$. Moreover, this test is particularly relevant for the comparison with AGN (Section~\ref{sec:relation_XRB_vs_AGN}), for which the disk is observed in a different energy band and background-dominated coronae would be either non detected, or poorly constrained thus excluded from any quality selection.

\begin{figure}[tb]
	\centering
	\includegraphics[width=0.77\columnwidth]{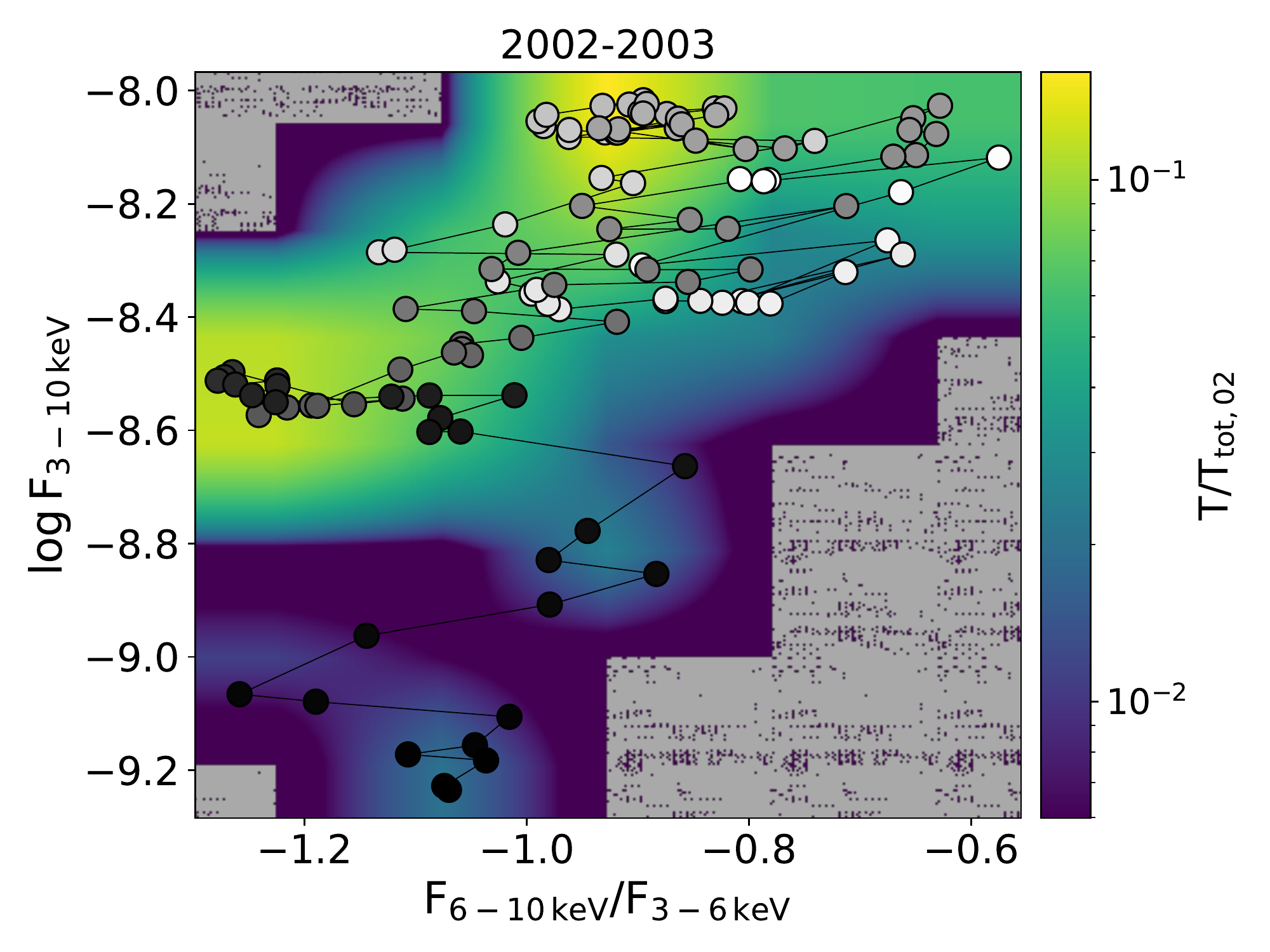}
	\includegraphics[width=0.77\columnwidth]{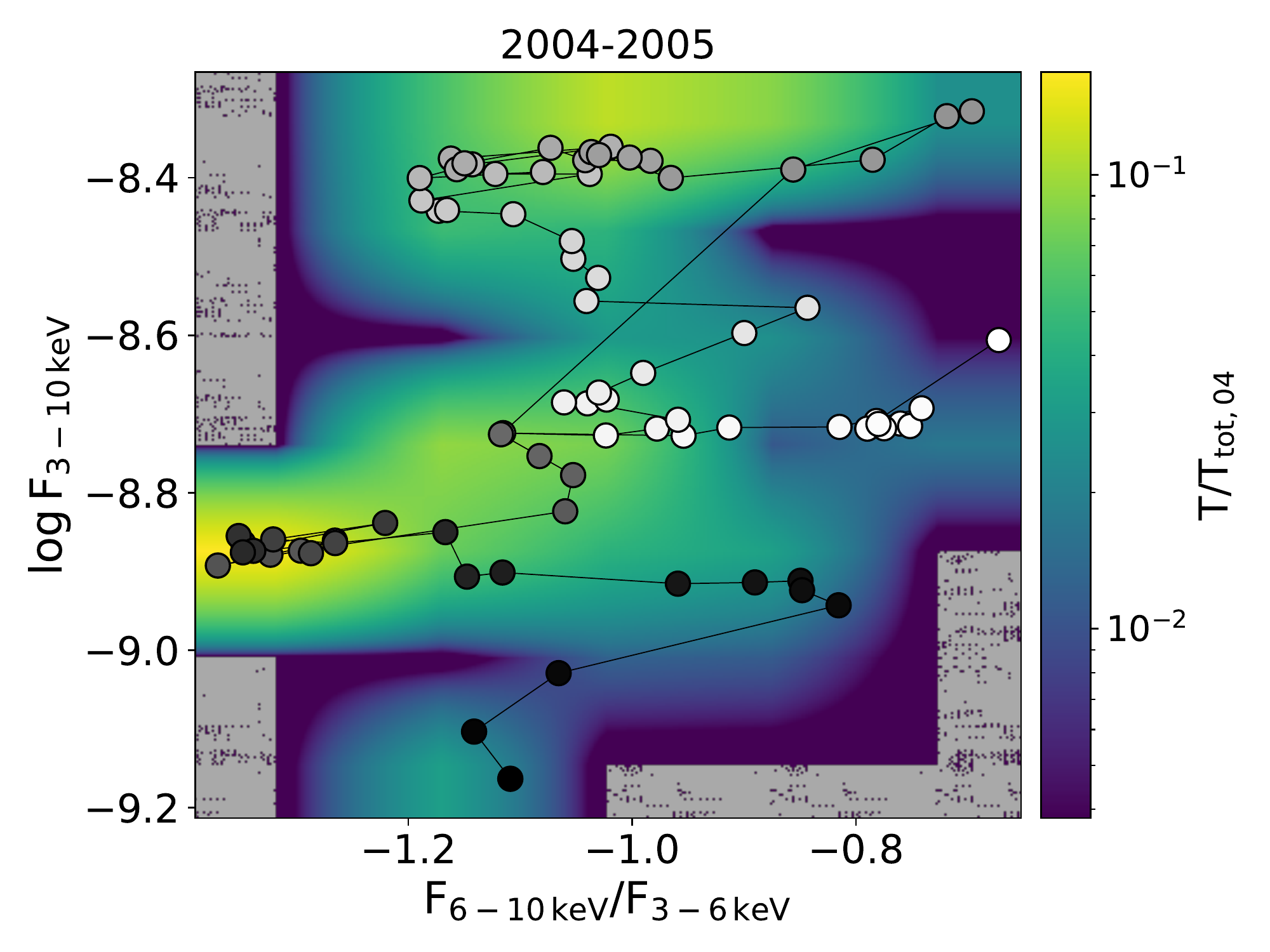}
	\includegraphics[width=0.77\columnwidth]{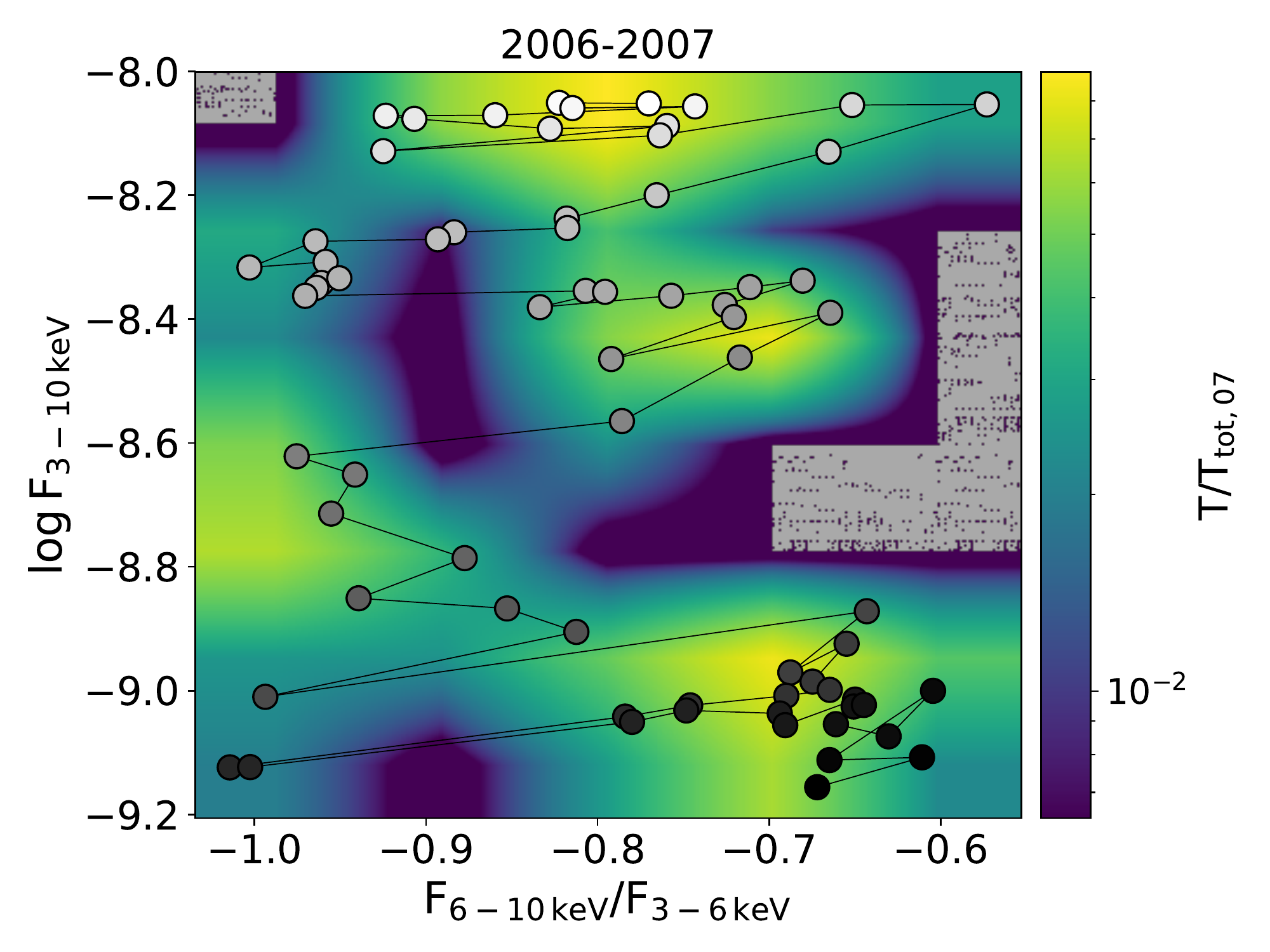}	
	\includegraphics[width=0.77\columnwidth]{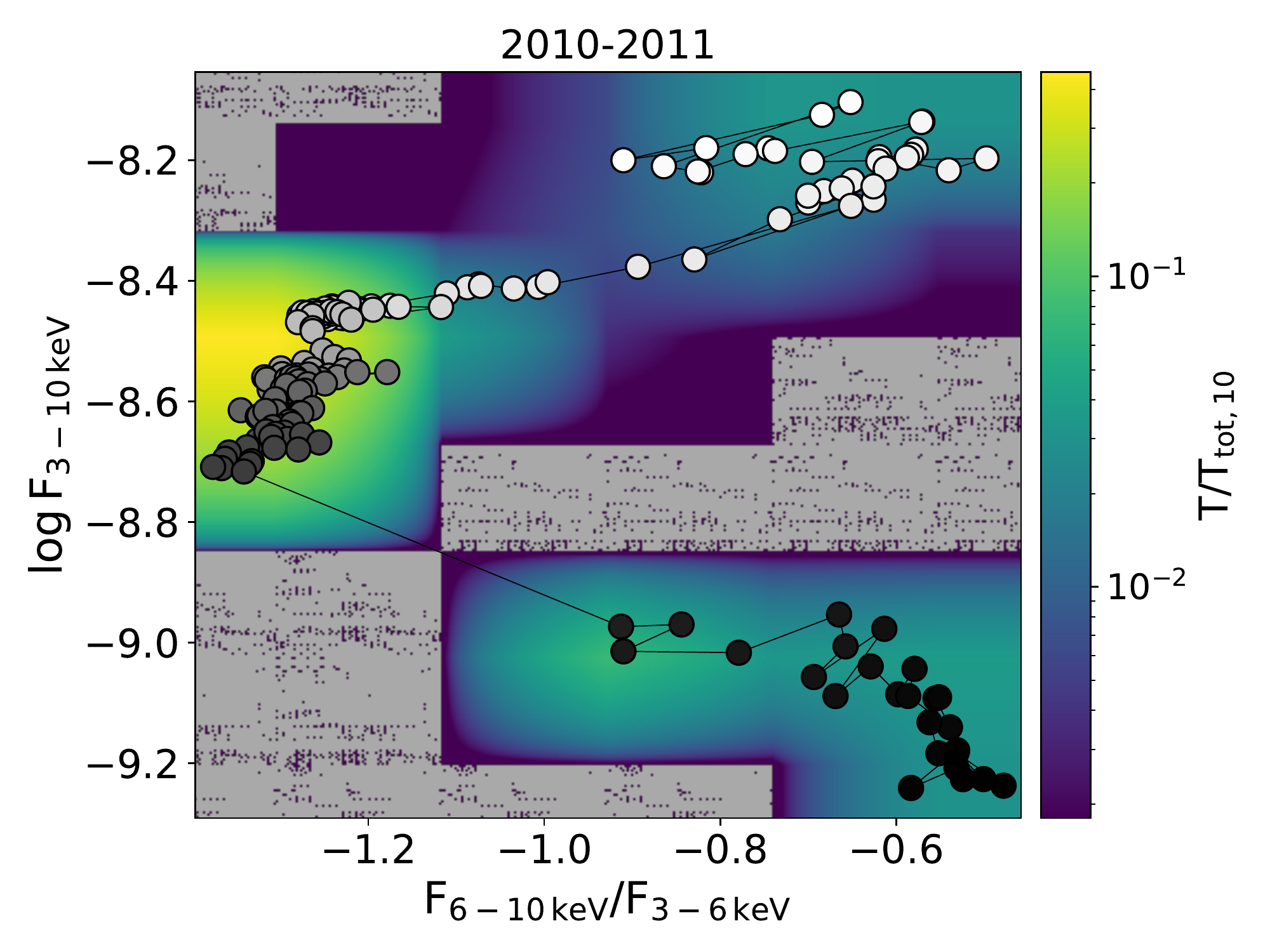}	
	\caption{HID of the four outbursts color coded, from white to black, with MJD values, superimposed to a smoothed map showing the fraction of time spent in each region, including only SS and SIMS.}
	\label{fig:time_frac_cut}
\end{figure}

\begin{figure}[tb]
	\centering
	\includegraphics[width=\columnwidth]{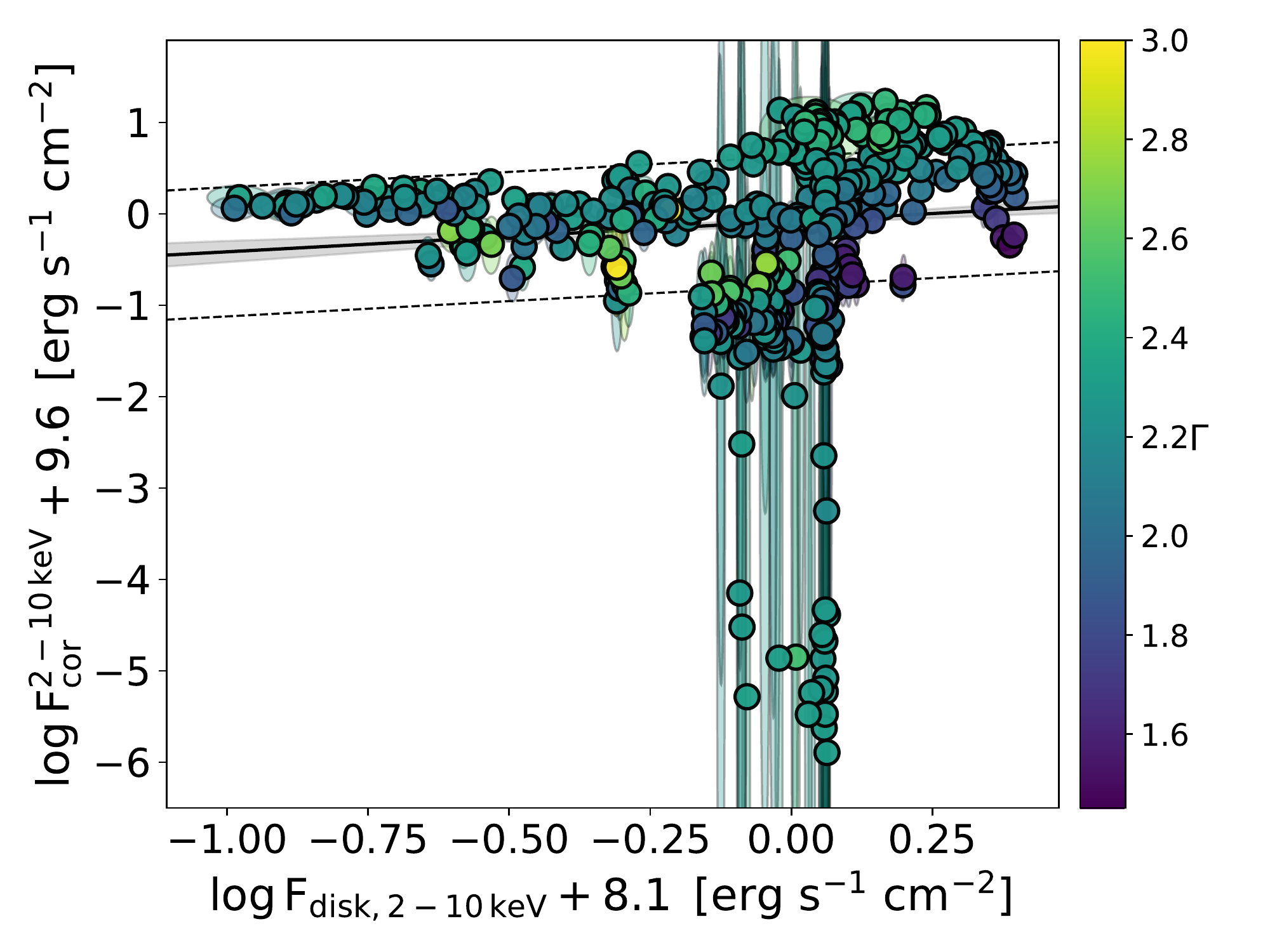}
	\includegraphics[width=\columnwidth]{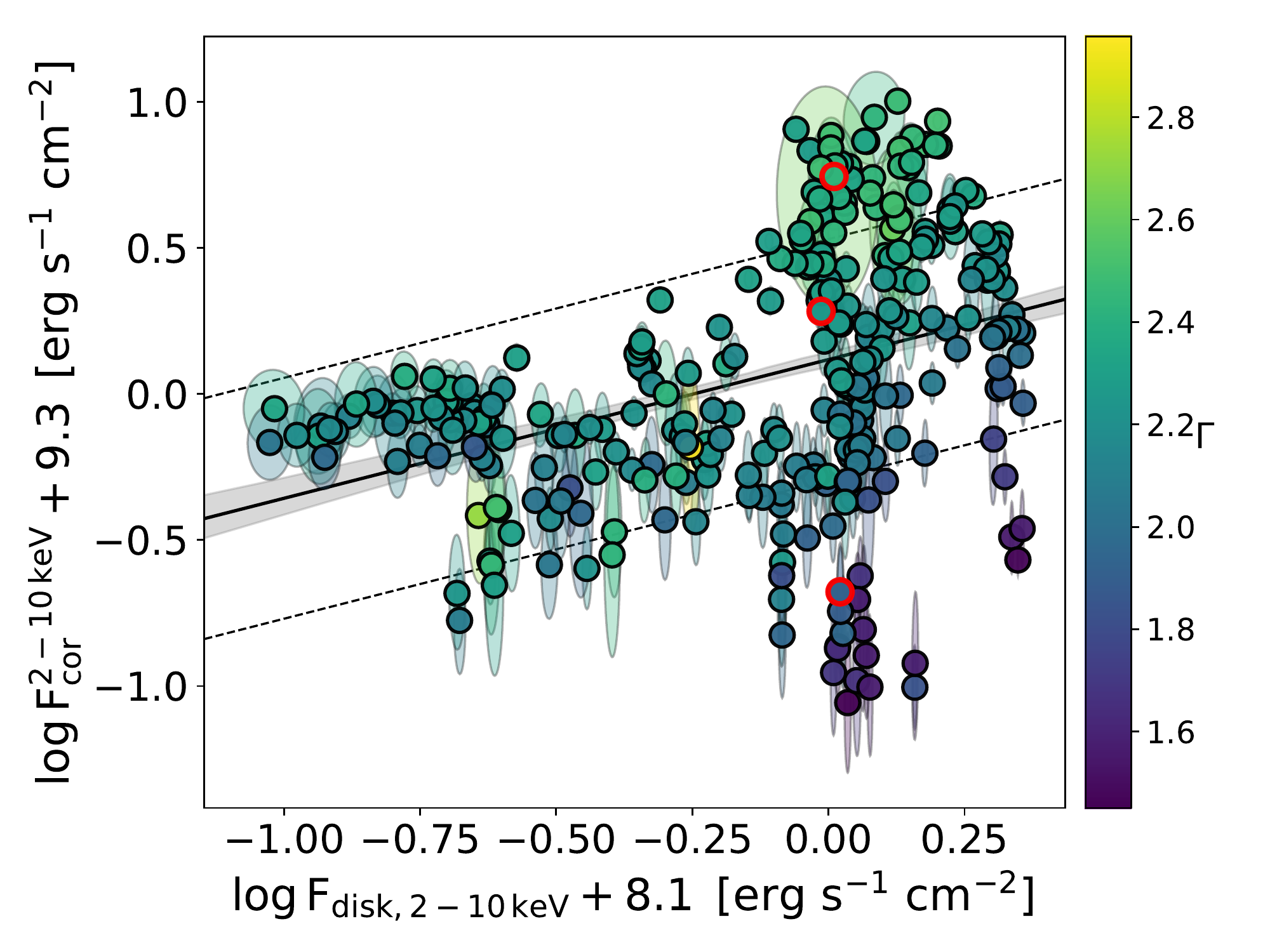}
	\caption{$F_{disk}-F_{cor}$ plane as described in Fig.~\ref{fig:results_Fdisk_Fcor}, but shown here for all outbursts combined (\emph{top panel}) and for the subset used for the comparison with AGN (\emph{bottom panel}), as described and motivated in Section~\ref{sec:scatter}. In the bottom panel, points with a red contour highlight three states taken as examples for Fig.~\ref{fig:seds}.}
	\label{fig:results_Fdisk_Fcor_all}
\end{figure}

We show in the top panel of Fig.~\ref{fig:scatter_vs_bkgcut} how the scatter of the $F_{disk}-F_{cor}$ changes as a function of a cut performed on the ratio between the total (source plus background) and background-only $10-25\,$keV count rates (see, e.g., white and grey symbols in the top panel of Fig.~\ref{fig:stuff_vs_mjd}). In Fig.~\ref{fig:scatter_vs_bkgcut} grey symbols show results obtained excluding SS10, stating that the enormous scatter is mostly due to some SS10 states in which the hard component is background-contaminated. From spectral simulations (see Appendix~\ref{sec:appendix_robustness}) we conservatively obtained a value of $\sim1.3$ for this count-rate ratio above which all spectral fits can be considered robust. Above this threshold, the fraction of states that in the simulations did not retrieve the input $\Gamma$ within the 16th-84th inter-quantile range is below $\sim4\%$ and it stays roughly constant. Moreover, this value is also approximately where the scatter with and without SS10 share the same trend (top panel of Fig.~\ref{fig:scatter_vs_bkgcut}), namely where the critical states seem to be excluded.

Furthermore, an accurate comparison between XRBs and AGN should take their different sampling and evolution timescales into account. Building the $F_{disk}-F_{cor}$ plane with multi-epoch observations of a single super-massive AGN requires too large efforts, although the first test cases at the low-mass end are being now explored \citep[e.g.,][]{Ruan+2019:individual_AGN}. Therefore, typically large AGN samples are used to trace the evolution of one (or few) XRB(s), assuming a putative scaling between the two classes. However then, AGN would be preferentially found in periods of their evolution that broadly correspond to regions of the HLD where XRBs spend the most of their time. Then, we tried to compute a rough but motivated estimate of the fraction of time spent by GX339-4 in each portion of its SS and SIMS phases combined, for each outburst separately. We first verified that the observations duration and cadence were fairly uniform, which is often the case with monitoring instruments like RXTE. Then we computed a modified duration adding to each observation exposure half of the unobserved time fraction, both before and after it, to sample the whole SS and SIMS duration. We then built a grid in HR (six bins) and luminosity (eight bins) in the HLD and summed this extended duration in each bin. The resulting smoothed maps, obtained normalizing for the total time spent in the SS and SIMS for each outburst separately, are shown in Fig.~\ref{fig:time_frac_cut}. 

Then, we show in the bottom panel of Fig.~\ref{fig:scatter_vs_bkgcut} how the scatter of the $F_{disk}-F_{cor}$ changes as a function of a cut performed on this fraction of time $T/T_{tot}$ (which color codes Fig.~\ref{fig:time_frac_cut}). Since each SS-SIMS outburst did not last the same amount of time, we cut the data subsets selecting above a given percentile (e.g. from the 10th to the 80th) of $T/T_{tot}$, with the actual value then changing among the outbursts accordingly. As it can be seen in Fig.~\ref{fig:scatter_vs_bkgcut}, the scatter changes as a function of the cut in the time fraction only if SS10 is included and this is a spurious effect driven by the background contamination, described above: the scatter jumps to lower values around the cut with the 50th percentile of $T/T_{tot}$ simply because the low $F_{cor}$ data points in SS10 are cut out of the data set; as a matter of fact, there is no evolution if SS10 is left out from the exercise (grey points in the bottom panel of Fig.~\ref{fig:scatter_vs_bkgcut}). Hence, the scatter in XRBs is not high because of the frequent sampling and the shorter variability timescales.

\begin{figure*}[tb]
	\centering
	\includegraphics[width=\columnwidth]{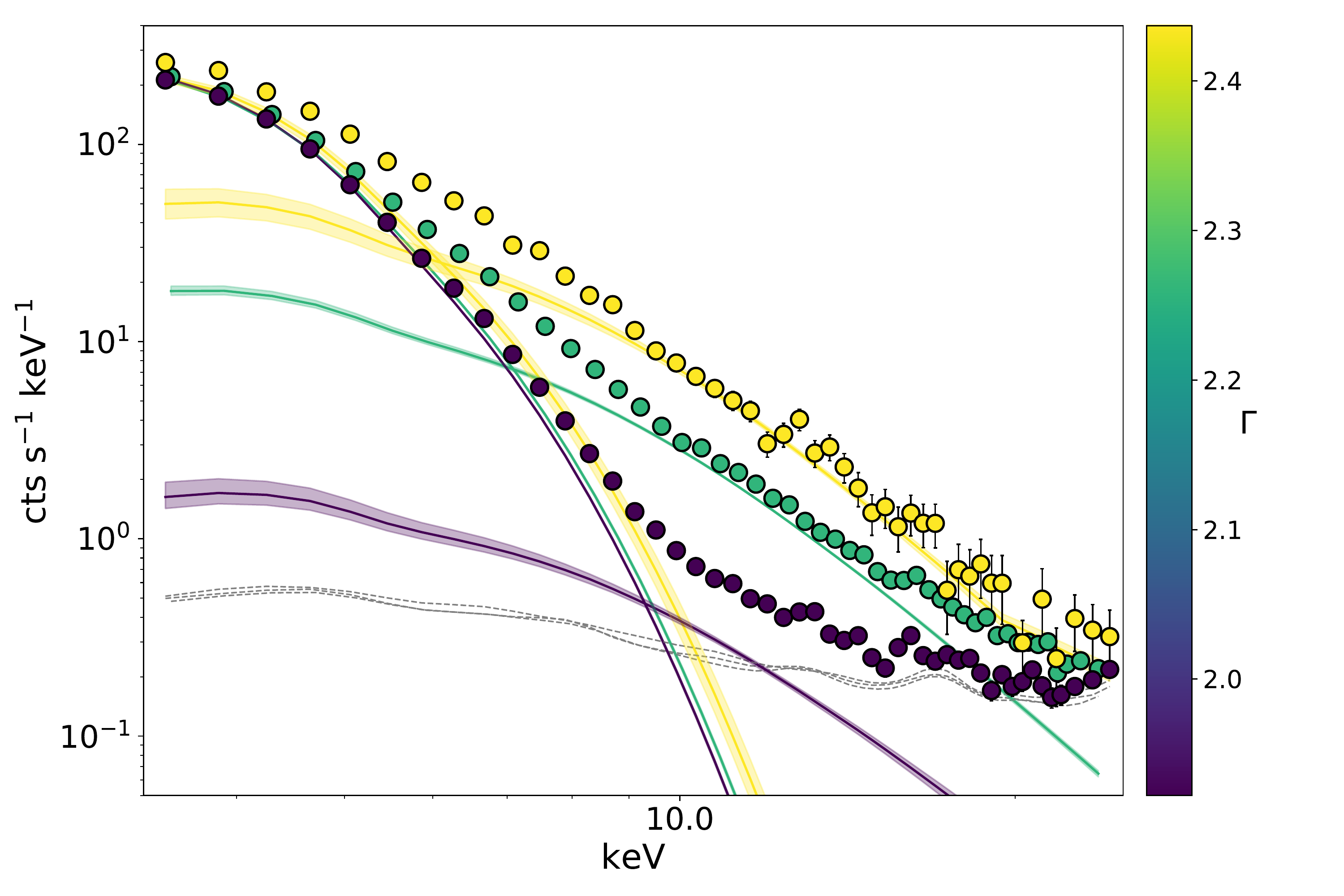}		\includegraphics[width=\columnwidth]{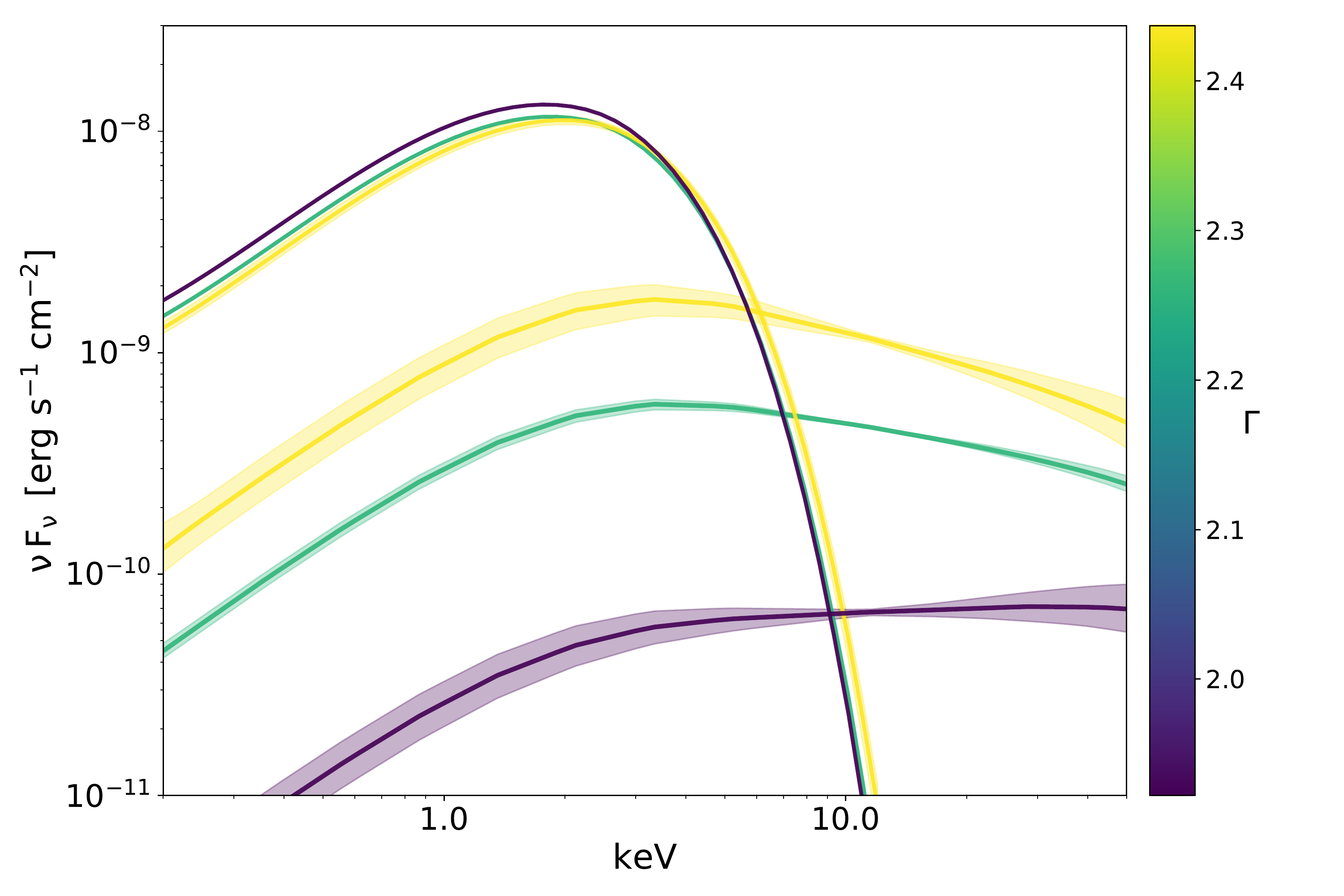}
	\caption{Examples of observed spectra with best-fit models (\emph{left}, including background, shown with a grey-dashed line) and modeled spectral energy distributions (\emph{right}, without background) of three states, selected taking the 10th, 50th and 90th percentile of $\log F_{cor}$ in a narrow range of $F_{disk}$ (namely around a $\pm0.05$\,dex of the median $\log F_{disk}$). Both data and models are color-coded with the fit $\Gamma$, highlighting a softer-when-brighter trend in $\log F_{cor}$. These three states are represented by red-contour data points in the bottom panel of Fig.~\ref{fig:results_Fdisk_Fcor_all}.}
	\label{fig:seds}
\end{figure*}

Summarizing, the take-home message from this section is that the scatter in the $F_{disk}-F_{cor}$ plane for XRBs is very high mostly because of a subset of spectra in which the hard band count rate ($\gtrsim10\,$keV) is comparable to the background level (see also Appendix~\ref{sec:appendix_robustness}). Nonetheless, even excluding the most critical states (e.g. a ratio of $\gtrsim1.3-1.5$ in the top panel of Fig.~\ref{fig:scatter_vs_bkgcut}) the scatter of the relation is between $\sim0.30-0.45\,$dex, still higher that what is claimed to be the physical scatter of the $L_{disk}-L_{cor}$ in AGN \citep[$\lessapprox0.19-0.20$\,dex;][]{Lusso&Risaliti2016:LxLuvtight,Chiaraluce+2018:dispandvariab_Lx_Luv}. Furthermore, we stress that in XRBs all disk-corona data come simultaneously and from the same source, namely from constant mass, distance and inclination, even if the estimates are uncertain in an absolute sense. Thus, the source of this scatter cannot be due to these factors, which makes the high observed scatter even more puzzling. This result is thus important, since a higher scatter for XRBs would either argue against a common scale-invariant accretion paradigm, or would state that the physical scatter in AGN is not necessarily as low as we think.

We here showed that different outbursts are not intrinsically homogeneous, the main differences being both observational (a different background contamination of the hard component) and physical (different $\Gamma$ distribution spanned during the outbursts). Based on the above arguments and on spectral simulations (see Appendix~\ref{sec:appendix_robustness}), we solved the former selecting a subset of XRB states which are above a ratio between the total (source plus background) and background-only $10-25\,$keV count rates of $\sim1.3$. We show in the top and bottom panel of Fig.~\ref{fig:results_Fdisk_Fcor_all} the $F_{disk}-F_{cor}$ relation of the full XRB sample and the one of this subset, respectively. The related slope and scatter in the last two rows of Table~\ref{tab:emcee_single_outb}. 

In Fig.~\ref{fig:seds} we show counts spectra and the related spectral energy distributions (SEDs) of three states, selected taking the 10th, 50th and 90th percentile of $\log F_{cor}$ in a narrow range of $F_{disk}$ (namely around a $\pm0.05$\,dex of the median $\log F_{disk}$). This allows us to visualize more clearly how the scatter in the $F_{disk}-F_{cor}$ actually looks like as a variety of observed spectra and modeled SEDs. Data and models are color-coded by the fit $\Gamma$ to highlight the softer-when-brighter trend (where both softer and brighter here refer to $\log L_{cor}$ alone in this context) also visible from the color coding of Fig.~\ref{fig:results_Fdisk_Fcor}. These three states are represented by red-contour data points in the bottom panel of Fig.~\ref{fig:results_Fdisk_Fcor_all}. 

\begin{figure*}[tb]
	\centering
	\includegraphics[width=0.86\columnwidth]{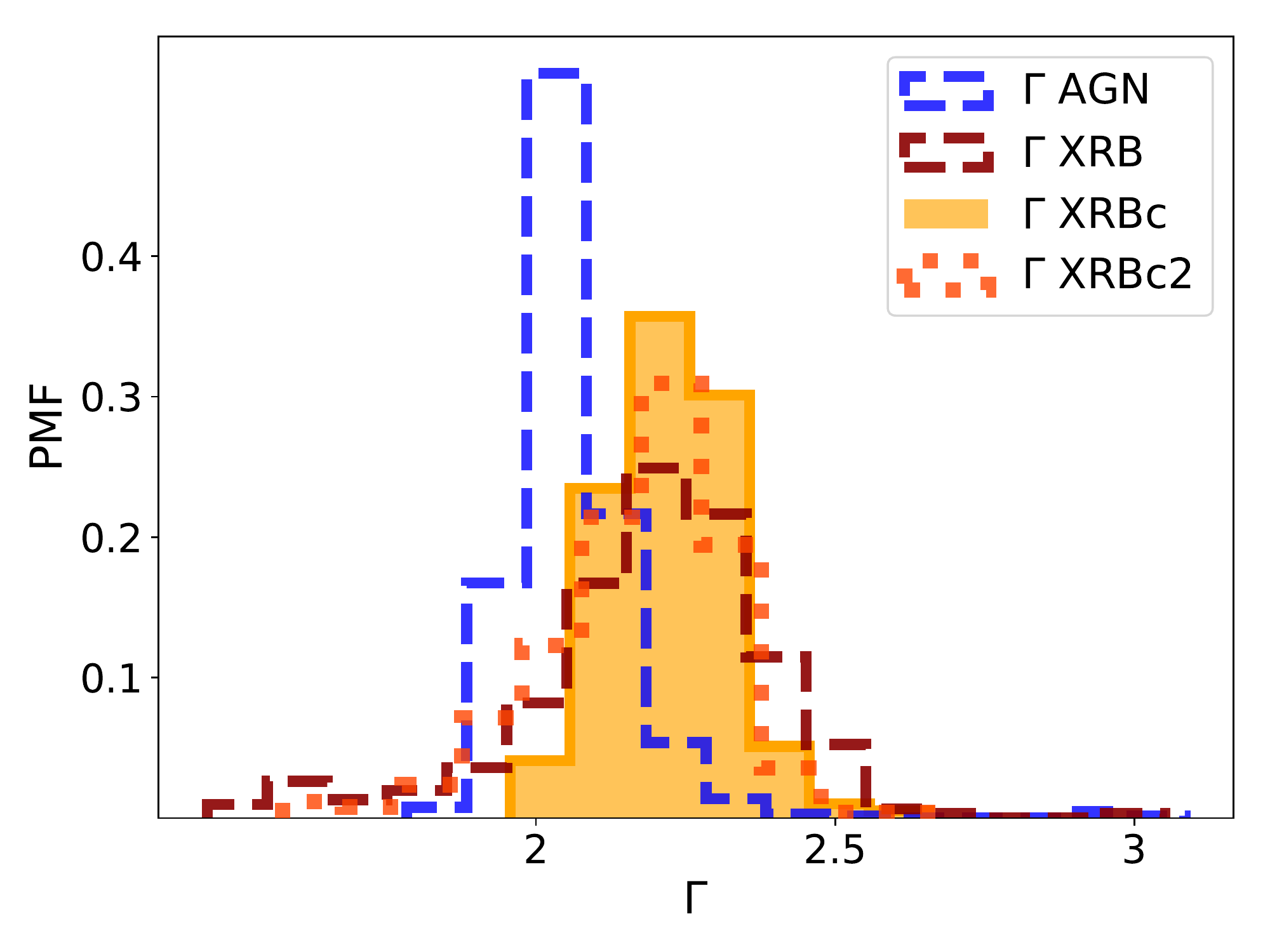}	
	\includegraphics[width=0.86\columnwidth]{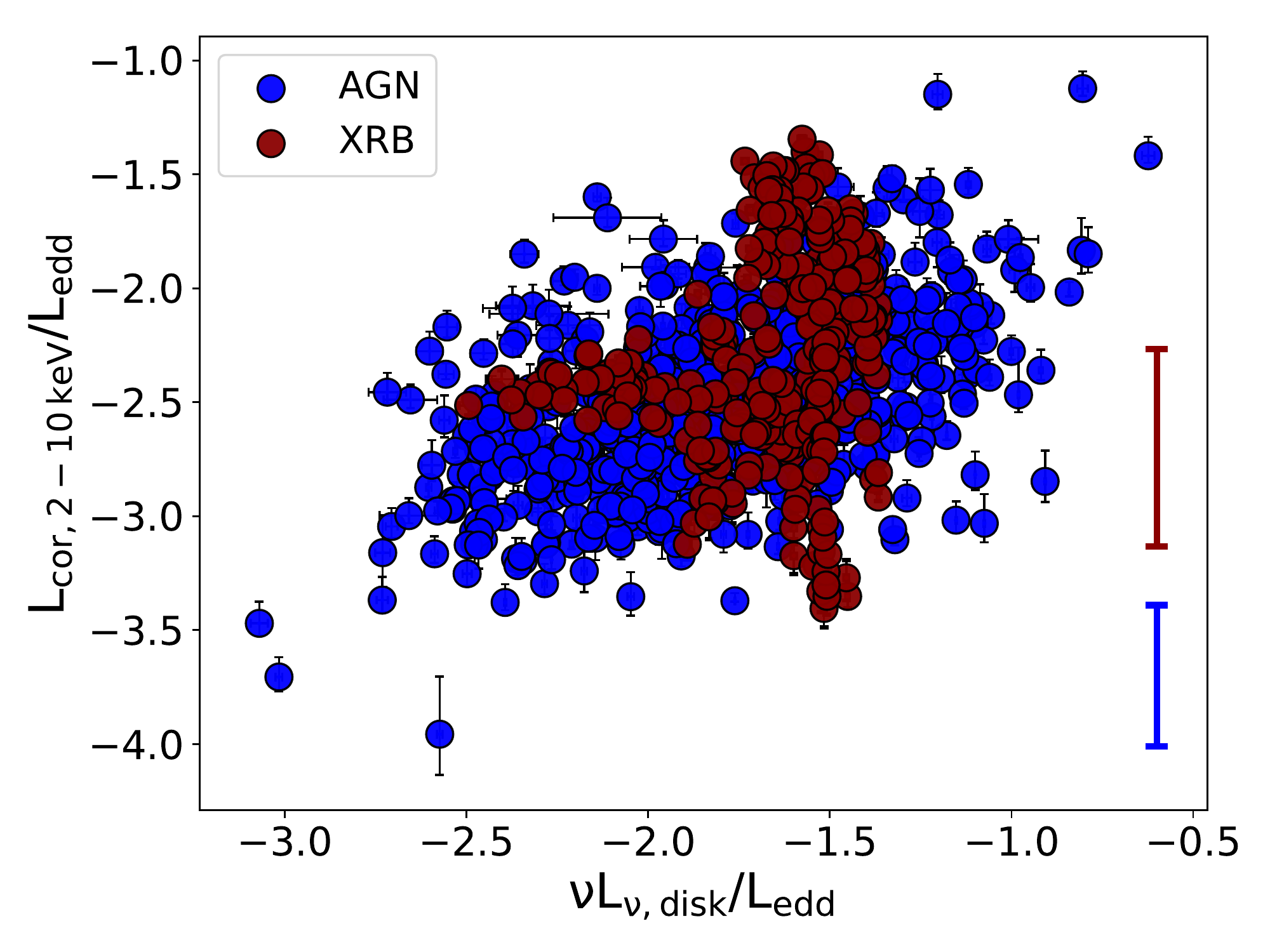}
    \includegraphics[width=0.86\columnwidth]{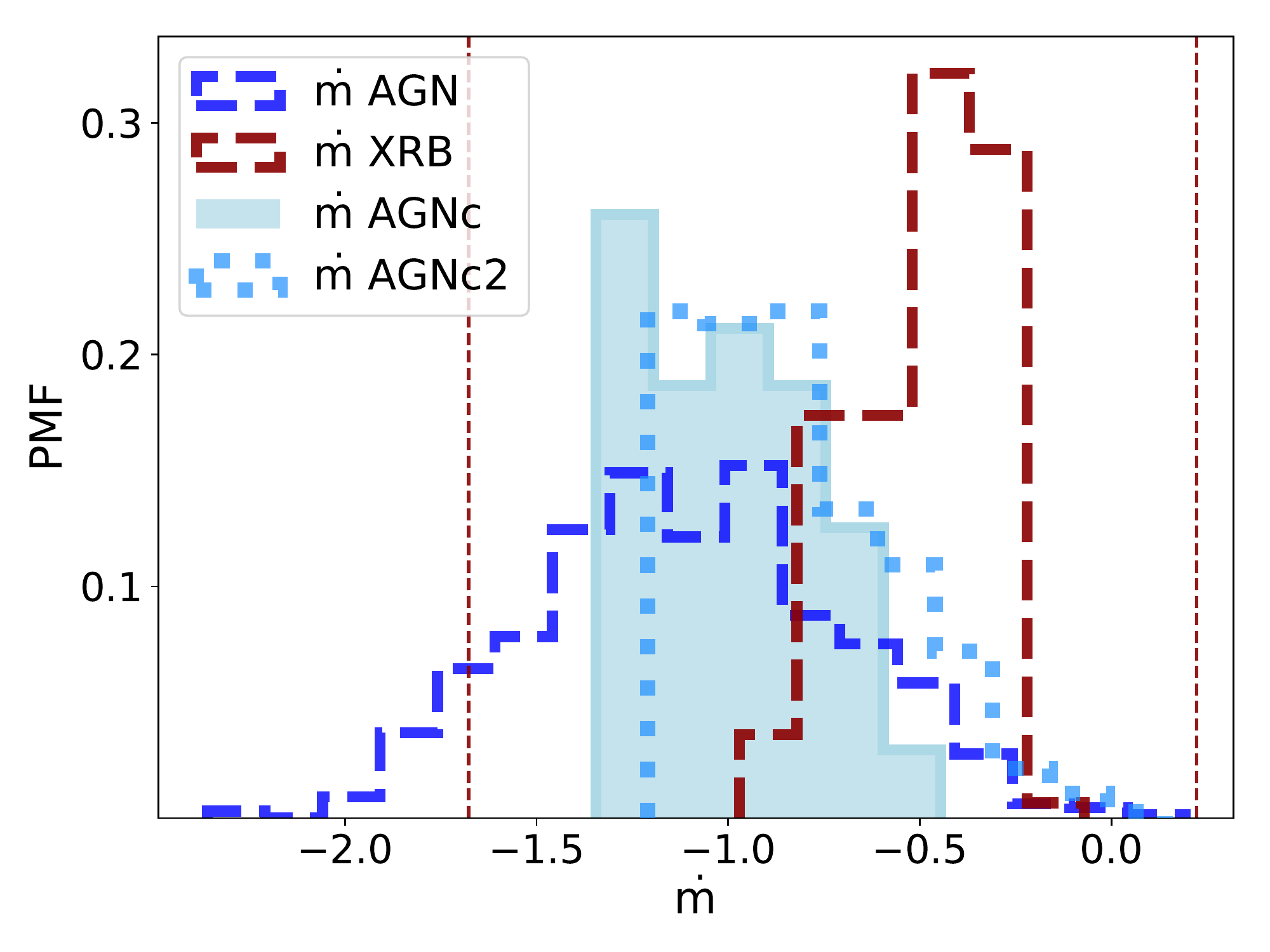}
	\includegraphics[width=0.86\columnwidth]{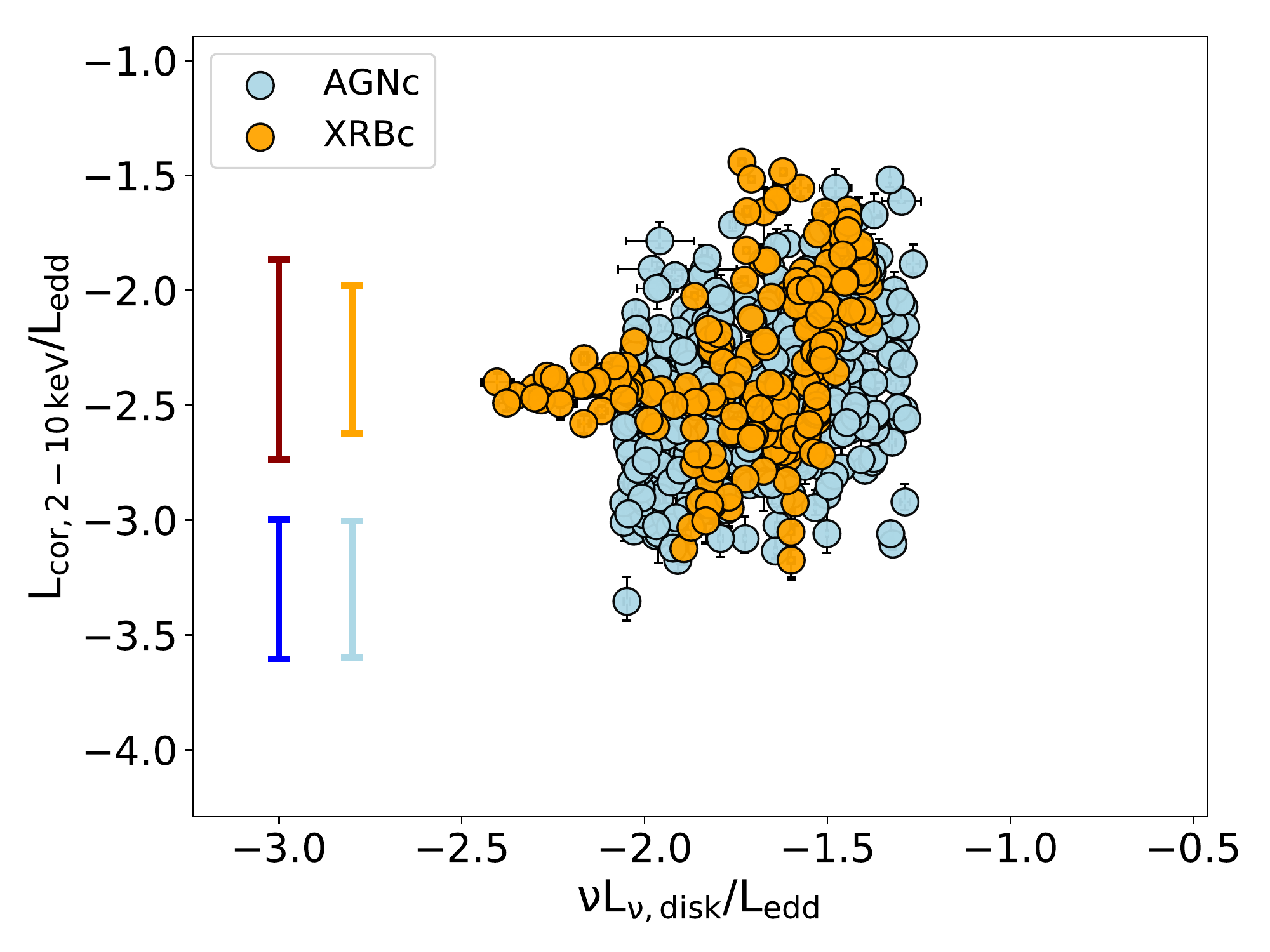}
	\includegraphics[width=0.86\columnwidth]{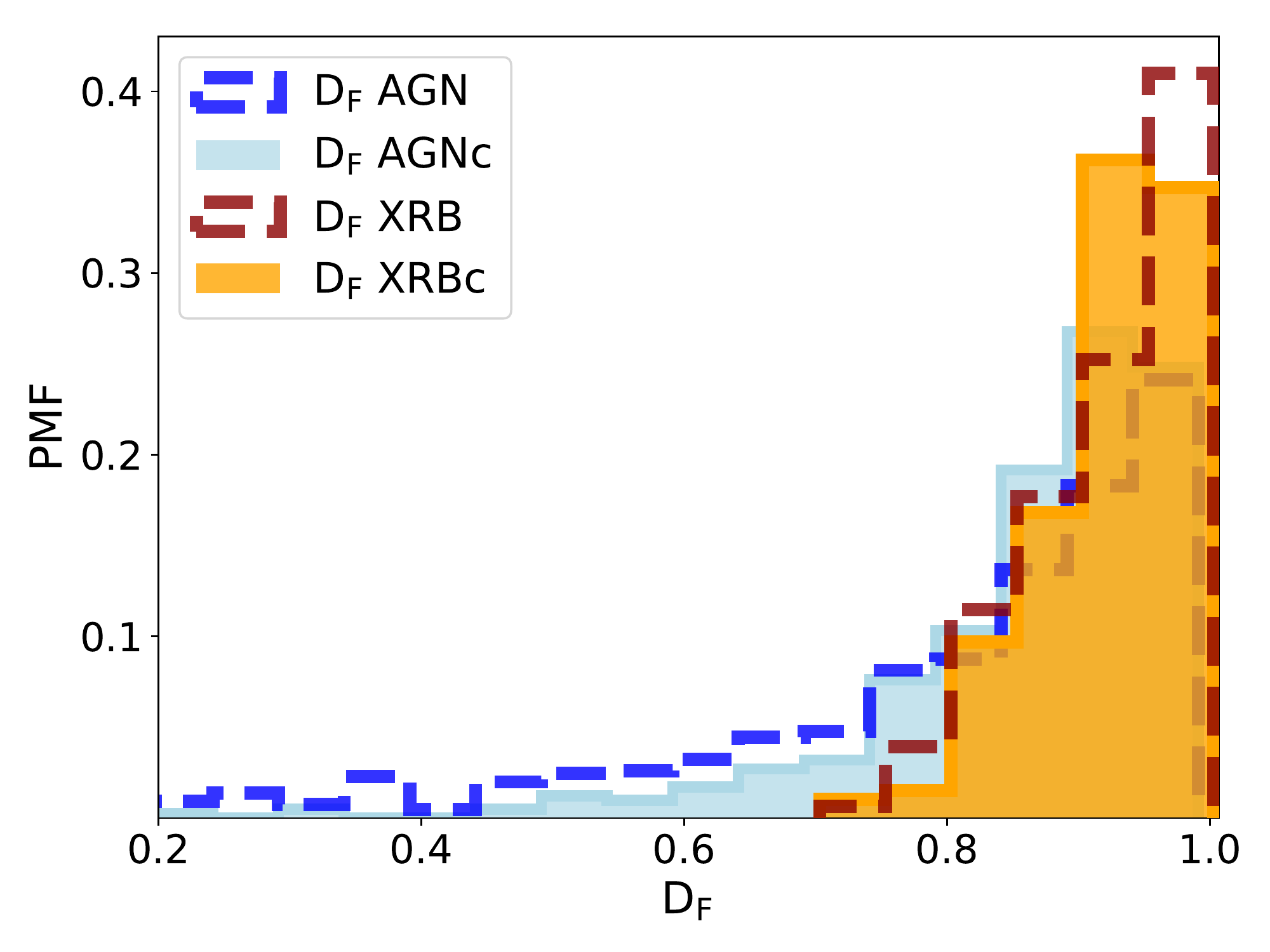}
	\includegraphics[width=0.86\columnwidth]{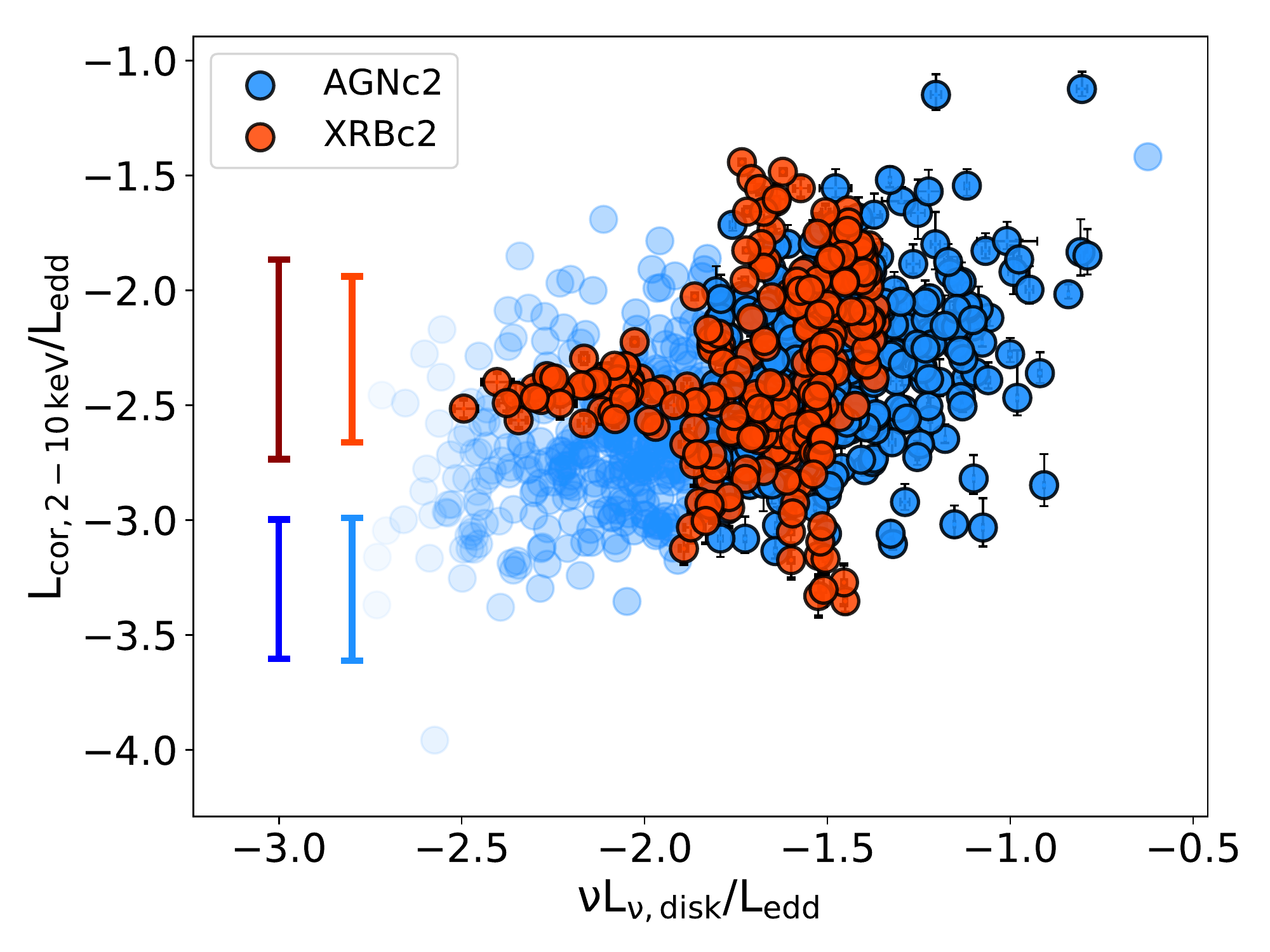}
	\caption{\emph{Left panels}: we show $\Gamma$, $\dot{m}$ and $D_F$ histograms of the full AGN and XRB samples (blue and red-dashed lines) with additional sub-samples: "AGNc" (lightblue) is obtained reshaping the original $\dot{m}$ distribution to be as narrow as the one in XRBs (using its 16th and 84th percentiles), albeit keeping the same median as in the full "AGN" sample; the same reasoning is applied with the $\Gamma$ distributions to select "XRBc" (orange) from "XRB"; instead, "AGNc2" (dark azure) is obtained selecting sources with $\dot{m}$ compatible within 0.4\,dex with the 5th-95th inter-quantile of "XRB", whereas "XRBc2" (dark orange) follows the same reasoning selecting $\Gamma$ compatible within errors with the "AGN" $\Gamma$ distribution. The vertical dashed lines in the middle-left panel represent the location of the 1st and 99th percentile of multiple XRBs $\dot{m}$ distributions, obtained converting the fit $T_{in}$ spanning $a_*=0-0.98$ and $m=5-10$, while we highlight with a red-dashed histogram the one obtained with $a_*=0.5$ and $m=5.8$ (see text). \emph{Right panels}: mass-normalised $\log L_{disk}-\log L_{cor}$ plane for the full AGN and XRB samples (top) and the above mentioned "c" (middle) and "c2" (bottom) sub-samples. In the latter case, the uncertainty in determining $\dot{m}$ for XRBs (vertical dashed lines in the middle-left panel) reverberated in the dark-azure points spreading in the $\log L_{disk}-\log L_{cor}$ plane, covering the same dynamic range of the full AGN sample. The computed observed scatter of each relation is shown on the side, with the same color coding as the data: it is, in dex, $0.30\pm0.01$ and $0.43\pm0.02$ for the full samples of AGN and XRBs, respectively; it then uniforms to $0.30\pm0.01$ and $0.33\pm0.02$ in the respective "c" sub-samples, while it becomes $0.31\pm0.01$ and $0.36\pm0.02$ for the "c2".
	}
	\label{fig:results_AGN_comparison}
\end{figure*}

\section{Comparisons between XRBs and AGN}
\label{sec:relation_XRB_vs_AGN}

For the rest of the AGN-XRB comparison, we adopt as disk proxy an Eddington-normalised (i.e. mass-normalised) monochromatic luminosity (i.e. $\nu L_{disk,\nu}/L_{edd}$), computed at $0.2\,$keV and $3000\AA$ for XRBs and AGN, respectively. The physical reason for adopting $0.2\,$keV is that this energy is roughly the XRB equivalent of what is $3000\AA$ for AGN, assuming $\nu\propto m^{-1/4}$ for a given Eddington ratio \citep[e.g.][]{Calderone+2013:BH_masses}. We note that, from an experimental point of view, at such a soft energy the required extrapolation of the RXTE response is big and there are surely covariances with the Galactic column density value. For the latter, we note that we left $N_H$ free to vary within a $\pm15\%$ of the tabulated value \citep[e.g.,][]{Arcodia+2018:nh15} in the spectral fits; thus this effect can be considered under control as we are marginalizing over this uncertainty interval in Galactic $N_H$. Then, we also tested in Appendix~\ref{sec:appendix_proxies} the impact of this change of disk emission proxy on the results discussed in the previous Sections. As a matter of fact, all the scatter values remain compatible within their 16th-84th inter-quantile range (see Table~\ref{tab:emcee_fd02}), stating that uncertainties in $N_H$ and in the RXTE response extrapolation are not significant. In addition, we added to the $0.2\,$keV fluxes an offset to correct for the known underestimation of soft fluxes in RXTE-like instruments by \texttt{DISKBB} (refer to the end Appendix~\ref{sec:bhspec_test} for a more detailed description).

Instead, the proxy for the corona is the Eddington-normalised broadband luminosity ($L_{cor}/L_{edd}$, also in erg\,cm$^{-2}$\,s$^{-1}$) computed in the $2-10\,$keV energy band, which is easily available for both XRBs and AGN. We adopted a black hole mass $m=5.8$ and a distance $d=7.8\,$kpc to estimate these luminosities for GX339-4 (see Appendix~\ref{sec:bhspec_test}). The estimates of mass and distance for GX 339-4 are very uncertain and debated, although the value is obviously the same for all data points and the resulting systematic error would be imprinted in the same way on both axis for all the points.

We show this mass-normalised $\log L_{disk}-\log L_{cor}$ plane in the top-right panel of Fig.~\ref{fig:results_AGN_comparison} and we report regression results in the top section of Table~\ref{tab:emcee_Ldisk_Lcor}. AGN data consist of a subset of 651 XMM-XXL broad-line AGN \citep[BLAGN;][]{Liu_zhu+2016:XMM-XXL,Menzel+2016:XXM-XXL_boss}, obtained excluding some objects to minimize the contamination from extinction in the UV \citep[selecting optical-UV continuum $\alpha'<-0.5$; see][]{Liu_teng+2018:Xobs_type1AGN} and obscuration in X-rays \citep[seclecting sources for which the 84th percentile of the $\log N_H$ posterior is $<21.5$; e.g.][]{Merloni+2014:obscuredAGN}. Also 44 radio-loud sources were excluded\footnote{Selected cross-matching the XMM-XXL sample with the FIRST survey \citep{Becker+1995:FIRST}, using as radio-loudness parameter both $R_X$ and $R_{uv}$ as defined in \citep{Hao+2014:RLdef}.}, which are thought to be scaled-up HIMSs \citep{Koerding:2006:HIMS_RL}, to validate our comparison with SSs and SIMSs only (see also Section~\ref{sec:discussion_states}). Thus, hereafter when referring to our AGN sample we will refer to radiatively-efficient radio-quiet AGN.

\begin{table}[tb]
	\footnotesize
	\caption{Summary of slope and scatter of the mass-normalised $\log L_{disk}-\log L_{cor}$, with a monochromatic disk proxy at $0.2\,$keV energy band, on the full AGN and XRBs samples and their subsets, as shown in Fig.~\ref{fig:results_AGN_comparison} and described in Section~\ref{sec:relation_XRB_vs_AGN}.}
	\label{tab:emcee_Ldisk_Lcor}
	\centering
	\begin{tabular}{C{0.07\columnwidth} C{0.4\columnwidth} C{0.25\columnwidth}}%
		\toprule
		\multicolumn{1}{c}{Sample} &
		\multicolumn{1}{c}{Slope} &		
		\multicolumn{1}{c}{Scatter} \\
		\midrule
		AGN & $0.49\pm0.03$ & $0.30\pm0.01$ \\
		XRB & $0.39\pm0.11$ & $0.43\pm0.02$ \\
		\midrule
		AGNc & $0.51\pm0.07$ & $0.30\pm0.01$ \\
		XRBc & $0.55\pm0.10$ & $0.33\pm0.02$ \\
		\midrule
		AGNc2 & $0.49\pm0.06$ & $0.31\pm0.01$ \\
		XRBc2 & $0.36\pm0.09$ & $0.36\pm0.02$ \\
		\bottomrule	
	\end{tabular}
\end{table}

The observed scatter for the AGN sample is $\sim0.31$\,dex. This is higher that the putative upper-limit on the real physical scatter of the relation, tentatively estimated at $\lessapprox0.19-0.20$ by controlling for non-simultaneity and variability \citep[e.g.,][]{Vagnetti+2013:variab_onalphaOX,Lusso&Risaliti2016:LxLuvtight,Chiaraluce+2018:dispandvariab_Lx_Luv} and potential instrumental calibration uncertainties \citep{Lusso2019:instr_effects}. This is partially because no further selections (i.e. on $\Gamma$ or X-ray counts) were performed in this work. Still, the observed scatter in AGN is incompatibly lower than in the XRBs data-set ($\sim0.43$\,dex). Furthermore, the dynamic range in disk luminosity is obviously larger for AGN (top-right panel of Fig.~\ref{fig:results_AGN_comparison}) and this is related to the wider $\dot{m}$ distribution (middle-left panel of Fig.~\ref{fig:results_AGN_comparison}). The accretion rate for GX339-4 is shown with a red-dashed line distribution and was estimated from the fit $T_{in}$ using the standard formulae of the multi-color black body used in \texttt{DISKBB} \citep{Mitsuda1984:diskbb} with the modifications of \citet{Kubota+1998:modificationDISKBB}, taking the radiative efficiency and innermost-stable circular orbit (ISCO) given a spin of 0.5 and assuming $m=5.8$ (see Appendix~\ref{sec:bhspec_test} for these estimates). The vertical dashed lines in the middle-left panel of Fig.~\ref{fig:results_AGN_comparison} represent the location of the 1st and 99th percentile of the same distribution spanning from spin 0 to 0.98 and $m$ from 5 to 10.

Also the $\Gamma$ distribution appears significantly different and it is narrower and peaked to harder values in AGN compared to XRBs (top-left panel in Fig.~\ref{fig:results_AGN_comparison}). Before thoroughly addressing the possible reasons why these distributions are different (see Section~\ref{sec:discussion}), it is intriguing that once these distributions are made equally narrow the overlap in the $\log L_{disk}-\log L_{cor}$ plane becomes remarkable. For instance, in the middle-right panel of Fig.~\ref{fig:results_AGN_comparison} lightblue points are a subset (named "AGNc") of the parent AGN sample obtained by selecting all sources with $\dot{m}$ values (taken conservatively with a $\sim0.4\,$dex systematic uncertainty, coming from the mass measurement) compatible with an inter-quantile range as wide as the 16th-84th range of the $\dot{m}$ distribution of the XRB sample; orange points are instead a subset (named "XRBc") of the XRB parent sample obtained by selecting all sources with $\Gamma$ values compatible with an inter-quantile range as wide as the 16th-84th range of the $\Gamma$ distribution of the AGN sample (see also the left panels in Fig.~\ref{fig:results_AGN_comparison}). In this case the distributions were kept at the same median values and simply narrowed accordingly to the other source class. However, we note that a very similar result for the observed scatter is obtained if the AGN and XRBs $\Gamma$ and $\dot{m}$ distributions are uniformed in a different way (defined with "c2", see Fig.~\ref{fig:results_AGN_comparison}), namely taking values of $\Gamma$ ($\dot{m}$) for XRBs (AGN) that are compatible within errors with the 5th-95th inter-quantile ranges of the analogous distribution for AGN (XRBs). Regression results on both sets of AGN and XRBs sub-samples are shown in Table~\ref{tab:emcee_Ldisk_Lcor}, all showing a compatible scatter within errors around $\sim0.33\,$dex. This is the reference value we attribute to the observed scatter in the $\log L_{disk}-\log L_{cor}$ plane for XRBs, using GX339-4 as a test case.

\subsection{On the faint-inefficient side}
The AGN disk-corona connection was studied for decades and, more recently, it was also tested in XRBs with an analogous proxy \citep{Sobolewska+2009:alphaox_GBH}. A comparison between the two BH classes was then performed using this observable, now more generically a "corona" loudness since in XRBs both components emit in X-rays, in \citet{Sobolewska+2011:simul},  which produced a set of simulated AGN spectral states scaling the luminosity ($\propto M_{BH}$) and the disk temperature ($\propto M_{BH}^{-1/4}$) from a selection of the XRB GROJ1655-40 spectral fits. The authors predicted an inversion of the corona loudness trend with $\lambda_{edd}$ to occur at low luminosity ($\lambda_{edd}\lessapprox0.01$), approximately where the accretion flow is thought to become radiatively inefficient \citep[e.g.,][]{Maccarone2003:Ltransition,Noda2018:CLAGN_mdot0.02}. This transition was recently confirmed by \citet{Ruan+2019:analogous_struct} using changing-look (or state) AGN \citep[or quasars, here referred to as CLAGN; e.g.,][]{LaMassa+2015:CLAGN,MacLeod+2016:CLAGN,Trakhtenbrot+2019:CLintheact} in their shut-down phase. 

However, with respect to the more well studied radio-to-X-ray correlations, one should be careful in testing the inefficient mode of accretion with UV-to-X-ray proxies in AGN and with scaled-up XRB spectra (i.e. not direct flux measurements). For instance, the simulated AGN spectral states were obtained by \citet{Sobolewska+2011:simul} for the complete hard-to-soft (i.e. inefficient to efficient) outburst using mass scaling laws that are however suitable only for efficient flows. Moreover, the monochromatic flux proxies for the putative disk-corona components were computed on the full model and not on the model components separately, thus in the hard state this corresponds to measuring two fluxes of the hard power-law component and the resulting corona loudness is the photon index itself. While it is true that in XRBs $\Gamma$ shows an inversion in the trend with the Eddington ratio, going from softer-when-fainter to softer-when-brighter \citep[e.g.][]{Corbel+2006:soft_chandra,Wu+2008:gamma_inversion,Russell+2010:softeningXTEJ1550,Homan+2013:soft_chandra,Kalemci+2013:soft_RXTE}, this inversion should be tested with $\Gamma$ also in AGN and not with the $\alpha_{OX}$ slope. Furthermore, RXTE results at low fluxes should be taken with care \citep[see, e.g., the discussion in][]{Homan+2013:soft_chandra}, as Chandra finds this putative softening at much lower fluxes and for a very narrow range in flux, below which $\Gamma$ saturates in a plateau \citep{Plotkin+2013:plat_in_quiesc,Plotkin+2017:heartbeat} rather than continuing in a v-shape pattern as claimed by earlier RXTE results. Finally, whatever the origin of the X-ray power-law is in XRBs hard states \citep[see, e.g.,][]{Heinz+2003:FP_theory,Markoff+2003:jets_FP}, it is not trivial to assess how this relates to what produces the UV emission in faint AGN, as one is forced to extrapolate assuming some QSO-like continuum and to rely on a very good handle on the host-galaxy subtraction \citep[but see][for some interpretations]{Ruan+2019:analogous_struct}. Our purpose here instead is to study the disk-corona relation in AGN and XRBs only in the bright end of (radiatively efficient) accretion states, where, despite the supposedly more secure observational proxies, there are still puzzling differences between the two source classes.

\section{Discussion}
\label{sec:discussion}

Results from Section~\ref{sec:relation_XRB_vs_AGN} show that AGN and XRBs overlap quite nicely in the $L_{disk}-L_{cor}$ plane, in terms of a compatible observed scatter ($\sim0.30-0.33$\,dex) and dynamic range on the x-axis, but only after similarly broad $\Gamma$ and $\dot{m}$ distributions, which represent the diversity in coronae and disks, respectively, were selected. This was of course merely a sanity check on the putative AGN-XRB analogy and it is indeed interesting to understand why the two distributions appear different.

\subsection{On our selection of AGN and XRBs accretion states}
\label{sec:discussion_states}

Our comparison performed in Section~\ref{sec:relation_XRB_vs_AGN} relied on the key assumption that radiatively-efficient AGN not dominated by the jet emisson are essentially scaled up XRBs in their SS and SIMS. This association is based on the fact that for both source classes the radio emission appears to be quenched with respect to their radio-loud phases, looking both at the fundamental plane of accretion \citep{Maccarone+2003:soft_vs_modAGN} and at the disk-fraction/luminosity diagram \citep{Koerding:2006:HIMS_RL}; and on the fact that the corona-loudness was already found to be broadly compatible \citep{Sobolewska+2009:alphaox_GBH}. It is true that the definition of XRBs accretion states based on both spectral (via $D_F$) or timing (via fractional rms) analysis is rather a continuum and the same should be true also for AGN. This being said, from timing analysis constraints one can have a fairly reliable grasp on the HIMS-SIMS and SIMS-SS transitions \citep[e.g.,][and references therein]{Belloni+2016:review} and this is indeed the criterion on which we mostly relied to select our SS-SIMS sample (see Section~\ref{sec:gx339}). 

\begin{figure}[tb]
	\centering
	\includegraphics[width=0.99\columnwidth]{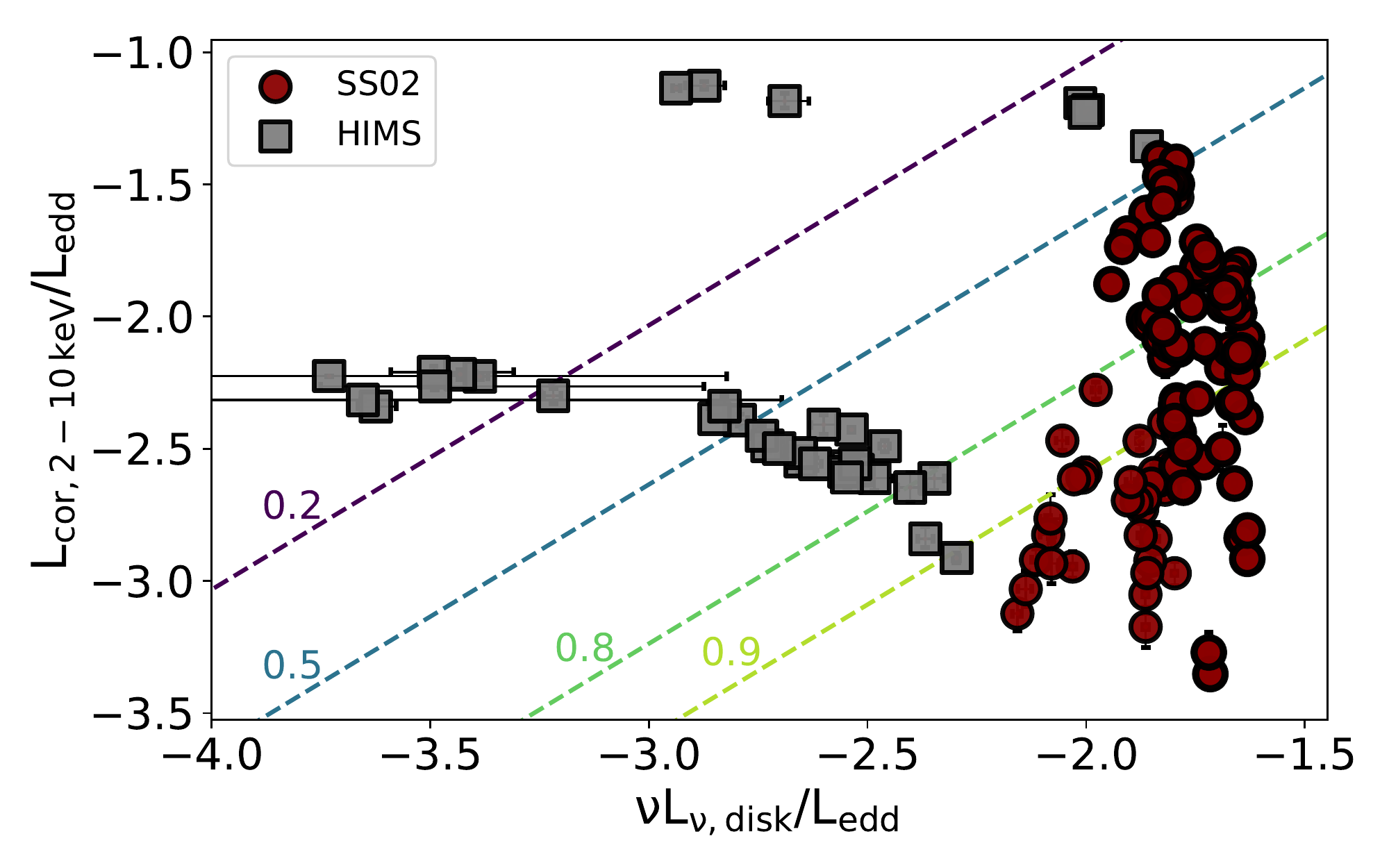}
	\includegraphics[width=0.95\columnwidth]{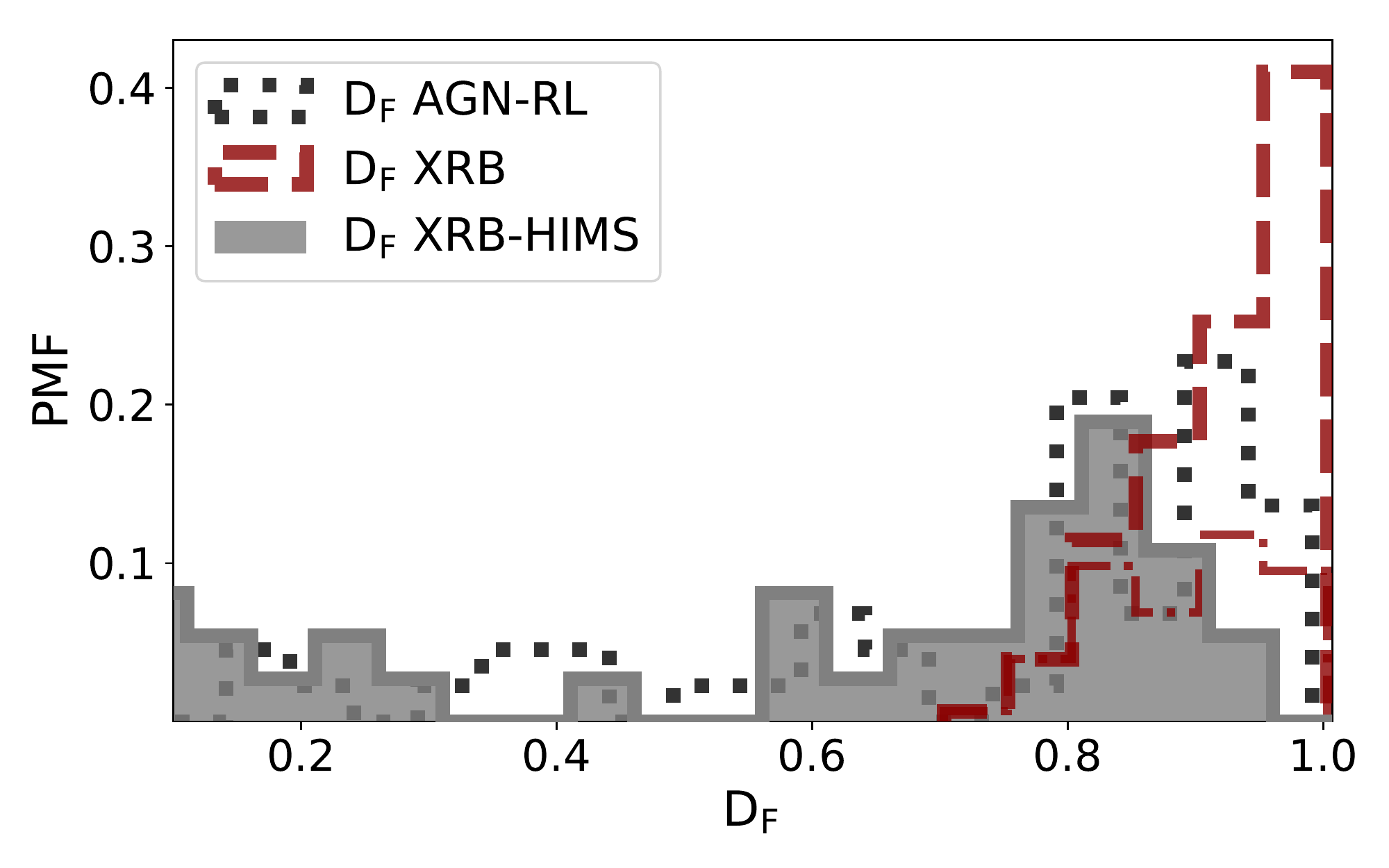}	
	\caption{\emph{Top panel}: same as the top right panel of Fig.\ref{fig:results_AGN_comparison}, with only the SSs and SIMSs of the 2002-2003 outburst (darkred circles, with SIMS highlighted by a thicker black contour) and its HIMSs (grey squares). Lines of approximately constant $D_F$ are show with dashed lines, the related values shown in the Figure. \emph{Bottom panel}: same as the bottom left panel of Fig.\ref{fig:results_AGN_comparison}, but with again the SS-SIMS (darkred dashed, with SIMS only highlighted by the dot-dashed line) and HIMS (grey) of the 2002-2003 outburst, plus the 44 radio-loud AGN excluded from our AGN sample (black dotted line).}
	\label{fig:himscheck}
\end{figure}

In this Section our aim is to further elaborate on the choice of including SSs and SIMSs and excluding HIMS from our XRB sample. We note that a comparably clear accretion state separation, which is nicely obtained in XRBs with fractional rms constraints, is more elusive for AGN where we can only more crudely rely on radio-loudness or spectral estimates. Then, we first tested whether the inclusion of SIMSs could be also motivated a posteriori comparing $D_F$ estimates for both our XRBs and AGN samples (see the bottom left panel of Fig.~\ref{fig:results_AGN_comparison}). For AGN $D_F$ was computed extrapolating the $2-10\,$keV catalog value to the bandwidth used in Eq.~\ref{eq:Df} for the corona emission and with the disk luminosity defined in the XMM-XXL AGN catalog \citep{Liu_zhu+2016:XMM-XXL}, that was approximated with standard thin-disk formulae from an optical monochromatic luminosity. The two distributions in the bottom left panel of Fig.~\ref{fig:results_AGN_comparison} do look quite similar with just a longer tail at low $D_F$ for AGN, although one must bear in mind that the full-band $L_{disk}$ estimate for AGN suffers from a much more uncertain extrapolation of the peak in the UV. From this comparison, there is no apparent reason to exclude SIMSs, which by definition seat at the lower end of the reported $D_F$ distribution of our XRBs sample (red dashed and dot-dashed histograms in Fig.~\ref{fig:himscheck}, see also Section~\ref{sec:soft}).

Then, we also tested the exclusion of HIMS from our analysis by fitting with the same disk-corona model (see Section~\ref{sec:model}) the HIMS states of the 2002-2003 outburst, as defined by \citet{Belloni+2005:OTB02} with timing analysis constraints, which we note are independent by $D_F$-based classifications. In the top panel of Fig.\ref{fig:himscheck} we show the mass-normalized $L_{disk}-L_{cor}$ plane (as in the top right panel of Fig.\ref{fig:results_AGN_comparison}) with the SS02 states (i.e. SSs plus SIMSs, darkred) plus the newly fit HIMSs (grey squares). SIMSs are highlighted with a thicker black contour and we also show lines of approximately constant $D_F$ with related values annotated in the Figure. HIMSs look indeed as different branches in the disk-corona emission plane, similarly to HLD diagrams, perhaps confirming that they are dominated by the jet emission processes while the SS-SIMS are not. In the bottom panel of Fig.\ref{fig:himscheck} we display instead the related $D_F$ distributions for SS-SIMSs (dashed red) and HIMSs (grey), which show again different properties with a small expected overlap, also visible in the top panel via the drawn $D_F$ lines. Then, we also report the radio-loud (but radiatively efficient) AGN distribution (black-dotted line) of the 44 such sources in XMM-XXL that were excluded from our analysis, which is remarkably similar to the HIMS confirming the analogy between the two source classes \citep{Koerding:2006:HIMS_RL}.

Hence, we verified that our analysis included a fairly pure sample of radiatively-efficient accreting systems which are not jet-dominated, selected among SS-SIMS in XRBs and (optical and X-ray) bright radio-weak AGN.

\subsection{On the different $\dot{m}$ distribution in AGN and XRBs}
\label{sec:discussion_mdot}

The different $\dot{m}$ distribution and dynamic range in $\log L_{disk}$ covered by the two samples (see Fig.~\ref{fig:results_AGN_comparison}) might indicate that the two systems do not follow the same accretion regimes. However, the almost four orders of magnitude span in AGN mass can play a role in enhancing this difference. As a matter of fact, predictions from disk-corona models do indicate that for the single mass the $L_{disk}-L_{cor}$ in AGN is in place but with a much lower dynamic range in $\log L_{disk}$ \citep[a bit more than one order of magnitude for the typical $m\sim10^8-10^9$ spanning $\dot{m}\sim0.03-1$; e.g.,][]{Kubota+Done2018:model_lx_luv,Arcodia+2019:LxLuv}, similarly to our XRB results (see Fig.~\ref{fig:results_Fdisk_Fcor} and~\ref{fig:results_Fdisk_Fcor_all}). Moreover, $\dot{m}$ values for both AGN and XRBs are not a secure measurement, particularly if compared to a quantity such as $\Gamma$: $\dot{m}$ values for the AGN sample were computed with a single bolometric correction \citep[5.15;][]{Shen+2008:BC_used} on the monochromatic luminosity at $3000\AA$, divided by a notoriously uncertain mass estimate \citep{Shen+2008:BC_used} which also hampers the selection of AGN within a narrow mass range; and $\dot{m}$ for XRBs was converted from the fit $T_{in}$ distributions assuming a $m=5.8$ and $d=7.8\,$kpc and changing these values, even within some reasonable intervals around spin and mass (e.g., $a_*=0-0.98$ and $m=5-10$), would significantly shift the $\dot{m}$ distribution (see middle-left panel in Fig.~\ref{fig:results_AGN_comparison}). However, its width would remain approximately the same and this validates our exercise in Section~\ref{sec:relation_XRB_vs_AGN}, where we relied on the width rather than the location of the XRBs $\dot{m}$ distribution (i.e. relying on the "c" sub-samples, see middle-right panel of Fig.~\ref{fig:results_AGN_comparison}). We conclude that AGN and XRB do appear with different $\dot{m}$ distributions and dynamic range in $\log L_{disk}$, although a significant role is played by the uncertainties and systematics on the $m$ and $\dot{m}$ estimates. Thus, we can not rule out that the intrinsic $\dot{m}$ distributions are instead broadly compatible.

\subsection{On the different $\Gamma$ distribution in AGN and XRBs}
\label{sec:discussion_gamma}

In principle, $\Gamma$ is a quantity that can be more securely estimated (see Appendix~\ref{sec:appendix_robustness}). In our work, we found evidence that XRBs have a distribution that is broader and shifted to softer values with respect to AGN (see Fig.~\ref{fig:results_AGN_comparison}). This result is puzzling and deserves a more in-depth analysis. 

The AGN sample used in this work has a $\Gamma$ distribution with mean and standard deviation of $\Gamma=2.06\pm0.11$. This is in line with diverse large samples of bright AGN (with either no jet or a non-jet-dominated emission) which show a fairly narrow distribution of $\Gamma$, typically centered between $\Gamma=1.9-2.1$ up to high redshift \citep[e.g.][and references therein]{Vito+2019:highz}, with a dispersion spanning $\approx0.10-0.40$, not always corrected for uncertainties and depending on the sample selection \citep[e.g.][]{Zdziarski+2000:sey_OSSE,Caccianiga+2004:HBS28sample,Piconcelli+2005:PGqso,Beckmann+2009:integral,Young+2009:sdss_xmm,Mateos+2010:XWASsample,Corral+2011:XBS,deRosa2012:hard,Liu_zhu+2016:XMM-XXL,Ricci+2017:BASS,Zappacosta+2018:Nustar,Ananna+2019:CXB}. Instead, from our analysis of GX 339-4 the fit $\Gamma$ values compile a distribution that is broader and shifted to softer values, with a mean and standard deviation of $\Gamma=2.19\pm0.21$. This is consistent with previous results of GX 339-4 in its SSs and SIMSs obtained with RXTE alone \citep{Zdziarski+2004:dist,Dunn+2008:SS04}, with XMM-Newton and INTEGRAL data \citep{Caballero-Garcia+2009:gx339_xmm_integral} and with simultaneous XMM-Newton and RXTE data \citep{Aneesha+2019:gx339_longterm}. For instance, cross-matching our MJD with \citet{Dunn+2010:global} we computed $\Gamma=2.26\pm0.47$ for 213 states, and from the SSs and SIMSs in \citet{Motta+2009:SS07} we computed $\Gamma=2.31\pm0.15$. These distributions are peaked at even softer energies, which is also in line with what is generally observed in disk-dominated states for all XRBs \citep[e.g.,][]{Remillard:review}. We note that this difference persists also after accounting for uncertainties in the $\Gamma$ values: we sampled for the mean and intrinsic deviation with \texttt{emcee} values using the likelihood defined in \citet{Maccacaro+1988:Einstein}, obtaining $2.07\pm0.08$ ($2.20\pm0.16$) for the AGN (XRB) sample. Hence, at least for our samples, uncertainties do not play a significant role in the difference between the two source classes.

In Section~\ref{sec:relation_XRB_vs_AGN} we showed that a compatible scatter ($\sim0.30-0.33$\,dex) of the $\log L_{disk}-\log L_{cor}$ between AGN and XRB is reached when both $\Gamma$ distributions were taken with the same 16th-84th inter-quantile width, leaving the median values unchanged (2.04 and 2.21, respectively). This hints that no matter where the preferred Comptonisation slope lies, a similar scatter in $\Gamma$ reverberates in a similar diversity in X-ray coronae for a given disk (see the middle-right panel of Fig.~\ref{fig:results_AGN_comparison}). As a matter of fact, in the above-mentioned literature of AGN samples there is fair concordance on where most of the observed $\Gamma$ values lie, although there is a variety of dispersion estimates according to the varying sample selections (i.e. soft or hard X-rays), instruments, analysis techniques and model degeneracies. Interestingly, in order to match the Cosmic X-ray Background shape, a diversity in photon indexes is needed \citep[e.g. with a dispersion of $\sim0.2$;][]{Gilli+2007:CXB} with an impact also on the complex parameters space involved, part of which includes $\Gamma$, the reflection fraction and the high-energy cutoff \citep[e.g.,][]{Ananna+2019:CXB}. In particular, \citet{Ananna+2019:CXB} explored the allowed regions of this very complex parameters space and showed that, independently from the luminosity function assumed, even a broad $\Gamma$ distribution with a dispersion of $\sim0.2-0.3$ can reproduce the CXB if the peak shifts to softer values, in order not to overestimate the production of high-energy photons. If the true intrinsic $\Gamma$ distribution of AGN followed this scenario, it would be somewhat closer to the one observed with XRBs. 

However, combining in a homogeneous picture several AGN samples with very different selections and characteristics is beyond the scope of this work. We here focus on addressing the role of possible contaminants shaping the observed $\Gamma$ distributions (Sections~\ref{sec:soft}, \ref{sec:bias_obs} and~\ref{sec:bias_refl}) and, if these are understood, explore possible similarities and differences in the physical process producing the observed distribution (Sections~\ref{sec:phys} and~\ref{sec:phys_sim}). 


\subsubsection{Possible biases: the soft-excess component}
\label{sec:soft}

We note that in our AGN sample only BLAGN were included, although the parent XMM-XXL sample contains also narrow-line AGN (NLAGN). However, the exclusion of NLAGN does not have an impact, as their $\Gamma$ distribution completely overlaps with the BLAGN one \citep{Liu_zhu+2016:XMM-XXL}. Then, an obvious objection is that we included SIMSs for XRBs, namely the brightest spectra with almost equally strong soft and hard components, although we did not include narrow-line Seyfert1 galaxies (NLS1), which might be the closest counterpart to SIMSs \citep[e.g.][]{Pounds+1995:nls1,Gierlinski+2008:periodicity}. This could indeed contribute in broadening the AGN $\Gamma$ distribution, since the X-ray emission in NLS1s is observed to be very soft \citep{Boller+1996:NLS1}. However, the overall emission is softer because an additional spectral component, broadly referred to as "soft-excess", is present \citep[e.g.,][]{Done+2012:soft_excess}. If this extra-component is taken into account, the emission from the hard component only would be compatible with the X-ray slopes from hot Comptonizing coronae in BLAGN: for instance, in NLS1 values around $\Gamma\sim1.8-2.1$ are obtained analysing spectra only above $2\,$keV \citep[e.g.][]{Ai+2011:NLS1} or looking only at the hard photon-index when a broken power-law is used \citep[e.g.][]{Grupe+2010:swift}. Hence, the $\Gamma$ distribution in BLAGN can be considered representative of the observed (i.e. not necessarily the intrinsic) properties of hot coronae in AGN. However, since the hard X-ray emission in soft NLS1 is matched to the one in BLAGN only when an extra-component is added to account for the soft-excess, the question is then whether one should expect the same to happen also in XRB SIMSs. As a matter of fact, if the different spectral states of Mrk 1018 \citep{Noda2018:CLAGN_mdot0.02} are compared to XRBs on a HLD, the brightest ones with a strong soft-excess component would broadly overlap with bright SIMSs and not with SSs (H. Noda, private communication). Moreover, there has been evidence of intermediate states requiring an additional spectral component to the thermal disk and hot Comptonisation alone \citep[e.g.][]{Kubota+2001:3comp,Kubota+2004:3comp,Kubota+2004:VHS_extra,Abe+2005:anomalous,Yamada+2013:extra_c,Hjalmarsdotter+2016:extracomp,Kawano+2017:extracomp,Oda+2019:extracomp}, also in GX339-4 itself \citep[e.g.][]{Kubota+2016:model_3comp_gx339}.

In order to test the impact of these states on our results, we excluded SIMSs as defined in the earlier literature (see Section~\ref{sec:gx339}) and the mean and standard deviation values change from $\Gamma=2.19\pm0.21$ to $2.16\pm0.21$. Thus, even including only previously defined SSs the XRB $\Gamma$ distribution is still broader and peaked at softer slopes with respect to our AGN sample. Instead, the scatter in the $F_{disk}-F_{cor}$ plane would go from $0.43\pm0.02$ to $0.38^{+0.03}_{-0.02}$\,dex, thus it would be compatible within errors but slightly smaller. Moreover, we conservatively tested a different selection, excluding states below $D_F\sim0.8$ \citep[e.g.,][]{Dunn+2010:global}: the resulting mean with standard deviation is $\Gamma=2.18\pm0.21$, thus almost identical; the scatter in the $F_{disk}-F_{cor}$ plane would be $0.40\pm0.02$\,dex, thus again compatible within errors but slightly smaller. Hence, we here simply highlight that the role of SIMSs, which possibly include a soft-excess component, is not trivial and it may contribute in broadening the observed $\Gamma$ distribution and slightly increasing the scatter in the $F_{disk}-F_{cor}$ plane, although not to the extent needed to conciliate XRBs to AGN samples. Moreover, we showed in Section~\ref{sec:discussion_states} the AGN $D_F$ distribution, for which the threshold $D_F=0.8$ is actually the $\sim35$th percentile, with then a significant tail of lower $D_F$ values. This would argue against the exclusion of SIMSs from the comparison.

\subsubsection{Possible biases: X-ray absorption and continuum models}
\label{sec:bias_obs}

The AGN sample used in this work was compiled from the parent XMM-XXL BLAGN sample \citep{Liu_zhu+2016:XMM-XXL,Menzel+2016:XXM-XXL_boss} minimizing the extinction in the UV \citep[selecting optical-UV continuum $\alpha'<-0.5$; see][]{Liu_teng+2018:Xobs_type1AGN} and obscuration in X-rays \citep[seclecting sources for which the 84th percentile of the $\log N_H$ posterior is $<21.5$; e.g.][]{Merloni+2014:obscuredAGN}. We then tested the impact of the latter selection criterion on the observed $\Gamma$ distribution, since in AGN obscuration plays an important role within a complicated mixture of orientation and evolution effects \citep[e.g.][and references therein]{Klindt+2019:evolution}. Further, fitting for both absorption and continuum emission in X-ray spectra within the typical $\sim0.5-10\,$keV energy band leads to well-known covariances between the two parameters, which enhance or hampers a possible intrinsic correlation merely for observational and/or instrumental reasons. However, we tested this and the effect of selecting very un-obscured objects is minimal and the observed $\Gamma$ distribution is equally narrow: for the totality of 1659 objects in the XMM-XXL BLAGN sample the mean is $\Gamma=2.01\pm0.10$, with respect to 
$\Gamma=2.06\pm0.11$ of the adopted sub-sample. We also conservatively tested the impact of leaving the Galactic absorption free to vary within a $\pm15\%$ of the value tabulated in \citet{Willingale+2013:nhmol}, since this may broaden the spread in $\Gamma$ artificially. We used SS02 as representative of all outburts and kept the Galactic absorption column fixed, producing an almost identical $\Gamma$ distribution. 

Finally, since the continuum was modeled with a simple power-law for AGN \citep{Liu_zhu+2016:XMM-XXL} and with a Comptonisation model for XRBs in this work, we verified that adopting a simple power-law as well has negligible effects: we fit states in SS02 and SS07 and obtained a compatible distribution, even shifted even more to softer values with a slightly larger width. Hence, the use of the \texttt{NTHCOMP} model had a minor impact.

\subsubsection{Possible biases: X-ray reflection}
\label{sec:bias_refl}

\begin{figure}[tb]
	\centering
	\includegraphics[width=0.8\columnwidth]{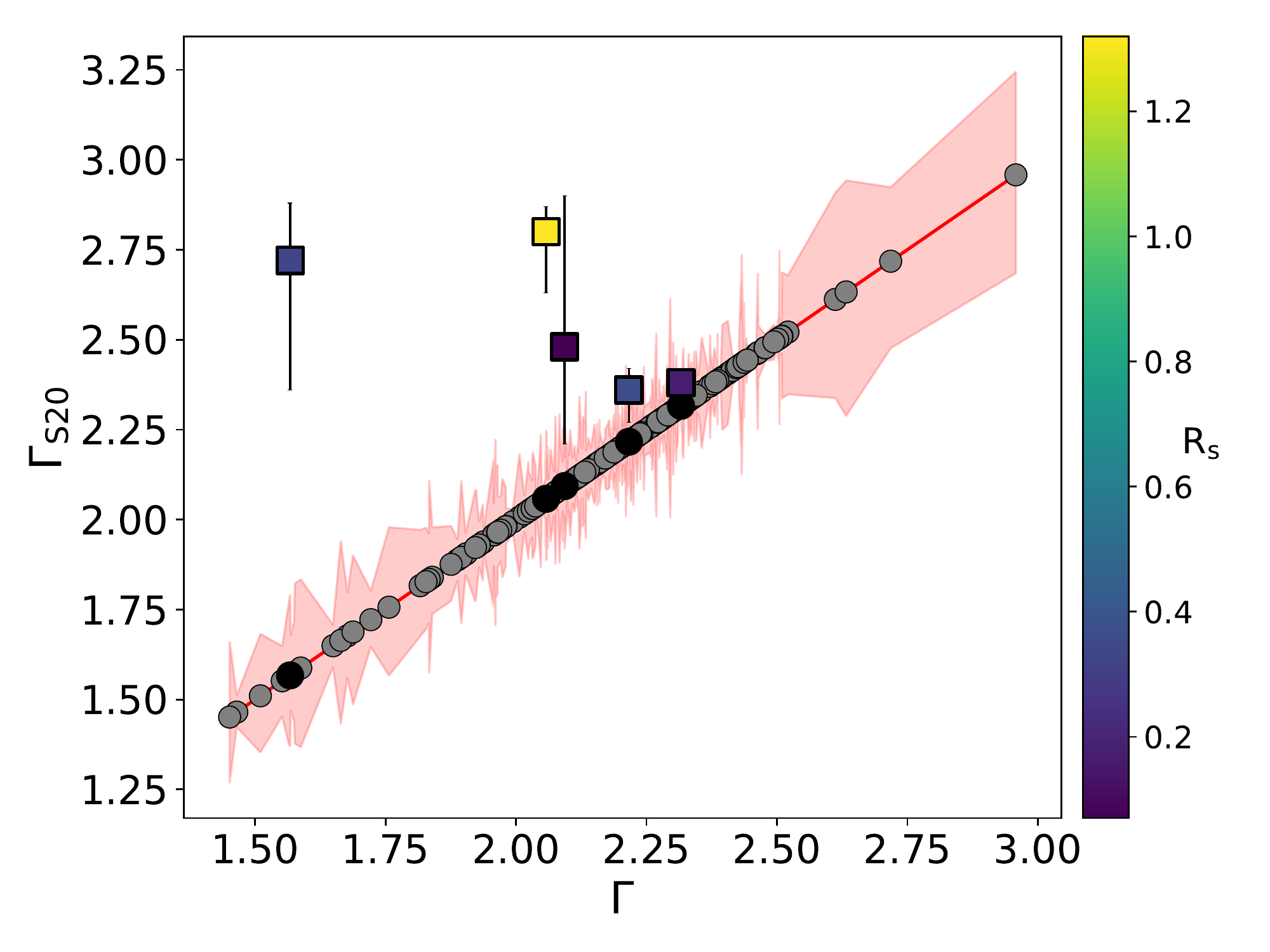}
	\includegraphics[width=0.82\columnwidth]{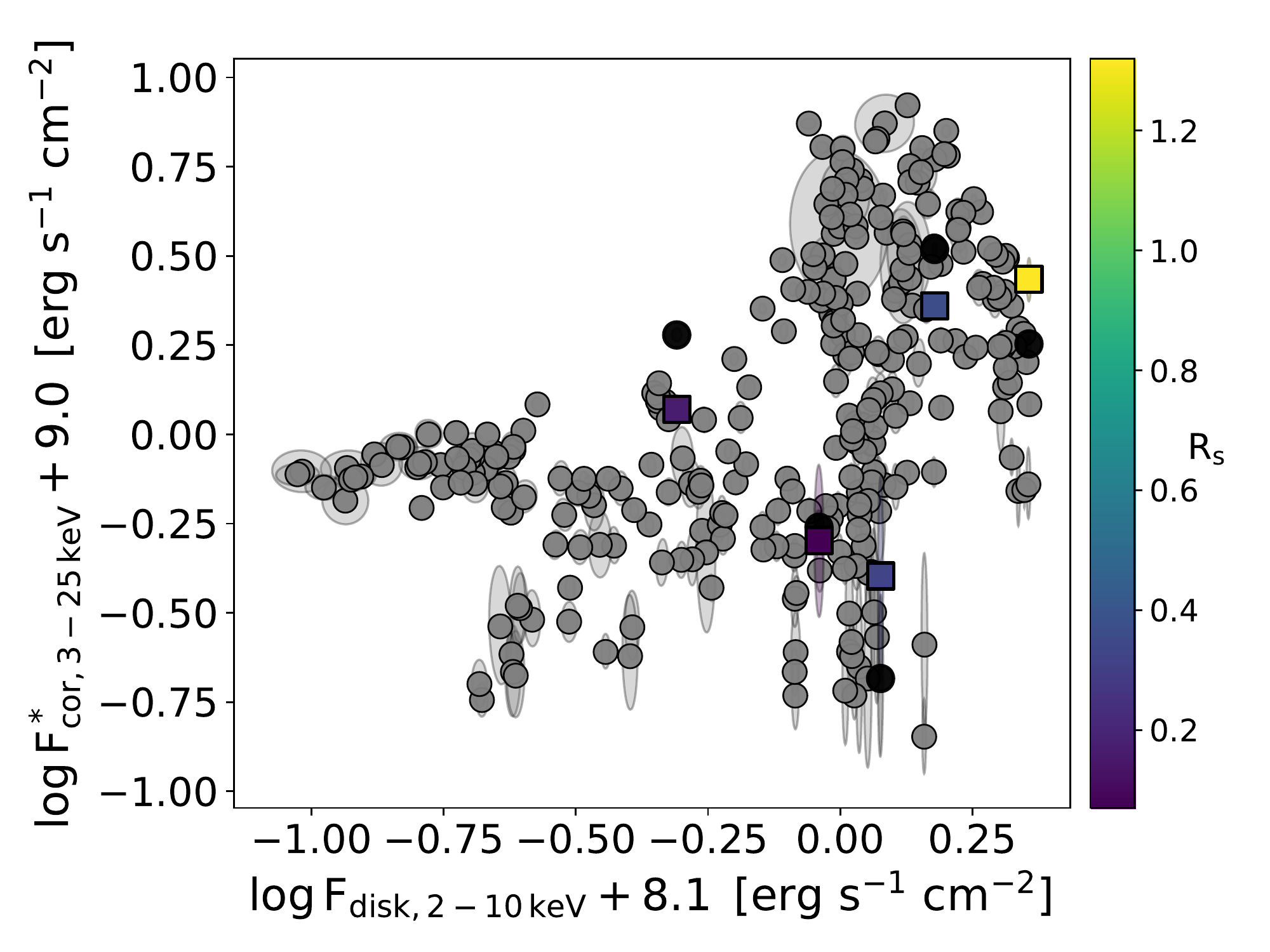}	
	\caption{\emph{Top panel}: comparison between our fit $\Gamma$ (black circles) and the values obtained by \citet{Sridhar+2019:reflGX339} for the five observations in common (squares, color coded by the reflection strength as defined by Sridhar et al.). The grey points with red contours refer to our whole sample. \emph{Bottom panel}: similar comparison in the $L_{disk}-L_{cor}$ plane where we highlight the difference between our original $F_{cor,\,3-25\,keV}$ (black circles) and the reflection-corrected $F_{cor,\,3-25\,keV}^*$ obtained with the reflection strength estimates computed by \citet{Sridhar+2019:reflGX339}.}
	\label{fig:reflcheck}
\end{figure}

Furthermore, X-ray reflection was included in the analysis of the AGN sample done by \citet{Liu_zhu+2016:XMM-XXL}. We here simply empirically tested whether excluding objects with the strongest X-ray reflection (including only sources in which the 16th percentile of $\log R$ was $<-0.2$ and the 84th was $<0.5$, where $R$ is the ratio of the normalisation of the reflection component with respect to the power-law component) had an impact on the observed $\Gamma$ distribution: the mean and standard deviation values become $2.07\pm0.11$, thus almost identical to $\Gamma=2.06\pm0.11$ of the sub-sample adopted here. 

For XRBs, we note that the reflection contribution was approximated in all spectral states with a Gaussian line, bound to be centered between 6.4 and 6.97\,keV, so its flux did not contaminate our $L_{cor}$ estimates. Moreover, fitting in the $3-25\,$keV band avoided most of the contamination from the Compton hump, although the reflection spectrum can be quite complex and its contribution should be at least tentatively quantified. A thorough treatment is beyond the scope of this paper, although we performed a few tests to exclude that our simplified treatment of the reflection features had an impact on our main results. We fit SS02 (which is representative of the range of observational properties explored with the complete XRB sample) with a \texttt{laor} model \citep{Laor1991:laor} and compared the newly obtained parameters (e.g. $\Gamma$, $T_{in}$) with the ones obtained with a simple Gaussian. The biggest effect resulted in a slight offset towards softer $\Gamma$ obtained with the \texttt{laor} model, although most of the parameters are compatible within 1-sigma uncertainty intervals and almost all of them within 3-sigma. Hence, we consider this not to imprint a significant impact on our results and, if anything, it would enhance the difference between the observed AGN and XRBs $\Gamma$ distributions strengthening our results. Moreover, recently \citet{Sridhar+2019:reflGX339} performed a detailed characterisation of the reflection features across the hard to soft transition of the 2002 and 2004 outbursts in GX339-4. With respect to the total 14 observations they used to sample SS02 and SS04, we only shared five, namely Obs. ID 40031-03-03-04 and 70110-01-33-00 for SS02 and Obs. ID 90704-01-03-00, 60705-01-76-00 and 90118-01-10-01 for SS04. The former two in SS02 were defined as SIMSs by \citet{Belloni+2005:OTB02} from timing analysis constraints thus included in our selection, while the latter in SS04 were included with our selection criterion following \citet{Belloni+2006:begSS04}. The remaining in \citet{Sridhar+2019:reflGX339} were instead defined either as hard or as hard-intermediate states by Belloni et al., thus they were not analysed in this work.

In the top panel of Fig.~\ref{fig:reflcheck} we compare our fit $\Gamma$ with the values from \citet{Sridhar+2019:reflGX339}, which apart from the reflection component shared the same model configuration for \texttt{DISKBB} and \texttt{NTHCOMP}. It is evident that including the reflection the incident $\Gamma$ becomes steeper, although our median source has $\Gamma\sim2.2$ so the typical displacement would be small. This also appears from other works that included a reflection component in the fit of SSs and SIMSs \citep[e.g.][]{Plant+2014:reflectionGX339} and it would increase even more the difference between the observed $\Gamma$ distributions in AGN and XRBs, also in line with the above-mentioned \texttt{laor} model test. Furthermore, in order to check the impact on our broadband fluxes for the corona emission, we tentatively corrected them using the reflection strength defined by \citet{Sridhar+2019:reflGX339}, namely the ratio between the observed reflected component and the incident continuum component in the $20-40\,$keV band. Our original fluxes $F_{cor,\,20-40\,keV}$ were then turned into reflection-corrected $F_{cor,\,20-40\,keV}^*$ and then the reflection-corrected $F_{cor,\,3-25\,keV}^*$ was extrapolated using the asymptotic \texttt{NTHCOMP} Photon Index computed by \citet{Sridhar+2019:reflGX339}. We show in the bottom panel of Fig.~\ref{fig:reflcheck} how this correction affected the five observations in common in the $L_{disk}-L_{cor}$ plane. We want to stress that this test on five sources can not be taken as conclusive, but since the change in flux is not dramatic we can exclude a huge impact of the reflection component on our results in the $L_{disk}-L_{cor}$ plane. A proper treatment of the reflection should be done directly via spectral fitting and even then, in SSs and SIMSs it is unclear to what extent the prominent disk emission contributes to the incident radiation, both self-illuminating the outer radii from the inner ones and with returning radiation due to General relativistic effects \citep[see, e.g.,][]{Connors+2020:self_irr}.

\subsubsection{Possible physical reasons for the different $\Gamma$ distributions}
\label{sec:phys}

In the previous Sections we investigated some possible reasons for which the observed $\Gamma$ distribution of the AGN and XRBs samples might have been biased narrow or broad, respectively. Nonetheless, none of them alone 
seems to play a major role and even a conspired combination of all is unlikely to explain all the differences. Thus, we can assume that the observed $\Gamma$ distributions appear different for AGN and XRBs in their radiatively-efficient phase and at least some or most of the differences is likely intrinsic to the physical mechanisms of the hard component emission. Of course the observed differences in $\Gamma$ could be due to different emission mechanisms being responsible for the coronal emission in the two source classes. However, and in line with essentially all past observational evidence, throughout this discussion we assume that the $\Gamma$ distributions arise from hot electrons \citep[thermally and/or non-thermally distributed;][]{Coppi+1999:hybrid,Gilfanov2010:phys} Compton (up)scattering the seed photons emitted by the thermal disk. 

In this framework, the observational evidence we presented here is that XRBs produce a preferentially softer emission than AGN. As a matter of fact, results from MONK, a general relativistic Monte Carlo code of Comptonised spectra in the Kerr space-time \citep{Zhang+2019:MONK}, indicate the opposite, namely that X-ray spectra in XRBs would appear harder if, apart from the different mass and seed photons temperature, the two source classes share the same geometry and extent of the corona, spin, inclination and accretion rate in Eddington units (W. Zhang, private communication). Regarding the inclination, the un-obscured AGN sample used here is likely composed by a mixture of objects below $\approx30^{\circ}-40^{\circ}$, and although in GX339-4 the inclination is still debated, even a large difference would have a small impact on the X-ray slope above $\sim2\,$keV \citep{Zhang+2019:MONK}. Regarding the accretion rate, we note that in Section~\ref{sec:relation_XRB_vs_AGN} we tested an AGN sub-sample (labeled "c2") with accretion rate values compatible with the XRB distribution, for which $\Gamma$ values were found to be consistent with the ones of the parent AGN sample, thus harder than in XRBs. Moreover, for both AGN and XRBs $\dot{m}$ should not be considered a solid estimate and we can not exclude that the two distributions are compatible within the very large uncertainties (see Section~\ref{sec:discussion_mdot}).

The spin is another largely unconstrained and still lively debated quantity for both AGN and XRBs \citep[e.g., for GX339-4, ][]{Kolehmainen+Done2010:spin_spec,Parker+2016:dist,Ludlam+2015:spin,Garcia+2015:spin}, although flux limited AGN samples are likely biased in being preferentially populated by high spinning sources from several different lines of reasoning \citep{Brenneman+2011:high-spin_bias,Vasudevan+2016:high-spin_bias2,Baronchelli+2018:refl,Reynolds2019:obs_spin}, including modeling the $L_{disk}-L_{cor}$ itself \citep{Arcodia+2019:LxLuv}. Moreover, the effect of the spin on the corona luminosity is likely degenerate with its geometry and extent and a thorough treatment of these unknowns is beyond the scope of this paper. However, we tried to qualitatively discuss their effect on our results. Using a simplified but physically motivated model which couples the disk and corona energetically \citep{Arcodia+2019:LxLuv}\footnote{Code available here: \href{https://github.com/rarcodia/DiskCoronasim}{https://github.com/rarcodia/DiskCoronasim}}, we were able to infer that for a given accretion rate\footnote{Intended in units of g\,s$^{-1}$, as the Eddington-normalized $\dot{m}$ is proportional to $\dot{M}$ times the radiative efficiency, which increases with the spin.} the mass-normalized corona luminosity increases with the spin (i.e. a factor $\approx2$ between $2-10\,$keV from $a_*=0$ to 0.998). Further, the corona luminosity also appears to increase with the radial \citep{Zhang+2019:MONK} and vertical extent of the corona \citep{Kara+2019:vert_cor, Alston+2020:dyn_cor}. Thus, it is remarkable that we observed a compatible normalization in the $L_{disk}-L_{cor}$ plane for the AGN population and GX339-4 (see right panels of Fig.~\ref{fig:results_AGN_comparison}). This might suggest that either the spin distributions and/or the extent of the X-ray coronae, both in mass-normalized units, are not too far apart; or that the compatible normalization in the $L_{disk}-L_{cor}$ plane is just a conspiracy of these multiple unknowns (i.e. one of the two source classes has lower spin but wider extent of the corona or viceversa). 

Finally, in case all the above quantities were found to be broadly compatible between the AGN and XRBs, a remaining possibility is that the $\Gamma$ distributions are different because the typical values for temperature and (or) optical depth are (is) not the same. As a matter of fact, the Comptonisation slope depends on both \citep[e.g.][]{Pozdnyakov+1983:compt_MC} and both can not be constrained in this work, since we do not have a good handle on the high-energy cut-off with RXTE-PCA (see footnote~\ref{note:Ecut}), which is in general the case for SSs \citep[e.g.][]{Grove+1998:gammaSS,Gierlinski+1999:soft_cyx1,Motta+2009:SS07}. Alternatively, the energy distribution of the hot scattering electrons might not be the same in the two source classes. However, in both radiatively-efficient AGN and XRBs we have a photon-rich environment and we are (relatively) far from the tails of the emitted spectrum, thus the underlying electron distribution is not necessarily a major concern \citep{Coppi+1999:hybrid}.

Hence, despite the different environmental conditions (a single star versus a galactic centre) and characteristics of the matter reservoir (different density, temperature, ionisation and pressure support), the phenomenology of the disk-corona energetic emission is radiatively-efficient AGN and XRBs seems indeed very similar (see Figure~\ref{fig:results_AGN_comparison}). What might not be then entirely understood is whether the physics of the disk-corona emission is also the same. Based on all the arguments in the discussion, our results are consistent with having disk-corona systems in AGN and XRBs undergoing the same physical processes under different conditions (e.g. temperature, optical depth and electron distribution in the corona, spin regime and/or heating-cooling balance..) and/or geometry (radial and vertical extent of the corona). They are also however consistent with a scenario in which the physical processes are not the same and the mass-normalized disk-corona energetics are comparable by chance, although we consider this less favorable and against decades of past results \citep[e.g.,][and references therein]{Merloni+2003:fund_plane,Maccarone+2003:soft_vs_modAGN,Falcke+2004:unify,Uttley+2005:variabSey,McHardy+2006:variab,Koerding:2006:HIMS_RL,Sobolewska+2009:alphaox_GBH,Sobolewska+2011:simul,Svoboda+2017:simult,Ruan+2019:analogous_struct}.

\subsubsection{On the similarities between AGN and XRBs despite the differences}
\label{sec:phys_sim}

In the previous section we outlined that, contrary to the observed difference in $\dot{m}$ for which we can not securely claim that the intrinsic distributions are actually compatible, $\Gamma$ values seem to be really intrinsically different for (radiatively-efficient and not jet-dominated) AGN and XRBs. However, we showed in Section~\ref{sec:relation_XRB_vs_AGN} that, when the two $\Gamma$ distributions are taken with the same width, independently on where the peak lies, both AGN and XRBs show a similar scatter of $\sim0.30-0.33\,$dex in the $\log L_{disk}-\log L_{cor}$ plane. Thus, as far as the disk-corona relation is concerned, it seems more important how similarly diverse (i.e. $\sigma_{\Gamma}$) X-ray coronae are rather than how different the typical one (i.e. <$\Gamma$>) is between the two source classes. As a matter of fact, another similarity is that in XRBs there is a clear dependence of $\log L_{cor}$ from $\Gamma$ (see Fig.~\ref{fig:results_Fdisk_Fcor} and~\ref{fig:results_Fdisk_Fcor_all}) in a softer-when-brighter pattern (where both softer and brighter refer to $\log L_{cor}$ alone in this context, see Fig.~\ref{fig:seds}). This trend is absent in our sample of BLAGN \citep[see also][]{Beckmann+2009:integral,Corral+2011:XBS,deRosa2012:hard} or hidden among the various mass, distance and inclination effects, although steeper $\Gamma$ values for brighter sources have often been noticed in AGN samples \citep{Sobolewska+2009:spec_variab,Mateos+2010:XWASsample,Gibson+2012:spec_var,Serafinelli+2017:spec_variab,Zappacosta+2018:Nustar}, provided they lied in the radiatively-efficient regime \citep{Gu+2009:gamma_vs_Edd,Connolly+2016:spec_variab_lowEdd,Peretz+2018:HR_variab}.

Assuming there is indeed not only a phenomenological, but also a physical connection between radiatively-efficient (not jet-dominated) AGN and XRBs, we can exploit the high-cadence monitoring on single XRBs to obtain a less-biased and more comprehensive analysis on the possible co-evolution of the disk-corona spectral components. However, our results then imply that the physical scatter of the $\log L_{disk}-\log L_{cor}$ cannot be $\lessapprox0.19-0.20$\,dex \citep{Lusso&Risaliti2016:LxLuvtight,Chiaraluce+2018:dispandvariab_Lx_Luv}. Indeed, this estimate might be contaminated by the adopted $\Gamma$ distribution, that is likely biased narrow in flux-limited AGN samples. As a matter of fact, a common procedure in AGN samples is indeed to cut the extreme $\Gamma$ values as a selection criterion for more robust sources. Alternatively, a standard photon index of $\sim1.8-1.9$ is typically attributed to faint spectra that do not allow to constrain it, with the obvious consequence of an artificial narrowing of the observed $\Gamma$ distribution. Including all these extreme $\Gamma$ values would results in a larger scatter of the $\log L_{disk}-\log L_{cor}$ and, possibly, an increased fraction of X-ray weak sources \citep[e.g.][]{Nardini+2019:brightQSO_xfaint}. Hence, assuming a priori that extreme $\Gamma$ values come from unreliable spectral fits and, then, looking for physical correlations involving that parameter itself is circular and might be misleading. Here, we simply selected spectra above a ratio of $\sim1.3$ between source-plus-background and background-only $10-25\,$keV count rates (see Section~\ref{sec:relation_binaries} and Appendix~\ref{sec:appendix_robustness}). This approach was purely observational and resulted in a slight narrowing in the $\Gamma$ distribution (see color-coding in Fig.\ref{fig:results_Fdisk_Fcor_all}) only as a secondary consequence. This being said, we want to highlight that increasing the observed scatter of the $\log L_{disk}-\log L_{cor}$ relation in AGN was shown with detailed simulations to imprint a minor effect on the cosmology, only slightly enlarging the uncertainty in the slope and, consequently, on the cosmological contours (D. Coffey, PhD Thesis and private communication).

Finally, still working under the assumption that a unified prescription between radiatively-efficient XRBs and AGN is present, it is of interest to find out what the AGN counterparts of the excluded background-contaminated XRB states (i.e. with a very weak hard component) would look like. Some AGN counterparts could be the optically-bright X-ray weak quasars found at $z\sim3.0-3.3$ by \citeauthor{Nardini+2019:brightQSO_xfaint} \citep[\citeyear{Nardini+2019:brightQSO_xfaint}; see also][]{Martocchia+2017:wissh} showing unusually flat slopes, which would be in accord with the softer-when-brighter trend we discussed above. This could be indeed an interesting science case for the now launched extended ROentgen Survey with an Imaging Telescope Array (eROSITA; Predehl et al., in prep.).

\section{Conclusions}

Through the last two decades, several attempts to connect accretion in AGN and XRBs in a BH mass scale-invariant fashion have been made \citep[e.g.,][and references therein]{Merloni+2003:fund_plane,Maccarone+2003:soft_vs_modAGN,Falcke+2004:unify,Uttley+2005:variabSey,McHardy+2006:variab,Koerding:2006:HIMS_RL,Sobolewska+2009:alphaox_GBH,Sobolewska+2011:simul,Svoboda+2017:simult,Ruan+2019:analogous_struct}. Besides the more or less understood differences in the composition of their matter reservoir (i.e. density, temperature, ionisation and consequently pressure support) and their environmental surroundings (a single star with respect to the center of a galaxy) their timing and spectral phenomenology have always been found to be comparable. The simplistic but commonly accepted picture emerged from decades of multi-wavelength efforts connects strong radio-emitting low-luminosity AGN to hard-state XRBs, both showing a prominent jet component and a radiatively-inefficient accretion flow \citep[e.g.,][]{Merloni+2003:fund_plane,Falcke+2004:unify}, strong radio-emitting high-luminosity AGN to hard-intermediate states in XRBs, both showing an efficient accretion flow in coexistence with a jet \citep[e.g.,][]{Koerding:2006:HIMS_RL,Svoboda+2017:simult}, and (very) weak radio-emitting moderately- to high-accreting AGN (both combined spanning $\lambda_{edd}=L/L_{edd}\simeq0.0x-1$) to XRBs in the soft states and soft-intermediate states \citep[e.g.,][]{Maccarone+2003:soft_vs_modAGN,Koerding:2006:HIMS_RL,Sobolewska+2009:alphaox_GBH}.

In this work we attempted to improve on this AGN-XRBs connection in the radiatively-efficient (and non- or weakly-jetted) end of accretion mode. Motivated by the tight relationship observed between the disk and coronal luminosities in AGN \citep[e.g.,][and references therein]{Lusso&Risaliti2016:LxLuvtight}, we analysed 458 RXTE-PCA archival observations of the XRB GX339-4, using this object as an test case for the XRBs properties in general (Section~\ref{sec:gx339}). We focused on soft and soft-intermediate states, which have been suggested to be analogous to radiatively efficient (and non- or weakly-jetted) AGN \citep[e.g.][]{Maccarone2003:Ltransition, Koerding:2006:HIMS_RL,Sobolewska+2009:alphaox_GBH}, modeling the emission with a thermal accretion disk and a Comptonising corona (Section~\ref{sec:analysis}). We then populated the $\log L_{disk}- \log L_{cor}$ plane with a quantitative focus on the physics hidden in the scatter, which represents the diversity of X-ray coronae emission given a narrow range in accretion disks (Section~\ref{sec:relation_binaries}). 

The observed scatter in the $\log L_{disk}- \log L_{cor}$ plane of XRBs is high ($\sim0.43\,$dex) and significantly larger than in our control sample of radiatively-efficient (non- or weakly-jetted) broad-line AGN ($\sim0.30\,$dex). This would appear contrary to the hypothesis that the systems simply scale with mass. However, we also found that our AGN and XRBs samples appear to have very different observed $\dot{m}$ and $\Gamma$ distributions. In particular, while we were not able to exclude that the intrinsic $\dot{m}$ distributions are compatible, $\Gamma$ is arguably a more robust estimate and appeared to be directly linked to the observed scatter (Fig.~\ref{fig:results_Fdisk_Fcor_all}). Even after accounting for the measured uncertainties the XRB $\Gamma$ distribution was estimated to be broader (dispersion of $\sim0.16$ with respect to $\sim0.08$) and shifted to softer slopes (mean value of $\sim2.20$ with respect to $\sim2.07$). 

It is nonetheless remarkable that once similarly broad $\Gamma$ and $\dot{m}$ distributions were selected (i.e. compatible $\sigma_{\Gamma}$ and $\sigma_{\dot{m}}$, regardless of <$\Gamma$> and <$\dot{m}$>), AGN and XRBs overlapped quite nicely in the mass-normalised $\log L_{disk}- \log L_{cor}$ plane both showing an observed scatter of $\sim0.30-0.33\,$dex (Section~\ref{sec:relation_XRB_vs_AGN}). This indicates that a mass-scaling between the properties of the two might indeed hold, with our results being consistent with the disk-corona systems in AGN and XRBs exhibiting the same physical processes, albeit under different conditions, for instance in terms of temperature, optical depth and/or electron energy distribution in the corona, heating-cooling balance, coronal geometry and/or black hole spin (see Section~\ref{sec:discussion_gamma}). 

The amplitude of this common scatter ($\sim0.30-0.33\,$dex) is still significantly higher than $\lessapprox0.19-0.20$\,dex, namely what was claimed to be the physical intrinsic (i.e. not due to variability and non-simultaneity) scatter in the $L_X-L_{UV}$ (or $\alpha_{OX}-L_{UV}$) relation in AGN \citep[e.g.,][]{Vagnetti+2013:variab_onalphaOX,Lusso&Risaliti2016:LxLuvtight,Chiaraluce+2018:dispandvariab_Lx_Luv}. On the other hand, it is worth stressing that when a single XRB is used, like in our case, any possible issue arising from non-simultaneity of the data probing the two components is avoided and there is no additional scatter coming from a mixed bag of masses, distances and inclinations, which is instead typical of AGN. Hence, under the assumption of a mass-scaling paradigm, one would expect the scatter in XRBs to be lower. We conclude that
, as the results of the past decades and in this work suggested, since the two systems are similar both in their phenomenology and physical processes the physical scatter of the disk-corona emission in AGN is likely not as low as we think, with important implications for both accretion physics and quasar cosmology.

\begin{acknowledgements}
	We thank the referee for his/her helpful comments. RA warmly thanks the organizers and participants of the workshop "AGN Spectral States - Unification of Black Holes across the Mass Scale" held in Prague between 27th-29th of November 2019 for stimulating discussions and conversations. RA is particularly grateful to Wenda Zhang for kindly sharing MONK calculations. We acknowledge the work done by Mickael Coriat in the data processing and we thank Robert Dunn for sharing results from Dunn et al. (2010). We thank Marat Gilfanov, Mara Salvato and Damien Coffey for useful discussions, Sergio Campana and Efrain Gatuzz for a careful read of the manuscript. RA also thanks Johannes Buchner and J. Michael Burgess for insightful comments on the statistical side of the analysis and hopes he did not misuse tools and/or terms. We acknowledge the use of the matplotlib package \citep{Hunter2007:matplotlib}.
\end{acknowledgements}

%
%
\bibliographystyle{aa} 
\bibliography{bibliography} 

\begin{appendix}	
\section{On the robustness of spectral analysis results}
\label{sec:appendix_robustness}

In X-ray spectral analysis, the outcome of a fit should not be blindly trusted without simulations, particularly in low-counts regime or when the background is at a level compatible with (part of) the source emission. Since in our science case the putative physics of the source is such that the hard component can be comparable to the RXTE background at energies above $\approx10\,$keV, a more thorough investigation is needed to validate our spectral fit results. One should bear in mind that this does not happen necessarily only when the total flux is low, as in spectra with a strong soft component and a weak hard component the total $3-25\,$keV emission is actually around the average value of the outburst. 


\begin{figure}[tb]
	\centering
	\includegraphics[width=0.70\columnwidth]{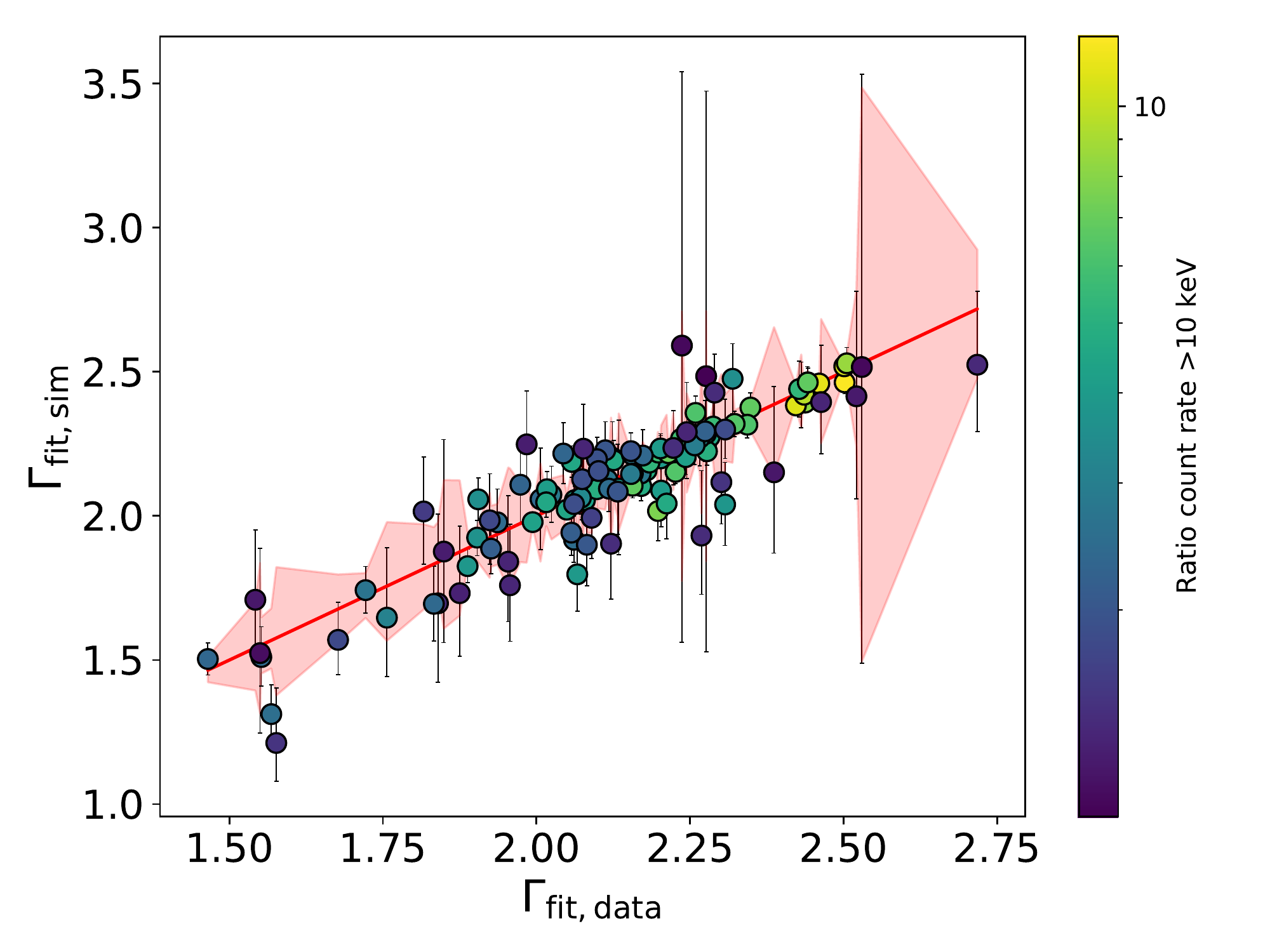}
	\includegraphics[width=0.70\columnwidth]{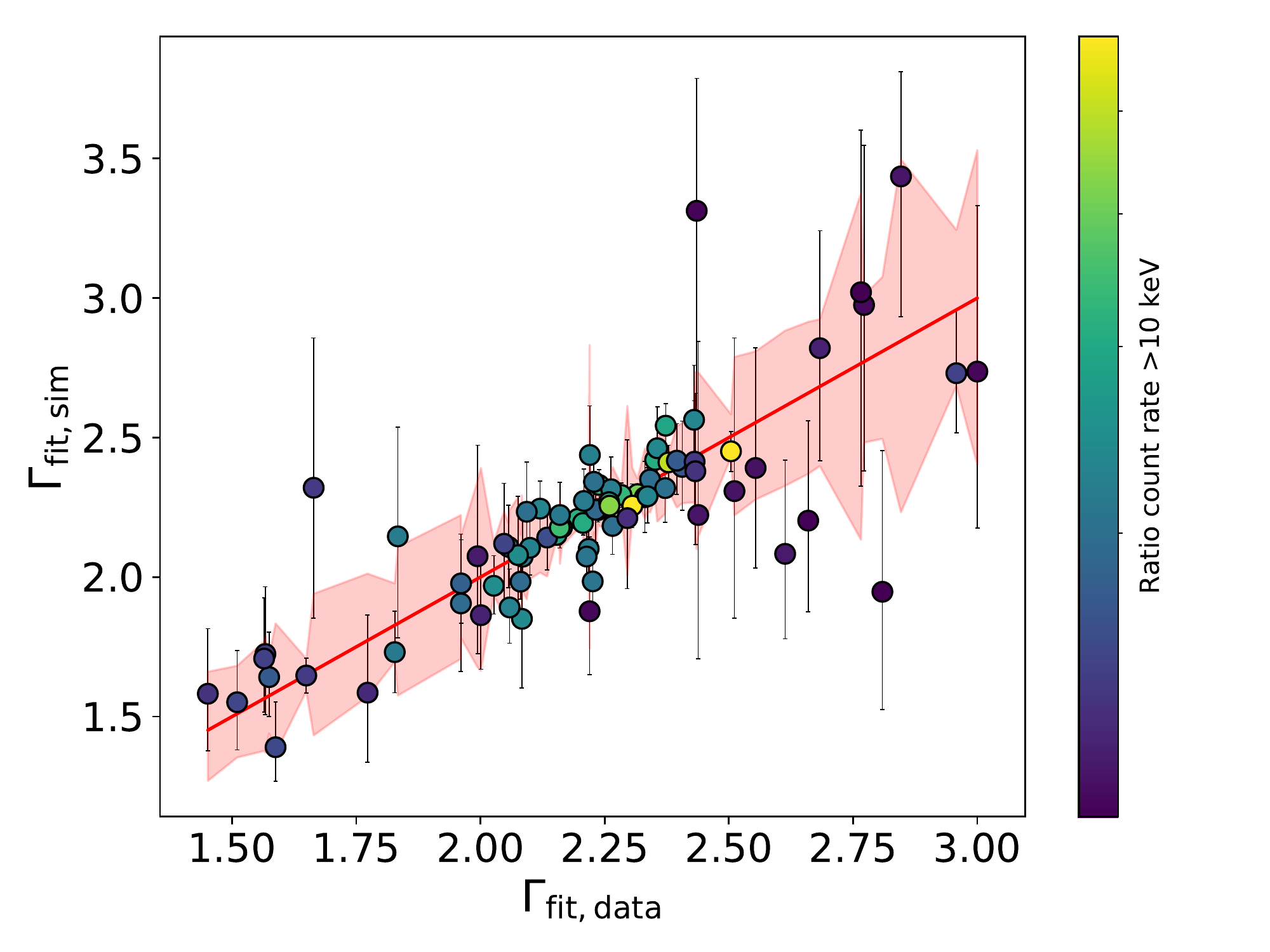}
	\includegraphics[width=0.70\columnwidth]{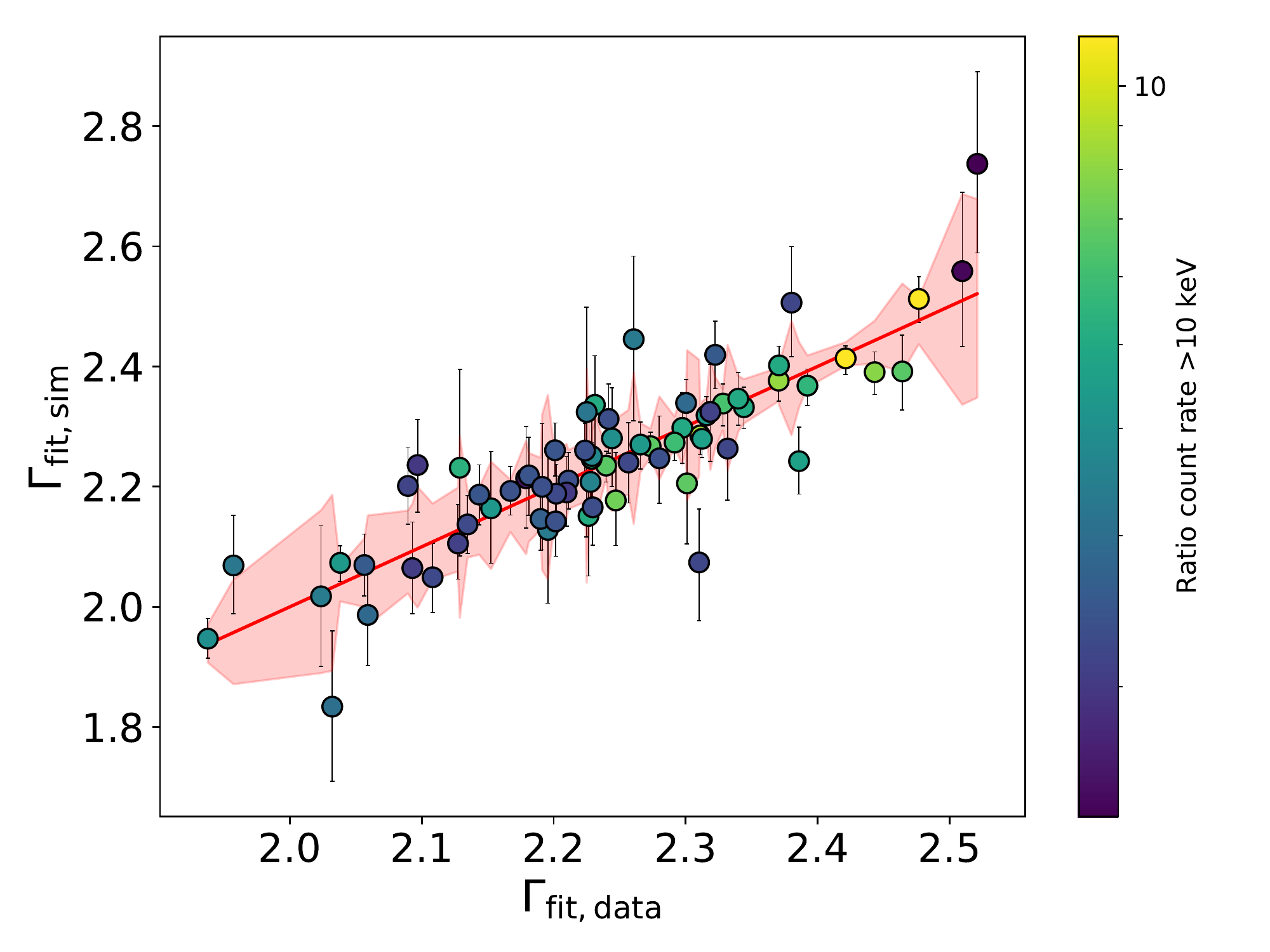}	
	\includegraphics[width=0.70\columnwidth]{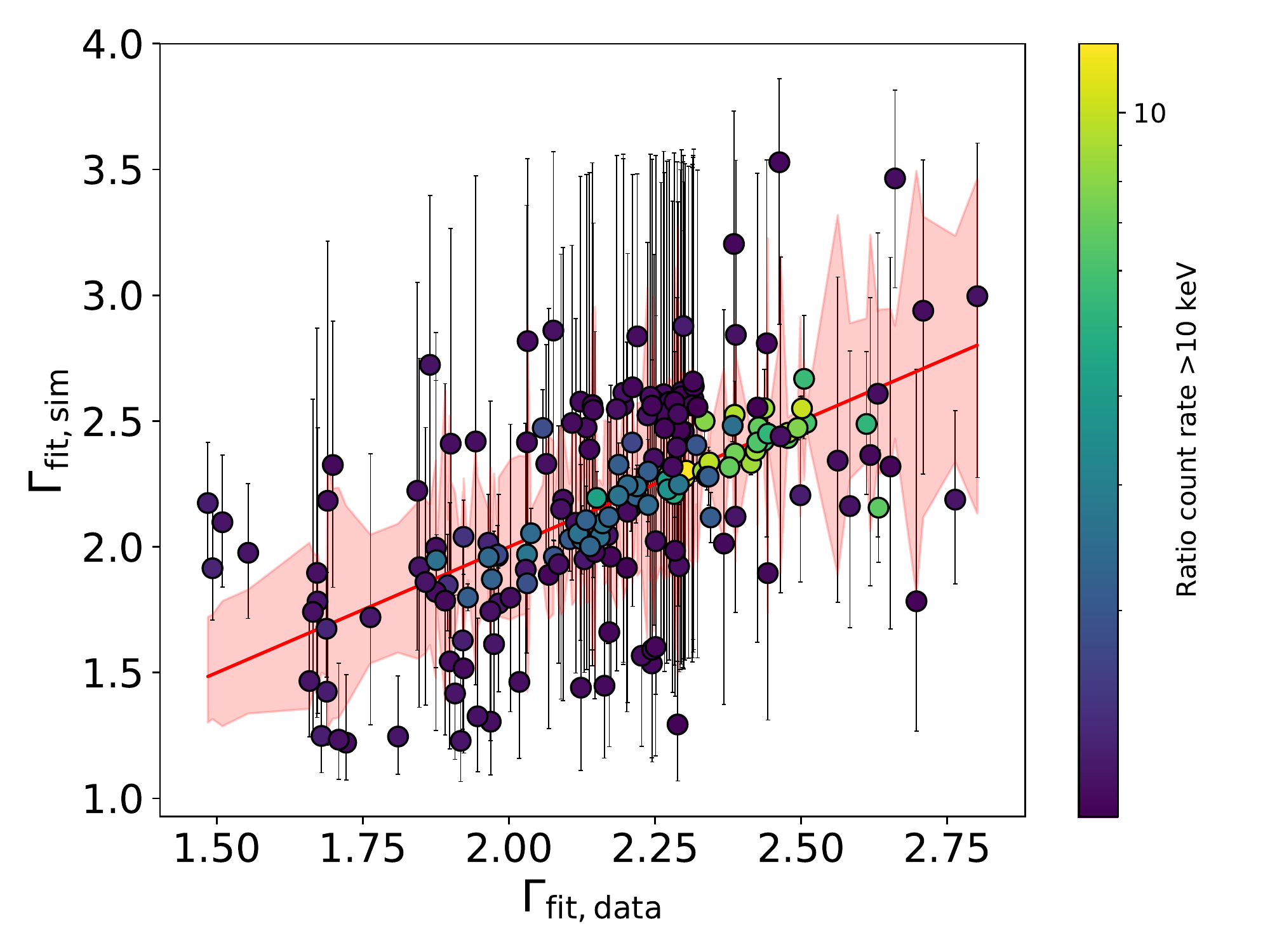}	
	\caption{Comparison between the Photon Index obtained in the spectral fit (see Section~\ref{sec:fit_results}; $\Gamma_{fit,data}$), which was then simulated and fit again ($\Gamma_{fit,simul}$). From top to bottom, results for SS02, SS04, SS07 and SS10 are shown. Error bars for $\Gamma_{fit,simul}$ are shown (16th-84th percentiles), whereas we show uncertainties (16th-84th percentiles) in $\Gamma_{fit,data}$ around the 1:1 relation. We color coded data with the ratio between the total (source plus background) and background-only $10-25\,$keV count rates (see white and grey symbols in the top panel of Fig.~\ref{fig:stuff_vs_mjd}).}
	\label{fig:simul_gamma}
\end{figure}

\begin{figure}[tb]
	\centering
	\includegraphics[width=0.75\columnwidth]{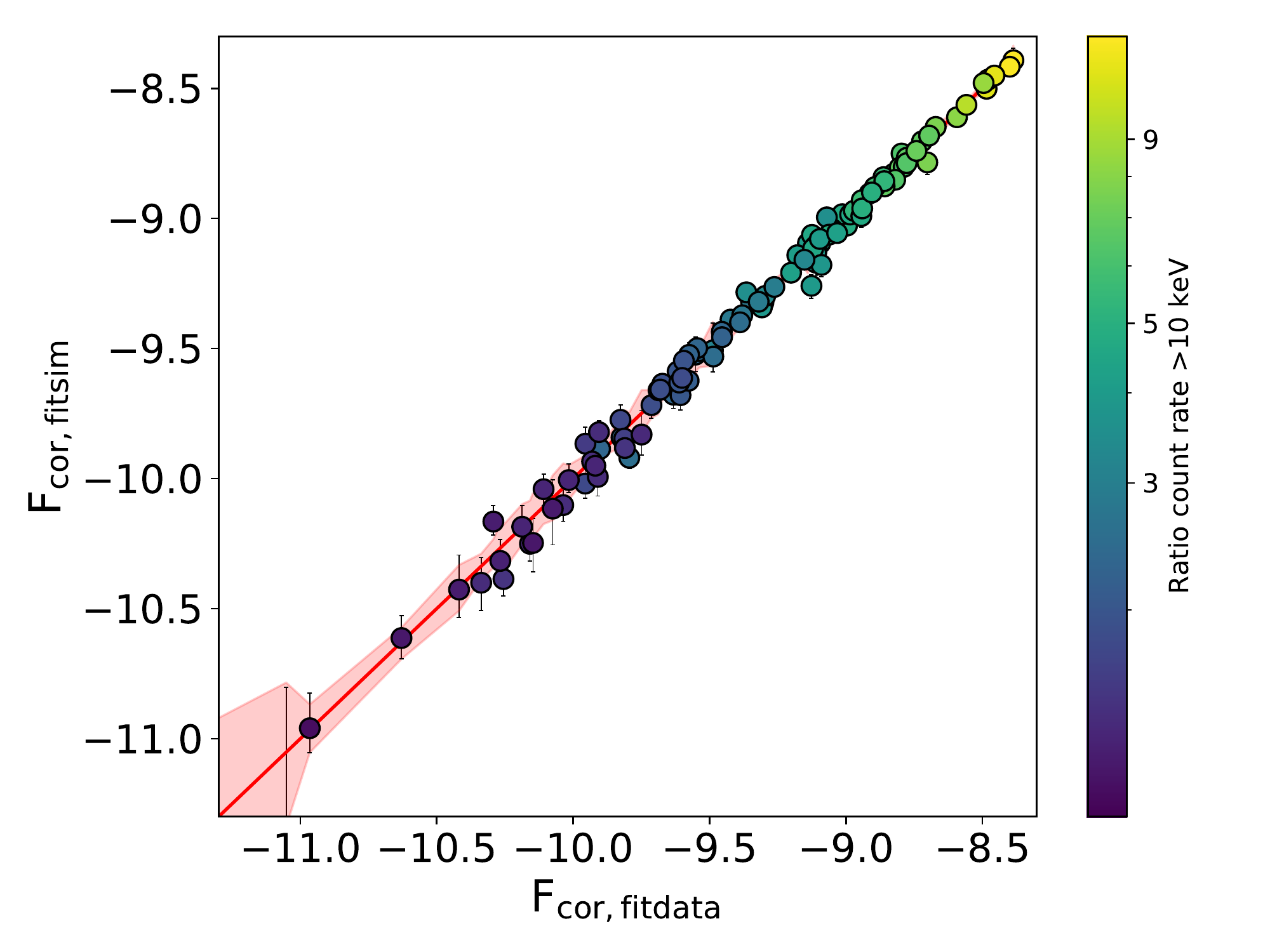}
	\includegraphics[width=0.75\columnwidth]{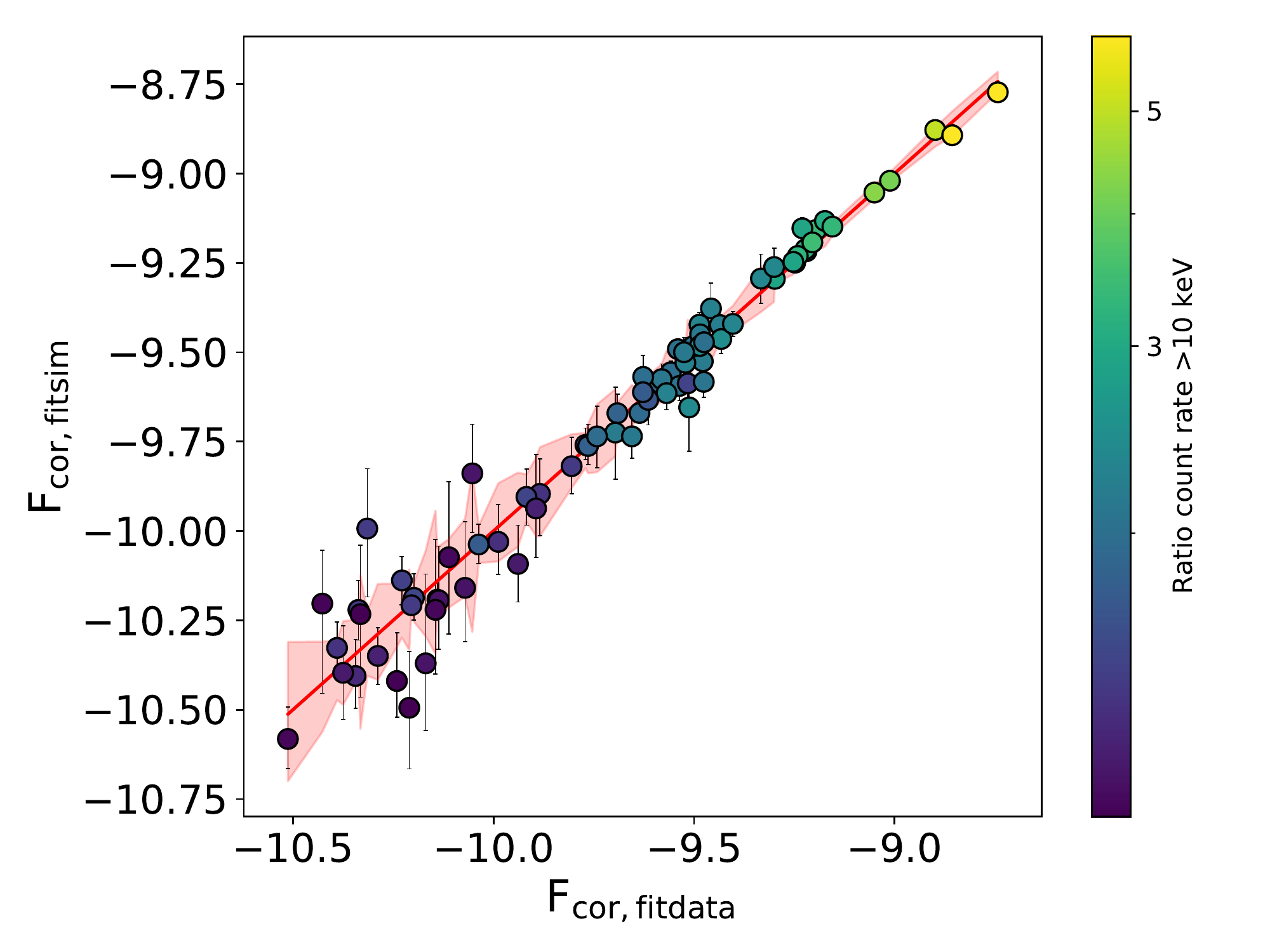}
	\includegraphics[width=0.75\columnwidth]{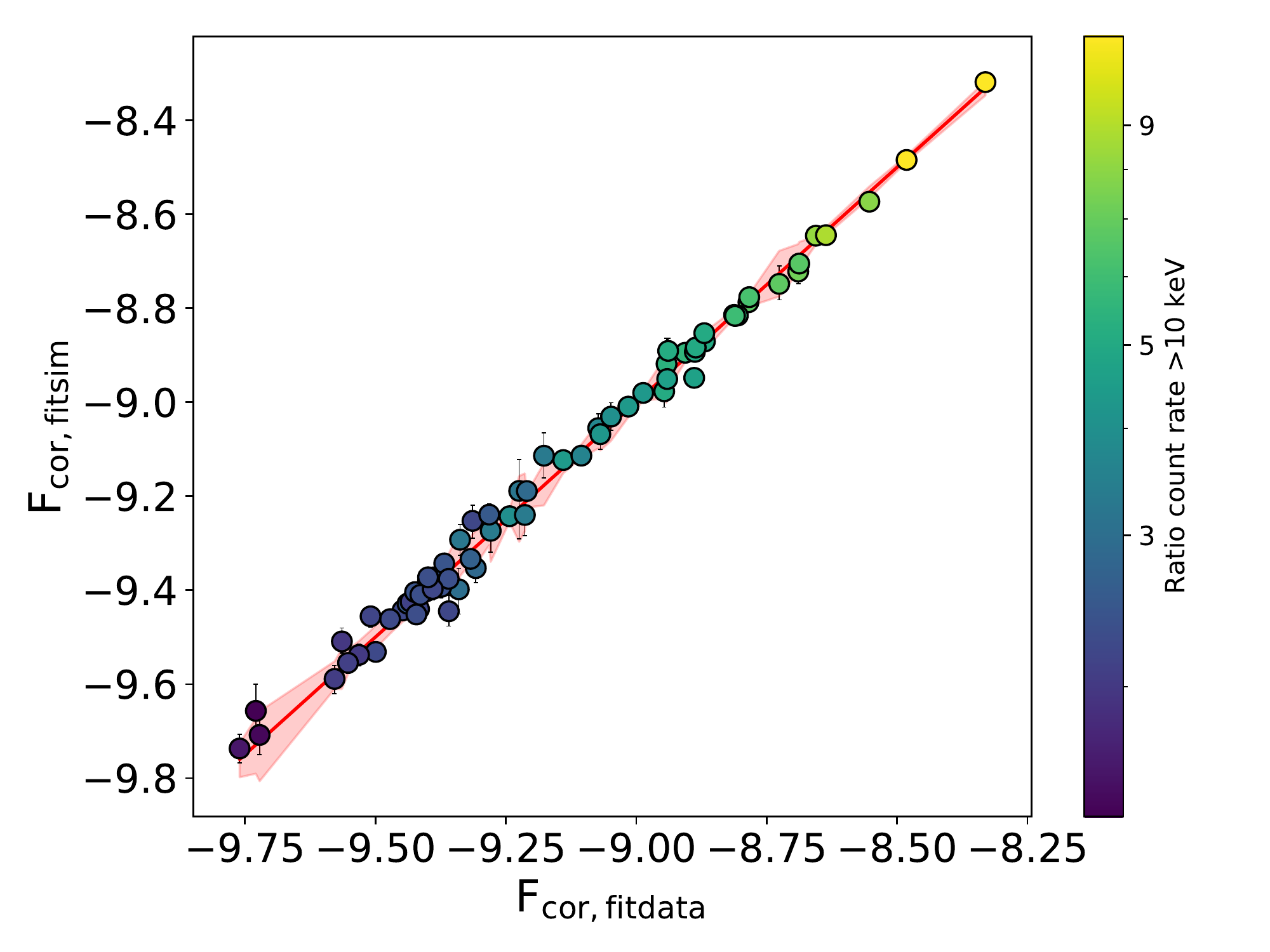}	
	\includegraphics[width=0.75\columnwidth]{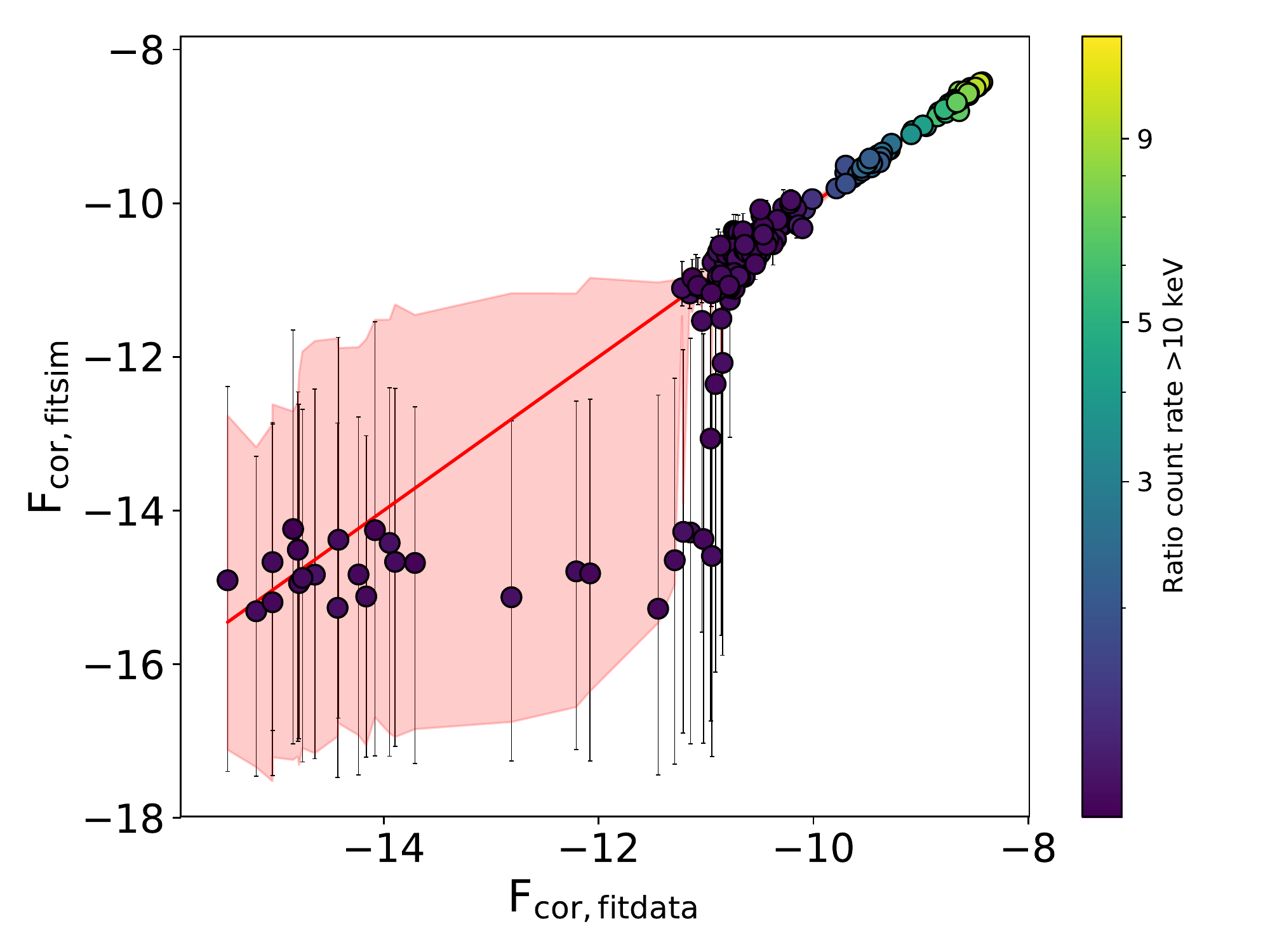}	
	\caption{Same as in Fig.~\ref{fig:simul_gamma} but with the $2-10\,$keV flux under the \texttt{NTHCOMP} model of the hard component. In the top panel, the figure was cut around $\log F_{cor} \sim -11$ for visualisation purposes and at lower fluxes simulations and fits are compatible within their very large errors.}
	\label{fig:simul_Fcor}
\end{figure}

\begin{figure}[tb]
	\centering
	\includegraphics[width=0.95\columnwidth]{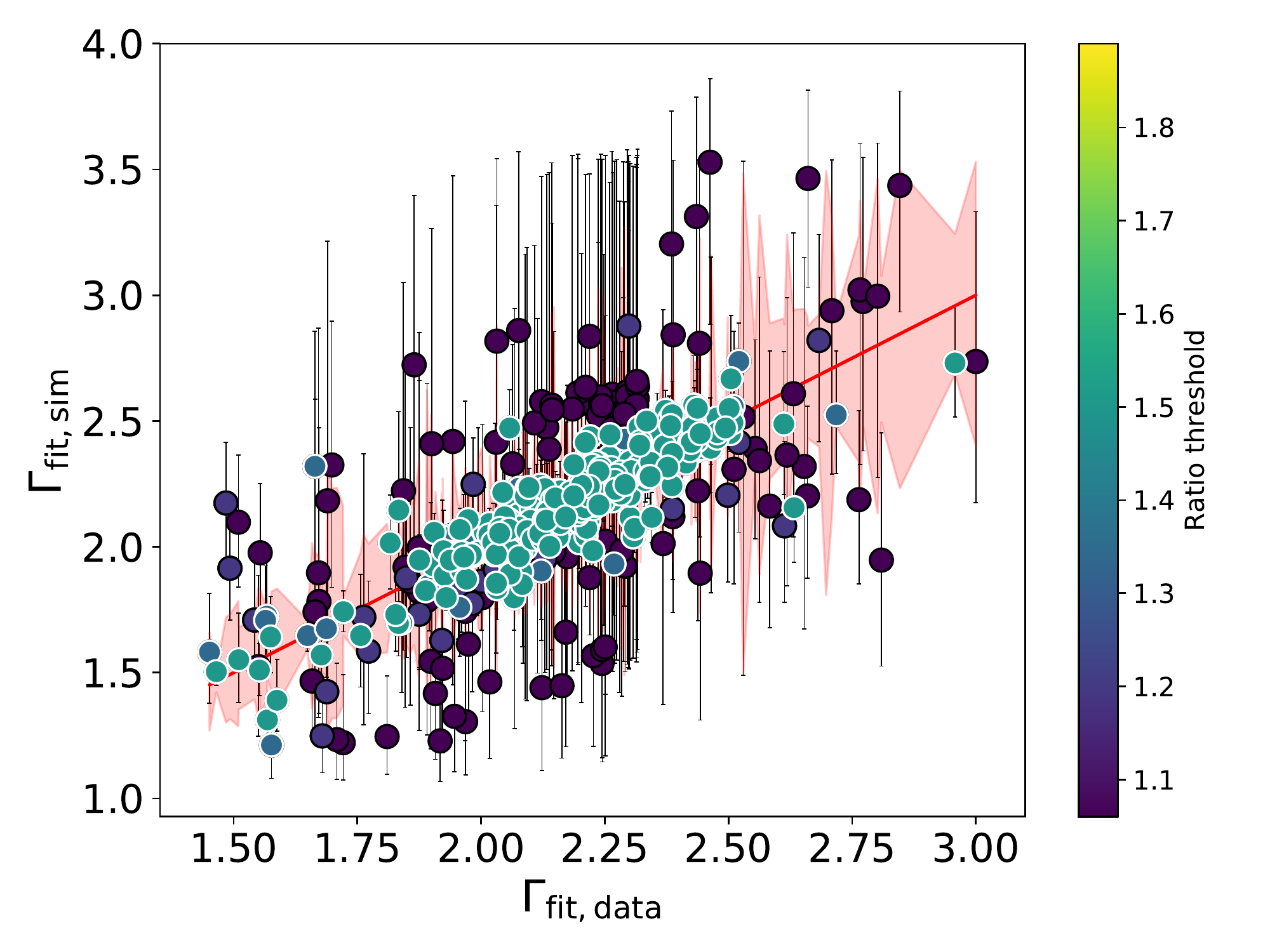}
	\caption{Same as Fig.~\ref{fig:simul_gamma}, but color coded with the threshold on the ratio between the total (source plus background) and background-only $10-25\,$keV count rates. The one we adopted (i.e. $\sim1.3$) is shown with white contours on data points.}
	\label{fig:simul_gamma_cut}
\end{figure}

\begin{figure}[tb]
	\centering
	\includegraphics[width=0.95\columnwidth]{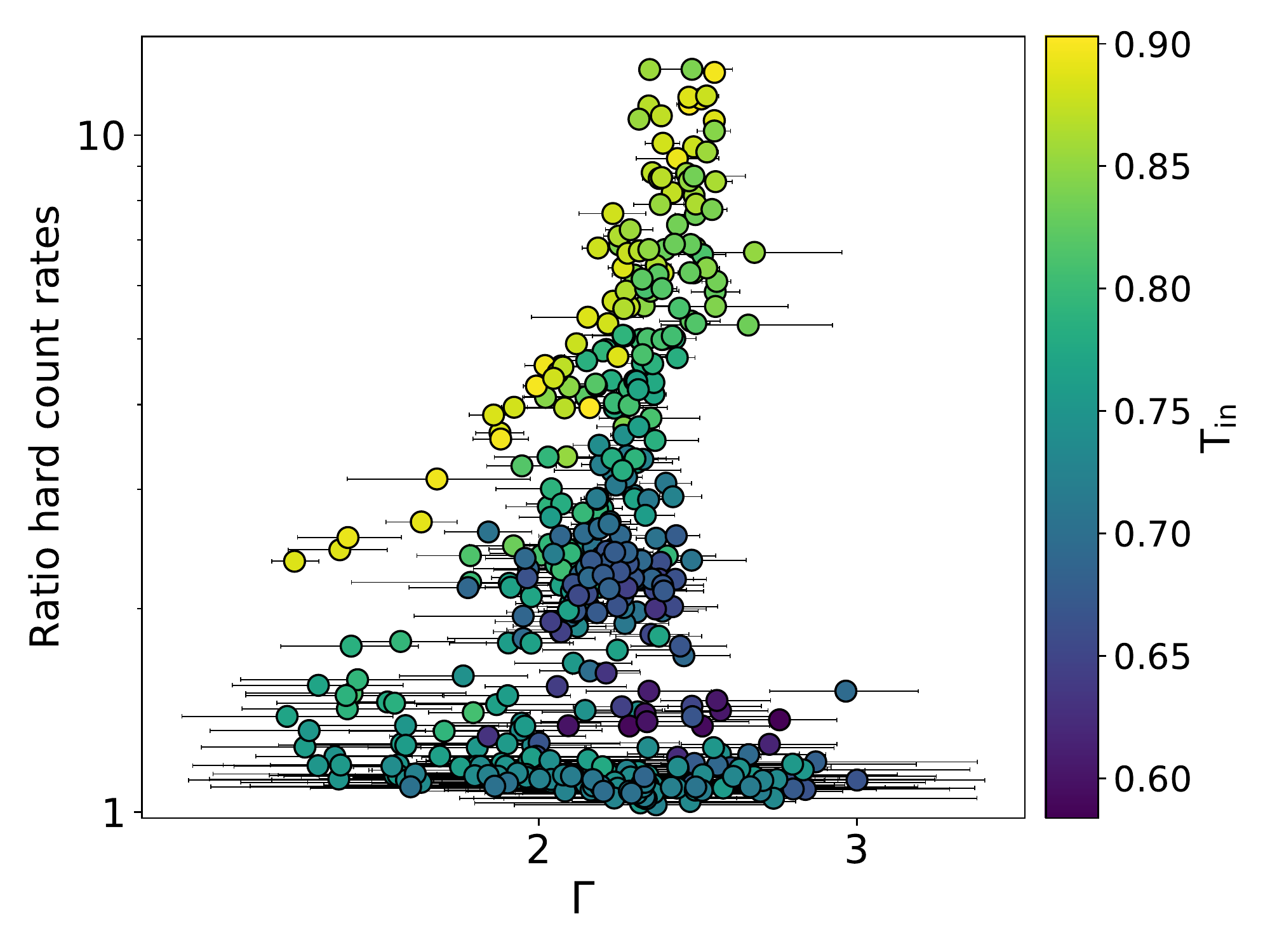}
	\caption{The radio between the total (source plus background) and background-only $10-25\,$keV count rates (see white and grey symbols in the top panel of Fig.~\ref{fig:stuff_vs_mjd}) as a function of the fit $\Gamma$, color coded with the fit disk temperature.}
	\label{fig:gamma_vs_bkg}
\end{figure}

\begin{figure*}[tb]
	\centering
	\includegraphics[width=0.8\columnwidth]{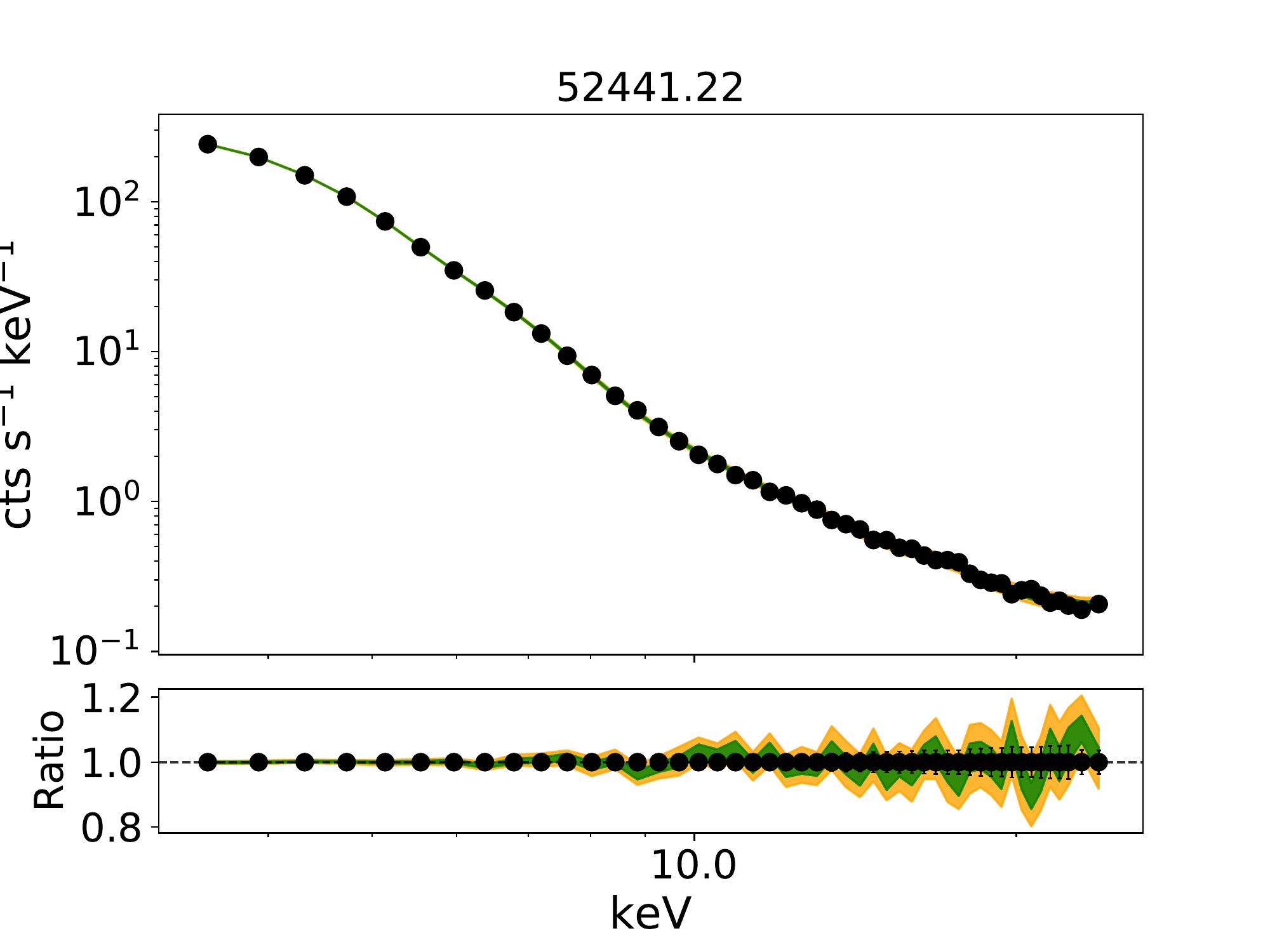}
	\includegraphics[width=0.8\columnwidth]{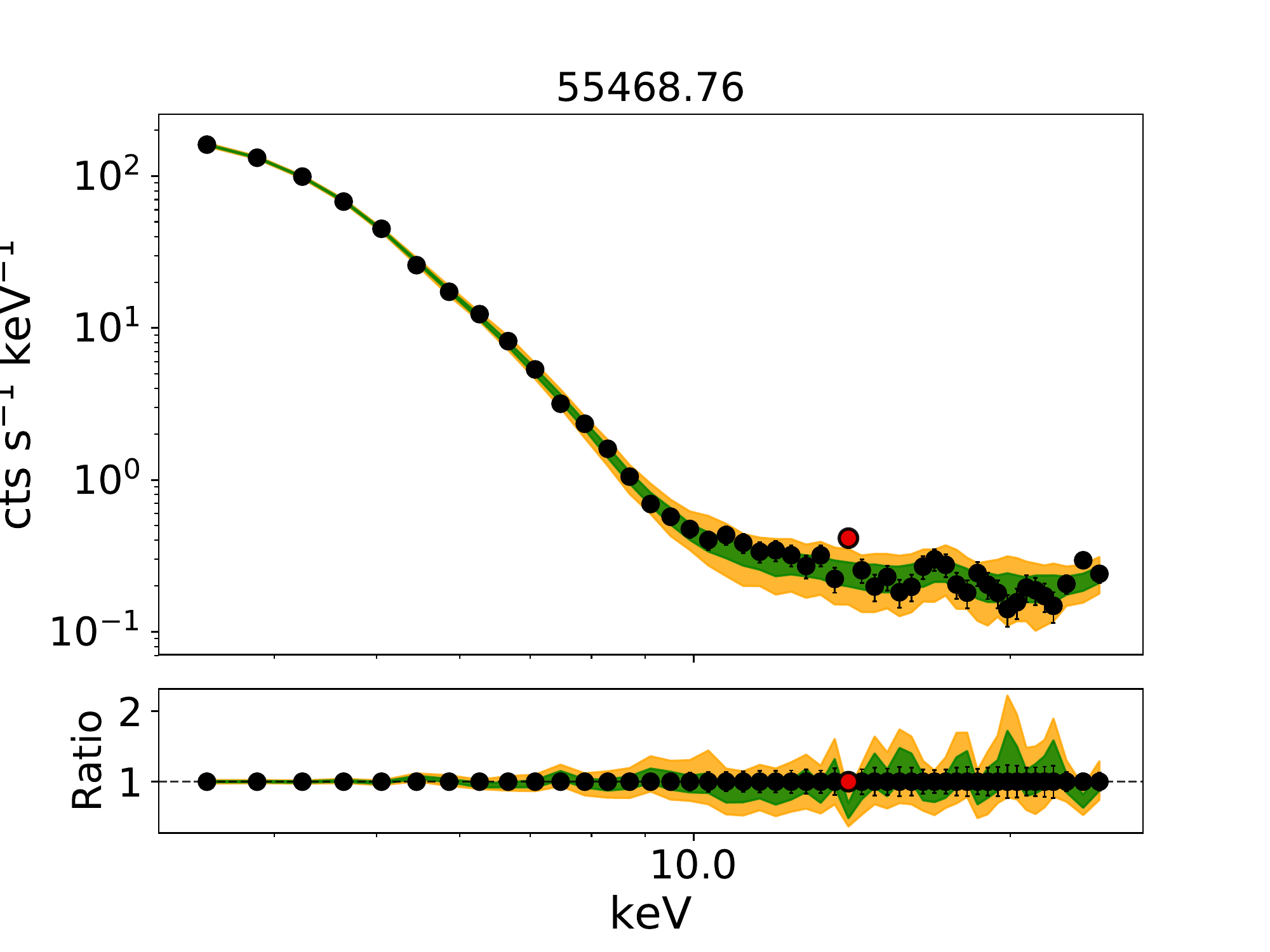}
	\caption{Two examples of the posterior predictive check performed simulating 300 count rate spectra starting from the best-fit parameters posterior distributions. Data points are the actual spectrum, with green and orange contours as 16th-84th and 1st-99th inter-quantile ranges representing the predictive power of the spectral model. Energy bins in which the observed data point was not compatible within the 1st-99th inter-quantile range of the predicted spectra from the best-fit models are shown in red.}
	\label{fig:PPC}
\end{figure*}

Two types of problems can arise in fitting the hard component that make a specific fit of a state questionable. Firstly, one may not be able to robustly fit a specific region of the source parameter space (e.g. the intrinsic photon index or hard flux) for instrumental or observational biases, for instance due to the background or to possible covariances; the fit parameters could be scattered in another region of the parameters space, physically reasonable, and the observer would have no way of knowing this from the fit alone, which can in principle appear robust. At first order, this can be tested simulating a synthetic spectrum from the best-fit parameters and fitting it again with the same model: if the input and output agree, that region of the parameters space is recoverable; if not, that spectral fit cannot be considered robust. Secondly, even if one is able to fit a specific parameters space region, one cannot be sure that biases have scattered a source, originally with other parameters, in the location of the parameter space where one has fit it. This is more subtle and would require a set of multidimensional spectral simulations, beyond the scope of this paper.

We try to address here the first problem with spectral simulations. We simulated each XRB state from the best-fit model with the \texttt{pyXSPEC} command \texttt{fake it} adding also statistical fluctuations. Then, we fit each simulated spectrum with the same model (see Section~\ref{sec:analysis}) and checked if the retrieved parameters were compatible, within errors, with the simulated spectrum (See Fig.~\ref{fig:simul_gamma} and~\ref{fig:simul_Fcor}). Simulations clearly show that for SS02, SS04 and SS07 all the spectral fits can be considered to be in reliable places of the explored parameters space. Instead, simulations of SS10 state that results from several observations are to be taken with caution (see bottom panels of Fig.~\ref{fig:simul_gamma} and~\ref{fig:simul_Fcor}): in particular, observations with relatively low count rates above $\sim10\,$keV, namely with a factor $\lessapprox1.5$ in ratio between the total (source plus background) and background-only $10-25\,$keV count rates (see white and grey symbols in the top panel of Fig.~\ref{fig:stuff_vs_mjd} and Fig.~\ref{fig:gamma_vs_bkg}). In most of these background-contaminated spectral states, the input and output values are compatible within their (very large) 1st-99th inter-quantile range, thus we are not overestimating our knowledge of the hard component even in these extreme cases. Nonetheless, it is evident that the median of the posterior distribution for $\Gamma$ strongly departs from the simulated value. In Fig.~\ref{fig:simul_gamma_cut} we show all the outbursts combined, color coding the threshold of the ratio between total (source plus background) and background-only $10-25\,$keV count rates, highlighting with white contours the sources above $\sim1.3$. This threshold ensures that $\lessapprox4\%$ of the input-output parameters are not compatible within their 16th-84th inter-quantile range. This represents the sub-sample selected for the comparison with AGN in Section~\ref{sec:relation_XRB_vs_AGN}. It is worth noting that in these background-contaminated states the presence of a disk only indirectly hampers the detectability of the hard component, as they do not host the brightest disks (see color coding in Fig.~\ref{fig:gamma_vs_bkg}). 

Moreover, we show in Fig.~\ref{fig:gamma_vs_bkg} how this count rate ratio is related to the fit $\Gamma$ in our spectral analysis. As it can be noted, our cut in ratio around 1.3 also consequently narrows the Photon Index distribution, although above it one can still note the softer-when-brighter behavior of our SSs and SIMSs. It is a typical habit to exclude extremes $\Gamma$ values from AGN samples before testing for correlations and physical interpretations, with the underlying assumption that they come from low-quality spectra. This can be true in most cases, although one can see that applying a vertical selection in Fig.~\ref{fig:gamma_vs_bkg} one could end up excluding not only the low-quality (or background-contaminated) states, but also the brightest-softest spectra and a few hard-faint (but still sufficiently-well constrained) spectra. Furthermore, excluding extremes X-ray slopes is a physical selection and applying it before looking for physical correlation and interpretations is a circular process. Conversely, a selection in count rate ratio is purely observational and for most of the sources results in an equivalent selection.

Finally, we also performed a posterior predictive check simulating 300 count rate spectra starting from the posterior distributions of the best-fit parameters. This is then visually compared with the original spectrum (examples in Fig.~\ref{fig:PPC}), flagging possible energy bins in which the observed data point was not compatible within the 1st-99th inter-quantile range of the predicted spectra from the best-fit models. This test highlights regions of the spectrum where our model predictions of future datasets significantly depart from the observed data. In general, we observe very few of these features, meaning that the uncertainties on our model are large enough not to overestimate the information drawn from the data. In particular, this check confirmed that results for most of the SSs in SS10 can be considered reliable only within the very large uncertainties and must be taken with caution. For instance, refrain from using only median values for $\Gamma$ from our SS10 results instead of considering the hard component as largely unconstrained. Alternatively, the more conservative option is to adopt the sub-sample of sources above a ratio between the total (source plus background) and background-only $10-25\,$keV count rates of $\sim1.3$.

\section{Testing different proxies for the disk and corona emission}
\label{sec:appendix_proxies}

\begin{table*}[tb]
	\footnotesize
	\caption{Same as Table~\ref{tab:emcee_single_outb}, but including additional columns (marked with \_3-25 and \_full) obtained within $3-25\,$keV and $F_{0.001-100\,keV,\,disk}- F_{1-100\,keV,\,cor}$, respectively.}
	\label{tab:emcee_bbproxies}
	\centering
	\begin{tabular}{C{0.05\columnwidth} C{0.3\columnwidth} C{0.15\columnwidth} C{0.3\columnwidth} C{0.15\columnwidth} C{0.3\columnwidth} C{0.15\columnwidth}}%
		\toprule
		\multicolumn{1}{c}{Outburst} &
		\multicolumn{1}{c}{Slope} &		
		\multicolumn{1}{c}{Scatter} &
		\multicolumn{1}{c}{Slope\_3-25} &		
		\multicolumn{1}{c}{Scatter\_3-25} &
		\multicolumn{1}{c}{Slope\_full} &		
		\multicolumn{1}{c}{Scatter\_full} \\
		\midrule
		SS02 & $1.04\pm0.18$ & $0.47\pm0.03$ & $1.21\pm^{+0.17}_{-0.16}$ & $0.40\pm0.03$ & $1.44\pm0.21$ & $0.41\pm0.03$ \\
		SS04 & $0.08\pm0.25$ & $0.42^{+0.04}_{-0.03}$ & $0.45^{+0.25}_{-0.27}$ & $0.39^{+0.04}_{-0.03}$ & $0.30^{+0.30}_{-0.31}$ & $0.38^{+0.04}_{-0.03}$ \\
		SS07 & $0.60\pm0.07$ & $0.21\pm0.02$ & $0.64\pm0.07$ & $0.19\pm0.02$ & $0.69^{+0.08}_{-0.09}$ & $0.20\pm0.02$ \\
		SS10 & $-0.45\pm0.22$ & $0.84\pm0.05$ & $-0.48^{+0.21}_{-0.22}$ & $0.78^{+0.05}_{-0.04}$ & $-0.74^{+0.24}_{-0.23}$ & $0.76\pm0.04$ \\
		\bottomrule	
	\end{tabular}
\end{table*}

\begin{table*}[tb]
	\footnotesize
	\caption{Same as Table~\ref{tab:emcee_single_outb}, but including additional columns (marked with \_0.2) obtained with $F_{0.2\,keV}$ as a disk proxy.}
	\label{tab:emcee_fd02}
	\centering
	\begin{tabular}{C{0.06\columnwidth} C{0.35\columnwidth} C{0.2\columnwidth} C{0.35\columnwidth} C{0.2\columnwidth}}%
		\toprule
		\multicolumn{1}{c}{Outburst} &
		\multicolumn{1}{c}{Slope} &		
		\multicolumn{1}{c}{Scatter} &
		\multicolumn{1}{c}{Slope\_0.2} &		
		\multicolumn{1}{c}{Scatter\_0.2} \\
		\midrule
		SS02 & $1.04\pm0.18$ & $0.47\pm0.03$ & $1.49\pm0.39$ & $0.51^{+0.04}_{-0.03}$ \\
		SS04 & $0.08\pm0.25$ & $0.42^{+0.04}_{-0.03}$ & $-0.73^{+0.42}_{-0.44}$ & $0.41^{+0.04}_{-0.03}$ \\
		SS07 & $0.60\pm0.07$ & $0.21\pm0.02$ & $0.74\pm0.12$ & $0.24\pm0.02$ \\
		SS10 & $-0.45\pm0.22$ & $0.84\pm0.05$ & $-1.28\pm0.27$ & $0.80^{+0.05}_{-0.04}$ \\
		\midrule
		All & $0.34\pm0.12$ & $0.71^{+0.03}_{-0.02}$ & $-0.21\pm0.17$ & $0.71^{+0.03}_{-0.02}$ \\
		All\_r1.3 & $0.47\pm0.07$ & $0.41\pm0.02$ & $0.39^{+0.11}_{-0.10}$ & $0.43\pm0.02$ \\
		\bottomrule	
	\end{tabular}
\end{table*}

Throughout the work, we used as proxy for the disk and corona component the $2-10\,$keV flux under the single \texttt{DISKBB} and \texttt{NTHCOMP} model, respectively. We here investigated whether our results change with a different proxy. For what concerns the $F_{disk}-F_{cor}$ in XRBs only, we tested the adoption of fluxes in the range $3-25\,$keV (the full energy range used in our spectral analysis) or a broader-band $F_{0.001-100\,keV,\,disk}- F_{1-100\,keV,\,cor}$ \citep[e.g.][]{Dunn+2010:global}. As shown in Table~\ref{tab:emcee_bbproxies}, the scatter values and slopes are compatible within errors.

Moreover, in Section~\ref{sec:relation_XRB_vs_AGN} we changed the disk proxy using a monochromatic flux for the disk emission at $0.2\,$keV, which is the rough low-mass equivalent of what is $3000\AA$ for AGN. We then tested if in the XRBs dataset the change in disk proxy affects what is described in Section~\ref{sec:fit_results} and~\ref{sec:relation_binaries}. As shown in Table~\ref{tab:emcee_fd02}, all the scatter and most of the slope values are compatible within errors. The fact that some slopes are not validates our choice of using the scatter of the $F_{disk}-F_{cor}$ to test the disk-corona relation in XRBs. It also argues that uncertainties in extrapolation of the RXTE response down to $0.2\,$kev and in $N_H$ (which we leave free to vary within a $\pm15\%$ of the tabulated value) are minimal or affect mostly the slope, if anything.

\section{Testing a different accretion disk model}
\label{sec:bhspec_test}

Throughout the work, we used \texttt{DISKBB} as disk model for its simplicity and better coupling with the Comptonisation model \texttt{NTHCOMP}, with respect to more accurate disk models \citep[e.g. \texttt{BHSPEC},][]{Davis&Hubeny2006:bhspec}. As a matter of fact, \texttt{DISKBB} simply fits for the temperature at the inner disk radius $T_{in}$ and for the normalisation, which is a function of the inner radius $R_{in}$, the distance $d$ of the source and inclination $i$ of the disk. Knowing $R_{in}$ and its evolution is not the purpose of this work, thus the large uncertainties on $d$ and $i$ \citep[see, e.g.,][]{Heida+2017:MF_gx339,Zdziarski+2019:dist_inc_binarypar} are not a big concern. Instead, the \texttt{BHSPEC} parameters are $\log L/L_{Edd}$, that is the luminosity of the accretion disk as a fraction of the Eddington luminosity ($L_{Edd}=1.3\times10^{38}\,m\,$erg s$^{-1}$), $d$, $i$, $m$ and the spin $a_*$. This model would in principle provide a more physical description of the accretion disk emission in terms of more useful source parameters with respect to \texttt{DISKBB}, although all these are very uncertain and extremely debated parameters \citep[e.g., ][]{Hynes+2003:dyn,Hynes+2004:dist,Zdziarski+2004:dist,Munoz-Darias+2008:mass,Kolehmainen+Done2010:spin_spec,Parker+2016:dist,Ludlam+2015:spin,Garcia+2015:spin,Heida+2017:MF_gx339,Zdziarski+2019:dist_inc_binarypar}. Fixing them, despite the large uncertainties, would be a much greater approximation than using \texttt{DISKBB}, with much less control on the many degeneracies and large dimensions involved in the problem; and the more proper approach of jointly fitting for $d$, $i$, $m$ and $a_*$ across all the 458 states would be prohibitive. 

We here explored the use of \texttt{BHSPEC} and tested if our data were good enough to try and directly constrain the disk parameters. We first separately fit 44 states which showed a ratio between the total (source plus background) and background-only $10-25\,$keV count rates below $\sim1.1$, ensuring minimum contamination from non-disk components. We left $m$, $d$, $i$ and $a_*$ free to vary (within $m=3-10$, $d=7-12\,$kpc, $i=5^{\circ}-77^{\circ}$ and $a_*=0-0.9$) in addition of $\log(L/L_{edd})$ and the other source parameters of \texttt{NTHCOMP} and the Gaussian model (see Section~\ref{sec:model}). Despite the good fits, the distribution of fit \texttt{BHSPEC} parameters was quite diverse and unconstrained and their errors quite large. Secondly, we simultaneously fit the three states with the lowest ratio between the total (source plus background) and background-only $10-25\,$keV count rates, keeping $m$, $d$, $i$ and $a_*$ tied together and the other source parameters free to vary. This yielded a good fit with median values of $m\sim5.8$, $d\sim7.8\,$kpc, $i\sim25^{\circ}$ and $a_*\sim0.47$. However, adding one or two more states to the simultaneous fit did not reach convergence, with the sampling far from the minima found with three sources. For this and the fact that among the 44 states separately fit there was generally poor agreement, we refrained from using any estimate of $m$, $d$, $i$ and $a_*$ coming from the simultaneous fit for more than just Eddington-ratio related calculations. 

Then, we fit all the states in SS02 with the same configuration as in Section~\ref{sec:model} (fixing the seed photons temperature to the value obtained with \texttt{DISKBB}) but with \texttt{BHSPEC} fixed at $m\sim5.8$, $d\sim7.8\,$kpc, $i\sim25^{\circ}$ and $a_*\sim0.47$ and free accretion rate. This allowed us to estimate whether, despite the non-robust source parameters for the disk model, the impact of a different accretion disk model on our results was significant. The slope and scatter of the $F_{disk}-F_{cor}$ plane are $0.88\pm0.13$ and $0.34\pm0.03$ respectively, to be compared with the \texttt{DISKBB} run\footnote{Both results refer to a SS02 sub-sample of 105 states above the threshold of $\sim1.3$ of counts ratio between source plus background and background only, see Appendix~\ref{sec:appendix_robustness}.} that yielded $0.88\pm0.17$ and $0.40\pm0.03$. Both values are compatible within errors. Instead, \texttt{BHSPEC} yields a mean $\Gamma\sim2.46$ with respect to $\Gamma\sim2.14$ obtained with \texttt{DISKBB}. Hence, the $\Gamma$ distribution would even more softer than in AGN, enhancing rather than contradicting our results of Section~\ref{sec:relation_XRB_vs_AGN}. 

Finally, we used this \texttt{BHSPEC} run on SS02 to compute a correction on the monochromatic $F_{disk}$ at 0.2\,keV, since \texttt{DISKBB} is known to underestimate the very soft emission in an RXTE-like instrument, even if above $\sim3\,$keV the two models produced the same flux \citep[see][]{Done&Davis2008:meanstress}. As a matter of fact, \texttt{DISKBB} has reportedly a narrower band-pass with respect to more physical models including radiative transfer in each disk annulus, of the order of a color correction which is however not constant in radius, and relativistic effects \citep[e.g.,][]{Davis&Hubeny2006:bhspec}. This narrower band-pass would result in underestimating soft fluxes in RXTE-like instrument or the hard-flux end of the disk emission in CCD-like instruments \citep[see][]{Done&Davis2008:meanstress}. The offset was quantified to be a fairly narrow distribution with a median of $\sim0.26\,$dex, for RXTE-PCA at 2\,keV.

\end{appendix}
			
\end{document}